\pdfoutput=1 

\documentclass[reprint, twocolumn, superscriptaddress,
               preprintnumbers, amsmath, amssymb, amsfonts]{revtex4-1}

\usepackage{graphicx, dcolumn, bm}
\usepackage{epstopdf}
\usepackage{color}
\usepackage{hyperref}

\usepackage{ulem}

\newcommand\redsout{\bgroup\markoverwith
      {\textcolor{red}{\rule[0.5ex]{2pt}{1.0pt}}}\ULon}
\newcommand\bluesout{\bgroup\markoverwith
      {\textcolor{blue}{\rule[0.5ex]{2pt}{1.0pt}}}\ULon}

\usepackage{multirow,fancybox,latexsym,float}

\begin{document}

\preprint{RBRC-1080}

\title{Neutral $B$ meson mixings and $B$ meson decay constants
with static heavy and domain-wall light quarks}

\newcommand{\KMI}{
  Kobayashi-Maskawa Institute for the Origin of Particle and the Universe (KMI),
  Nagoya University,
  Nagoya 464-8602,
  Japan}

\newcommand{\RBRC}{
  RIKEN BNL Research Center,
  Brookhaven National Laboratory,
  Upton, New York 11973,
  USA}

\newcommand{\BNL}{
  Physics Department,
  Brookhaven National Laboratory,
  Upton, New York 11973,
  USA}

\newcommand{\ADDRESS}{
  Addresses...
  }

\author{Yasumichi Aoki}
\affiliation{\KMI}

\author{Tomomi Ishikawa}
\affiliation{\RBRC}

\author{Taku Izubuchi}
\affiliation{\BNL}
\affiliation{\RBRC}

\author{Christoph Lehner}
\affiliation{\BNL}

\author{Amarjit Soni}
\affiliation{\BNL}


\date{June 27, 2015}

\begin{abstract}
Neutral $B$ meson mixing matrix elements and $B$ meson decay constants
are calculated.
Static approximation is used for $b$ quark and domain-wall fermion formalism is
employed for light quarks.
The calculations are carried out on $2+1$ flavor dynamical ensembles generated
by RBC/UKQCD Collaborations with lattice spacings $0.086$~fm
($a^{-1}\sim 2.3$~GeV) and $0.11$~fm ($1.7$~GeV),
and a fixed physical spatial volume of about $(2.7~{\rm fm})^3$.
In the static quark action, link-smearings are used
to improve the signal-to-noise ratio. 
We employ two kinds of link-smearings, HYP1 and HYP2, and
their results  are combined in taking the continuum limit.
For the matching between the lattice and the continuum theory,
one-loop perturbative $O(a)$ improvements are made
to reduce discretization errors.
As the most important quantity of this work,
we obtain SU(3) breaking ratio
$\xi=1.208(60)$,
where the error includes statistical and systematic one.
(Uncertainty from infinite $b$ quark mass is not included.)
We also find other neutral $B$ meson mixing quantities
$f_B\sqrt{\hat{B}_B}=240(22)$~MeV,
$f_{B_s}\sqrt{\hat{B}_{B_s}}=290(22)$~MeV,
$\hat{B}_B=1.17(22)$,
$\hat{B}_{B_s}=1.22(13)$ and
$B_{B_s}/B_B=1.028(74)$,
$B$ meson decay constants $f_B=219(17)$~MeV,
$f_{B_s}=264(19)$~MeV and $f_{B_s}/f_B=1.193(41)$,
in the static limit of $b$ quark,
which do not include infinite $b$ quark mass uncertainty.
\end{abstract}

\pacs{Valid PACS appear here}

\maketitle

\section{Introduction}

Standard Model (SM) of elementary particles is consistent with
all experimental data, so far.
The SM, however, does not still satisfy us, because it cannot
answer some of our basic questions, such as
the reason why the gauge group, constituents of particles and number of
generation in the model are chosen as they are,
hierarchical unnaturalness in mass scales between three generation of 
fermions, and so on.
While the existence of Higgs boson has been experimentally
confirmed at Large Hadron Collider (LHC),
expected new particles have not been discovered as yet.
Thus bottom-up approaches toward physics beyond Standard Model (BSM)
is becoming more and more important.
In order to address BSM, precision tests for SM are highly meaningful.
Combining theoretical predictions with the experimental results,
it would be possible to obtain hints for the BSM.
In such an attempt, the Cabibbo-Kobayashi-Maskawa (CKM) quark mixing
matrix elements~\cite{Kobayashi:1973fv} play a crucial role
to check the consistency of the SM.

In the SM, the transition of neutral $B$ ($B_s$) meson to its anti-meson
occurs via box diagrams involving exchange of
two $W$-bosons and this amplitude would provide a
clean determination for the matrix elements $V_{td}$ and $V_{ts}$
assuming $V_{tb}$ is known.
In the SM framework, dominant contribution to the mass difference
of the neutral $B$ meson mass eigenstates is related with
the CKM matrix elements by
\begin{eqnarray}
\Delta m_{B_q}=\frac{G_F^2m_W^2}{16\pi^2m_{B_q}}
\left|V_{tq}^{\ast}V_{tb}\right|^2
S_0\left(x_t\right)\eta_B\hat{\cal M}_{B_q},
\label{EQ:mass-defference}
\end{eqnarray}
where $q=\{d, s\}$.
In Eq.~(\ref{EQ:mass-defference}), both the Inami-Lim function
$S_0(x_t)$ ($x_t=m_t^2/m_W^2$)~\cite{Inami:1980fz} and
the QCD coefficient $\eta_B$ can be calculated perturbatively.
$\hat{\cal M}_{B_q}$ is a renormalization group invariant (RGI)
$\Delta B=2$ four-fermion operator matrix element
in an effective Hamiltonian of the box diagram at low-energy scale.
The mixing matrix element $\hat{\cal M}_{B_q}$ is
a highly nonperturbative quantity,
thus currently the only possible method for a precise determination is
via numerical lattice QCD simulations.
By taking a ratio~\cite{Bernard:1998dg} of Eq.~(\ref{EQ:mass-defference})
between $q=d$ and $s$, we obtain
\begin{eqnarray}
\left|\frac{V_{td}}{V_{ts}}\right|=
\xi\sqrt{\frac{\Delta m_B}{\Delta m_{B_s}}\frac{m_{B_s}}{m_B}},
\label{EQ:ratio_of_CKM}
\end{eqnarray}
where $\xi$ is called SU(3) breaking ratio
\begin{eqnarray}
\xi=\frac{m_B}{m_{B_s}}\sqrt{\frac{{\cal M}_{B_s}}{{\cal M}_B}}.
\label{EQ:xi}
\end{eqnarray}
The ratio constrains the apex of the CKM unitary triangle and
new quark-flavor-changing interactions from BSM would affect this quantity.
In the ratio many uncertainties get canceled and
precise determination of $\xi$ would lead to a tight constraint on the CKM
unitary triangle and hints for BSM physics
as inconsistency of the unitary triangle in the SM.

Lattice QCD simulations including $b$ quark are, however, quite challenging, 
because of the large scale difference between light quarks
($u$ and $d$) and $b$ quark.
While fine lattice spacings are needed to correctly treat the $b$ quark,
the large volume is required to accommodate pion dynamics.
Such a situation is difficult to achieve with the current computational ability.
Heavy Quark Effective Theory (HQET) provides one realistic solution
to this problem.
In this formulation, the heavy ($b$) quark dynamics is integrated out and
we may only treat the dynamics associated with light quarks.
The theory is described by systematic expansion of inverse of
heavy quark mass $m_Q$.
First attempt in this direction was carried out by Eichten and Hill
~\cite{Eichten:1989zv, Eichten:1989kb},
in which they used static approximation (leading order of heavy quark
mass expansion) and, for the static quark, they employed a standard
static action.
Soon after that attempt, however, it turned out that this approach leads
to a poor signal-to-noise ratio (S/N) in correlation functions,
because the static self-energy contains a notorious $1/a$ power divergence.
(On the other hand, in Non-Relativistic QCD (NRQCD),
another effective theory approach, the power divergence tends to be canceled
\cite{Lepage:1991ui}.)
This situation has been significantly improved
since ALPHA Collaboration introduced
link smearing technique in the static action, which partly cured
the difficulty~\cite{DellaMorte:2003mn, Della Morte:2005yc}.

In this paper, we calculate $B$ meson decay constants and
neutral $B$ meson mixing matrix elements using the static approximation.
The static approximation always has $O(\Lambda_{\rm QCD}/m_b)\sim O(10\%)$
uncertainty, since physical $b$ quark mass is not infinite.
For SU(3) breaking ratios like $\xi$ or
the ratio of $B$ meson decay constants, however,
the uncertainty coming from the static approximation
is down to around $2\%$ level.
This means the static limit could be
especially good approximation compared with other lattice approaches
that take into account $b$ quark mass dependence for such ratios.
To reduce the $O(\Lambda_{\rm QCD}/m_b)$ uncertainty in the HQET approach,
higher order operators in the $1/m_Q$ expansion need to be included.
Taking into account these contributions requires
nonperturbative matching with continuum using e.g.
Schr\"odinger functional scheme
with step scaling technique~\cite{Heitger:2003nj},
which requires considerable effort.
Instead, we stay in static limit assuming that the results can be valuable
for interpolation to physical $b$ quark mass combining with
lighter quark mass simulations,
for which high precision calculation is significantly important.
(We discuss the meaning of calculations at the static limit
in Sec.~\ref{SEC:static_limit}.)
This work is a first step toward the precise determination of
$B$ physics quantities in the static limit.

This paper is organized as follows.
In Sec.~\ref{SEC:static_limit}, we discuss the meaning of the calculations
at the static limit 
as an anchor point in the study of heavy quark physics.
In Sec.~\ref{SEC:physical_observables}, we summarize the physical
observables both in QCD full theory and HQET side,
which we address for the study of neutral $B$ meson mixing phenomena.
In Sec.~\ref{SEC:Lattice_actions}, the definition of lattice actions
and the gluon ensembles that we use in this study are explained.
In Sec.~\ref{SEC:matching}, we describe the matching procedure
between QCD full theory and HQET in continuum, as well as
between continuum and lattice in HQET.
The HQET matching is carried out by one-loop perturbation including
$O(a)$ lattice errors.
In Sec.~\ref{SEC:measurement}, details of the measurement, correlator
fits and formula for constructing physical quantities are shown.
In Sec.~\ref{SEC:chiral_extrapolation}, chiral and continuum extrapolation
formula (SU(2)$\chi$PT) are summarized and we show the fit results.
In Sec.~\ref{SEC:Systematic_errors}, we present the estimation of the
systematic uncertainties and summarize it in Tab.~\ref{TAB:error_budget}.
Finally, we present final results, compare them with other works
and discuss future direction of this project in Sec.~\ref{SEC:conclusions}.

\section{Static limit as a strong anchor point}
\label{SEC:static_limit}

We employ the static approximation as $b$ quark treatment in this study.
As discussed earlier, this approximation suffers from uncertainty of
$O(10\%)$ for primary quantities or $O(2\%)$ for flavor SU(3) breaking
ratios at the physical $b$ quark mass, which is heavy but finite.
The physical value of the approximation will eventually get
lost as one aims higher and higher precision.
The results at the static limit is, however, valuable
as an anchor point 
when combined with simulations in lower quark mass region.
In this section we clarify the meaning of our calculations at
the static limit.

We consider a heavy quark expansion of some heavy-light quantity
$\Phi_{\rm hl}$,
which has a finite asymptotic limit as $m_Q\rightarrow\infty$,
\begin{eqnarray}
\Phi_{\rm hl}(1/m_Q)=\Phi_{\rm hl}(0)
\exp\left[\sum_{p=1}^{\infty}\gamma_p
\left(\frac{\Lambda_{\rm QCD}}{m_Q}\right)^p\right],
\label{EQ:HQET_expansion_1}
\end{eqnarray}
where $m_Q$ is a heavy quark mass,
which is heavier than the QCD scale $\Lambda_{\rm QCD}$.
Equivalently, the expansion is written as
\begin{eqnarray}
&&\Phi_{\rm hl}(1/m_Q)=\Phi_{\rm hl}(1/m_{Q_A})\nonumber\\
&&\times
\exp\left[\sum_{p=1}^{\infty}\gamma_p
\left\{\left(\frac{\Lambda_{\rm QCD}}{m_Q}\right)^p
-\left(\frac{\Lambda_{\rm QCD}}{m_{Q_A}}\right)^p\right\}\right],
\label{EQ:HQET_expansion_2}
\end{eqnarray}
using some ``anchor'' point $m_{Q_A}$.
(In Eq.~\ref{EQ:HQET_expansion_1} the static limit
$m_Q\rightarrow\infty$ is regarded as an anchor point.)
Our task is to determine the expansion coefficients $\gamma_p$ and
the overall factor $\Phi_{\rm hl}(1/m_{Q_A})$
to reach a physical $b$ quark point.
There are several ways to the determination:
\begin{list}{}{}
\item[(i)]
HQET approach:
The anchor point is static limit $m_Q\rightarrow\infty$.
To treat the heavy quark expansion from the static limit,
the HQET is employed.
In addition to terms in the heavy quark action and operators
at the leading order of the expansion (static approximation),
those at $O(1/m_Q)$ are included.
To keep the theory renormalizable, the Boltzmann factor for the heavy quark
is expanded in $1/m_Q$, making operator insertions
in the expectation value evaluated with the static action.
The HQET must be matched with the original full theory.
An important point is that the matching beyond static approximation
cannot be carried out perturbatively,
because of the existence of $1/a$ power divergence in the HQET
~\cite{Maiani:1991az, Heitger:2003nj}.
\item[(ii)]
Relativistic approach:
The anchor point sits in lower mass region, typically $c$ quark mass
region.
The usual relativistic formulations can be applicable in that region,
while relatively finer lattices are required.
\item[(iii)]
Combination of (i) and (ii) above:
The anchor point is the static limit,
while $\gamma_p$s are explored by using usual relativistic formulations
in lower quark mass region, i.e., $c$ quark region.
(For example, Ref.~\cite{Bernard:1993zh}.)
\end{list}
Procedure (i) has been used by ALPHA Collaboration,
in which nonperturbative matching with QCD full theory can be implemented by
the step scaling strategy with Shcr\"{o}dinger functional scheme
~\cite{Heitger:2003nj}.
(See Ref.~\cite{Bernardoni:2014fva} for their recent achievements.)
In the procedure (ii),
relatively finer lattices with regular size of volume are required.
However, the lattices to treat $c$ quark are currently becoming
available and the approach (ii) is becoming feasible.
A recent sophisticated implementation in this direction is
``ratio method''~\cite{Blossier:2009hg}
by ETM Collaboration, which may be a viable option.
In this method, ratios of physical quantities at some heavy quark mass point
$m_Q$ and $m_Q/\lambda$ with a scale parameter $\lambda>1$,
are considered to separate
$\Phi_{\rm hl}(1/m_{Q_A})$ and $\gamma_p$s in the determination:
\begin{eqnarray}
\frac{\Phi_{\rm hl}(1/m_Q)}{\Phi_{\rm hl}(\lambda/m_Q)}
=\exp\left[\sum_{p=1}^{\infty}\gamma_p(1-\lambda^p)
\left(\frac{\Lambda_{\rm QCD}}{m_Q}\right)^p\right],
\end{eqnarray}
which enhances the precision of the $\gamma_p$s.
(See Ref.~\cite{Carrasco:2013zta} for their recent achievements.)
A combination of the ratio method and the static limit as an anchor point
would also be beneficial, which belongs to the category (iii).
In this sense, the static limit is not only of theoretical interest,
but also a valuable anchor point to explore physics at physical $b$ quark point.
The fact that ``the static limit is close to the physical $b$ quark mass
in terms of $1/m_Q$'' ensures usefulness of the static limit as
a ``strong'' anchor point.

\section{Physical observables}
\label{SEC:physical_observables}

\subsection{Observables in QCD full theory}
\label{SEC:observables_QCD_full}

Our main aim in this paper is to calculate the CKM matrix elements
$V_{td}$ and $V_{ts}$ to give constraints on the CKM unitary triangle.
The current accuracy of the mass difference (\ref{EQ:mass-defference})
from experiment is less than $1\%$, thus precise determination of
the hadronic matrix element ${\cal M}_{B_q}$ would give strong
constraints on the CKM matrix elements.
We here summarize current world average values related with
neutral $B$ meson mixing, 
which are quoted from Particle Data Group (PDG)~\cite{Beringer:1900zz}:
\begin{eqnarray}
m_b(\overline{\rm MS})
&=&4.18\pm0.03\;{\rm GeV},\\
m_{B^0}
&=&5279.58\pm0.17\;{\rm MeV},\\
m_{B_s^0}
&=&5366.77\pm0.24\;{\rm MeV},\\
\Delta m_{B^0}
&=&(0.510\pm0.004)\times10^{12}\;\hbar{\rm s}^{-1}\nonumber\\
&=&(3.337\pm0.033)\times10^{-10}\;{\rm MeV},\\
\Delta m_{B_s^0}
&=&(17.69\pm0.08)\times10^{12}\;\hbar{\rm s}^{-1}\nonumber\\
&=&(1.164\pm0.005)\times10^{-8}\;{\rm MeV}.
\end{eqnarray}
Thus, the ratio of the CKM matrix elements (\ref{EQ:ratio_of_CKM}) reads
\begin{eqnarray}
\left|\frac{V_{td}}{V_{ts}}\right|=
\xi\times(0.17071\pm0.00092),
\label{EQ:ratio_of_CKM_xi}
\end{eqnarray}
which indicates the determination of $\xi$ with high accuracy would yield
precise value of the ratio.

The $\Delta B=2$ mixing matrix element at a scale $\mu_b$
in the effective Hamiltonian is represented by
\begin{eqnarray}
{\cal M}_{B_q}(\mu_b)&=&
\langle\overline{B}_q^0|[\overline{b}\gamma_{\mu}(1-\gamma_5)q]
[\overline{b}\gamma_{\mu}(1-\gamma_5)q]|B_q^0\rangle_{\rm full}\nonumber\\
&\equiv&
\langle\overline{B}_q^0|O_L^{\rm full}|B_q^0\rangle_{\rm full},
\label{EQ:matrix_element_full}
\end{eqnarray}
where $b$ and $q$ represent $b$ quark and light ($d$ or $s$) quark fields,
respectively.
In Eq.~(\ref{EQ:matrix_element_full}), we put a superscript and a subscript
``full'' to indicate that the theory considered here is not HQET,
but rather QCD full theory.
In this paper, the standard PDG notation for the quark content of $B$ meson
is used; $B=(\overline{b}q)$ and $\overline{B}=(b\overline{q})$.
The matrix element is conventionally parametrized as
\begin{eqnarray}
{\cal M}_{B_q}(\mu_b)=\frac{8}{3}m_{B_q}^2f_{B_q}^2B_{B_q}(\mu_b),
\label{EQ:definition_B_parameter}
\end{eqnarray}
so that $B_{B_q}=1$ when vacuum saturation approximation (VSA) exactly holds,
where $B_{B_q}$ depicts a dimensionless hadronic $B$-parameter and $f_{B_q}$
denotes $B_q$ meson decay constant defined by
\begin{eqnarray}
if_{B_q}p_{\mu}&=&
\langle 0|\overline{b}\gamma_{\mu}\gamma_5q|B_q(p)\rangle_{\rm full}\nonumber\\
&\equiv&\langle 0|A_{\mu}^{\rm full}|B_q(p)\rangle_{\rm full},
\end{eqnarray}
with four-momentum of $B_q$ meson $p_{\mu}$.
An RGI definition of the $B$-parameters $\hat{B}_{B_q}$ is obtained from
$B$-parameters in some scheme and at some scale $\mu_b$ by:
\begin{eqnarray}
\hat{B}_{B_q}=[\alpha_s(\mu_b)]^{-\frac{\gamma_0}{2\beta_0}}
\left(1-\frac{\alpha_s(\mu_b)}{4\pi}Z_5\right)B_{B_q}(\mu_b),
\label{EQ:RGI_bag-parameter}
\end{eqnarray}
at next-to-leading order (NLO),
where $Z_5$ in naive dimensional regularization (NDR) with the modified
minimal subtraction ($\overline{\rm MS}$) scheme is written as
~\cite{Buras:1990fn}
\begin{eqnarray}
Z_{n_f}=\frac{\gamma^{(1)}}{2\beta_0}-\frac{\gamma^{(0)}\beta_1}{2\beta_0^2},
\end{eqnarray}
with
\begin{eqnarray}
\beta_0=11-\frac{2}{3}n_f,\;\;
\beta_1=102-\frac{38}{3}n_f,\\
\gamma^{(0)}=4,\;\;
\gamma^{(1)}=-7+\frac{4}{9}n_f.
\end{eqnarray}
In this study we use $\alpha_s(\mu_b)=0.2265$ obtained from strong coupling
at $Z$ boson mass scale
$\alpha_s({m_Z=91.1876(21)}~{\rm GeV})=0.1185(6)$~\cite{Beringer:1900zz}
using renormalization group (RG) evolution
(four-loop~\cite{Chetyrkin:1997dh, Vermaseren:1997fq})
with $n_f=5$.
Eq.~(\ref{EQ:RGI_bag-parameter}) thus becomes
$\hat{B}_{B_q}=1.516\times B_{B_q}(\mu_b)$.

One of the focal point of this paper is SU(3) breaking ratio (\ref{EQ:xi}),
which should be unity in the SU(3) light flavor symmetric case.
In this ratio most of the theoretical uncertainties
as well as statistical fluctuations are largely canceled out.
Using the parametrization of the matrix element
(\ref{EQ:definition_B_parameter}), the SU(3) breaking ratio is
represented as
\begin{eqnarray}
\xi=\frac{f_{B_s}}{f_B}\sqrt{\frac{B_{B_s}}{B_B}}.
\label{EQ:xi_from_b-para}
\end{eqnarray}
Because $B$-parameters are based on VSA by definition and
there is a suppression factor due to SU(3) light flavor symmetry,
the ratio of $B$-parameters in Eq.~(\ref{EQ:xi_from_b-para}) could be
close to one and a large fraction of the SU(3) breaking of $\xi$
likely to reside in the ratio of $B$ meson decay constants.

\subsection{Observables in static limit}

We regard the $b$ quark as a heavy quark and give it an on-shell velocity
$v_{\mu}=(1, 0, 0, 0)$, which leads to an on-shell momentum
$p_{\mu}=(m_b, 0, 0, 0)$.
Heavy quark field $h$ is introduced as a sum of heavy quark $h_+$ and
anti-heavy quark $h_-$:
\begin{eqnarray}
h=h_++h_-,\;\;\;
\overline{h}=\overline{h}_++\overline{h}_-=h_+^{\dagger}-h_-^{\dagger},
\end{eqnarray}
through
\begin{eqnarray}
h_{\pm}=e^{\mp im_bv\cdot x}\frac{1\pm\not{\!v}}{2}b
=e^{\mp im_bt}\frac{1\pm\gamma_0}{2}b,
\end{eqnarray}
where $b$ is a usual relativistic quark field.

In the static limit, the $B$ meson decay constant and the hadronic
matrix element behave like $f_{B_q}\propto 1/\sqrt{m_{B_q}}$ and
${\cal M}_{B_q}\propto m_{B_q}$, respectively.
Therefore it would be useful to introduce quantities:
\begin{eqnarray}
\Phi_{B_q}=\sqrt{m_{B_q}}f_{B_q},\;\;\;
M_{B_q}=\frac{{\cal M}_{B_q}}{m_{B_q}},
\label{EQ:Phi_M_HQET}
\end{eqnarray}
so that they behave as constants in the static limit.
Hadron states in the HQET are labeled by $v_{\mu}$ and a residual momentum
$k_{\mu}$, which satisfies $v\cdot k=0$.
They are defined in the static limit and differ
from those of the QCD full theory:
\begin{eqnarray}
|B_q\rangle_{\rm full}=\sqrt{m_{B_q}}\left\{|B_q\rangle_{\rm HQET}
+O(\Lambda_{\rm QCD}/m_b)\right\},
\end{eqnarray}
so that the HQET state normalization becomes
\begin{eqnarray}
\langle B_q(\vec{k})|B_q(\vec{k'})\rangle_{\rm HQET}=
2(2\pi)^3\delta^3(\vec{k}-\vec{k'}).
\end{eqnarray}
Using the HQET state, $\Phi_{B_q}$ in Eq.~(\ref{EQ:Phi_M_HQET}) is
simply written as
\begin{eqnarray}
\Phi_{B_q}&=&\langle 0|\overline{h}\gamma_0\gamma_5 q|B_q\rangle_{\rm HQET}
\nonumber\\
&\equiv&\langle 0|A_0^{\rm HQET}|B_q\rangle_{\rm HQET}
\label{EQ:def_Phi_B_HQET}
\end{eqnarray}
For $M_{B_q}$, we need two kinds of matrix element:
\begin{eqnarray}
M_L&=&\langle\overline{B}_q^0|[\overline{h}\gamma_{\mu}(1-\gamma_5)q]
[\overline{h}\gamma_{\mu}(1-\gamma_5)q]|B_q^0\rangle_{\rm HQET}\nonumber\\
&\equiv&\langle\overline{B}_q^0|O_L^{\rm HQET}|B_q^0\rangle_{\rm HQET},
\label{EQ:M_L}\\
M_S&=&\langle\overline{B}_q^0|[\overline{h}(1-\gamma_5)q]
[\overline{h}(1-\gamma_5)q]|B_q^0\rangle_{\rm HQET}\nonumber\\
&\equiv&\langle\overline{B}_q^0|O_S^{\rm HQET}|B_q^0\rangle_{\rm HQET},
\label{EQ:M_S}
\end{eqnarray}
owing to lack of four-dimensional Euclidean rotational symmetry
in the static limit,
where the $\Delta B=2$ four quark operator $O_L$ is decomposed into
spatial and time components:
\begin{eqnarray}
{\sum_{i=1,2,3}
[\overline{h}\gamma_i(1-\gamma_5)q][\overline{h}\gamma_i(1-\gamma_5)q],}&&\\
{[\overline{h}\gamma_0(1-\gamma_5)q][\overline{h}\gamma_0(1-\gamma_5)q],}&&
\end{eqnarray}
and they are renormalized differently.
As a consequence, operators (\ref{EQ:M_L}) and (\ref{EQ:M_S}) have mixings.
In the following, $B$ meson states $|B_q\rangle$ and operators represent
those in the static limit of $b$ quark unless stated otherwise.

\section{Lattice actions and gluon ensembles}
\label{SEC:Lattice_actions}

\subsection{Lattice action}

We perform lattice simulations in HQET side,
where lattice action comprises three pieces:
\begin{eqnarray}
S=S_{\rm static}+S_{\rm DWF}+S_{\rm gluon},
\end{eqnarray}
where $S_{\rm static}$ is the static quark action representing the heavy
($b$) quark,
$S_{\rm DWF}$ is the domain-wall fermion (DWF) action describing
the light ($u$, $d$, $s$) quarks and $S_{\rm gluon}$ is the gluon action.

\subsubsection{Standard static heavy quark action with link smearing}

The standard static quark action~\cite{Eichten:1989kb} is given by
\begin{eqnarray}
S_{\rm static}
&=&\sum_x\overline{h}(x)\nonumber\\
&&\times\left\{\frac{1+\gamma_0}{2}
\Bigl[h(x)-U_0^{\dag}(x-\hat{0})h(x-\hat{0})\Bigr]\right.\nonumber\\
&&\;\left.-\frac{1-\gamma_0}{2}
\Bigl[U_0(x)h(x+\hat{0})-h(x)\Bigr]\right\}.
\label{EQ:static_action}
\end{eqnarray}
The lattice derivatives used here are not symmetric for each heavy and
anti-heavy quark, thus fermion doublers do not arise.
The form of the action is technically the same as the Wilson quark action
with volume reduction into one dimension (time direction).
Therefore it has a Wilson term, which decouples from any low-energy physics
in the continuum limit and explicitly breaks the chiral symmetry
at finite lattice spacing.
This action suffers from huge $1/a$ power divergences, which results in
tremendous noise in correlators.
The solution to this problem is to introduce link smearing aiming at
a reduction of the power divergences~\cite{Della Morte:2005yc}.
The modification is simply to replace link variables $U_0(x)$ with
3-step hyper-cubic blocked~\cite{Hasenfratz:2001hp}
ones $V_0(x)$, which is defined by
\begin{eqnarray}
V_{\mu}(x)&=&{\rm Proj}_{\rm SU(3)}\biggl[
(1-\alpha_1)U_{\mu}(x)
\nonumber\\
&&\hspace*{-12mm}
+\frac{\alpha_1}{6}\sum_{\pm\nu\not=\mu}
\widetilde{V}_{\nu;\mu}(x)\widetilde{V}_{\mu;\nu}(x+\hat{\nu})
\widetilde{V}_{\nu;\mu}^{\dag}(x+\hat{\mu})\Biggr],\\
\widetilde{V}_{\mu;\nu}(x)&=&{\rm Proj}_{\rm SU(3)}\biggl[
(1-\alpha_2)U_{\mu}(x)
\nonumber\\
&&\hspace*{-12mm}
+\frac{\alpha_2}{4}\sum_{\pm\rho\not=\nu,\mu}
\overline{V}_{\rho;\nu\mu}(x)\overline{V}_{\mu;\rho\nu}(x+\hat{\rho})
\overline{V}_{\rho;\nu\mu}^{\dag}(x+\hat{\mu})\Biggr],\;\;\;\;\;
\end{eqnarray}
\begin{eqnarray}
\overline{V}_{\mu;\nu\rho}(x)&=&{\rm Proj}_{\rm SU(3)}\biggl[
(1-\alpha_3)U_{\mu}(x)
\nonumber\\
&&\hspace*{-12mm}
+\frac{\alpha_3}{2}\sum_{\pm\eta\not=\rho,\nu,\mu}
U_{\eta}(x)U_{\mu}(x+\hat{\eta})U_{\eta}^{\dag}(x+\hat{\mu})\Biggr],
\end{eqnarray}
where ${\rm Proj}_{\rm SU(3)}$ denotes an SU(3) projection and
$(\alpha_1, \alpha_2, \alpha_3)$ are hyper-cubic blocking
parameters ~\cite{Hasenfratz:2001hp}.
$(\alpha_1, \alpha_2, \alpha_3)=(0, 0, 0)$ corresponds to unsmeared link
($V_{\mu}=U_{\mu}$).
We use two parameter choices in this work:
\begin{eqnarray}
(\alpha_1, \alpha_2, \alpha_3)=\begin{cases}
(0.75, 0.6, 0.3) & :{\rm HYP1}
~\mbox{\cite{Hasenfratz:2001hp}} \\
(1.0 , 1.0, 0.5) & :{\rm HYP2}
~\mbox{\cite{Della Morte:2005yc}}.
                              \end{cases}
\end{eqnarray}

\subsubsection{Domain-wall fermion action}
\label{SEC:dwf_action}

The DWF action
~\cite{Kaplan:1992bt, Narayanan:1992wx, Shamir:1993zy}
is described by
\begin{eqnarray}
S_{\rm DWF}&=&\sum_{s, s'=1}^{L_s}\sum_{x, y}\overline{\psi}_{s}(x)
D_{ss'}^{\rm DWF}(x, y)\psi_{s'}(y)\nonumber\\
&&-\sum_{x}m_f\overline{q}(x)q(x),\\
D_{ss'}^{\rm DWF}(x, y)&=&D^4(x, y)\delta_{ss'}+D^5(s, s')\delta_{xy}
\nonumber\\
&&+(M_5-5)\delta_{ss'}\delta_{xy},\\
D^4(x, y)&=&\sum_{\mu}\frac{1}{2}
\left[(1-\gamma_{\mu})U_{\mu}(x)\delta_{x+\hat{\mu}, y}\right.\nonumber\\
&&\left.+(1+\gamma_{\mu})U_{\mu}^{\dagger}(y)\delta_{x-\hat{\mu}, y}\right],
\end{eqnarray}
\begin{eqnarray}
D^5(s,s')&=&\left\{
\begin{array}{ll}
 P_L\delta_{2, s'} & (s=1) \\
 P_L\delta_{s+1, s'} + P_R\delta_{s-1, s'} & (1<s<L_s) \\
 P_R\delta_{L_s-1, s'} & (s=L_s) \\
\end{array}
\right.,\;\;\;\;\;\;\;
\end{eqnarray}
where $\psi_s(x)$ are $4{+}1$-dimensional fermion fields.
The fifth dimension extends from $1$ to $L_s$ and is labeled by the
indices $s$ and $s'$.
The domain-wall height (fifth dimensional mass) $M_5$ is a parameter of
the theory which can be set between $0 < M_5 < 2$.
We use a setting of $M_5=1.8$.
The physical four-dimensional quark field $q(x)$ is constructed
from the fields $\psi_s(x)$ at $s=1$ and $L_s$:
\begin{eqnarray}
q(x)&=&P_L\psi_1(x)+P_R\psi_{L_s}(x),\\
\overline{q}(x)&=&\overline{\psi}_1(x)P_R+\overline{\psi}_{L_s}(x)P_L,
\end{eqnarray}
where $P_L$ and $P_R$ are left and right chirality projectors:
$P_L=(1-\gamma_5)/2$, $P_R=(1+\gamma_5)/2$.
In infinite $L_s$ limit, the right and left-handed modes are decoupled and
chiral symmetry is exactly restored.
The presence of the chiral symmetry plays a crucial role for reducing
unphysical operator mixing.
Note that the DWF is automatically $O(a)$ improved~\cite{Noaki:1999ij}.

\subsubsection{Gluon action}
\label{subsec:gluon} 

We consider a class of RG-improved gluon actions in this study:
\begin{equation}
S_{\rm gluon}=-\frac{2}{g^2_0}\left( (1-8c_1)\sum_P {\rm ReTr}[U_{P}]
+c_1\sum_R {\rm ReTr}[U_R]\right) , 
\end{equation}
where $g_{0}$ denotes the bare lattice coupling, $U_P$ and $U_R$ are
the path-ordered product of links along a $1\times 1$ plaquette $P$
and the path-ordered product of links along a $1\times 2$ rectangle $R$,
respectively.
Our choice of the parameter $c_1$ is $-0.331$ (Iwasaki gluon action)
~\cite{Iwasaki:1985we,Iwasaki:1983ck}.

\subsection{Gluon ensembles}

We use $2+1$ flavor dynamical DWF gluon configurations generated by
RBC and UKQCD Collaborations~\cite{Aoki:2010dy}.
A summary of the ensembles used in this work is listed
in Tab.~\ref{TAB:ensembles}.
\begin{table*}
\caption{\label{TAB:ensembles}
$2+1$ flavor dynamical domain-wall fermion ensembles
by RBC/UKQCD Collaborations.\cite{Aoki:2010dy}
Physical quark masses are obtained using SU(2)$\chi$PT
in the chiral extrapolation.
$m_{ud/s}^{\rm phys}=m_{l/h}^{\rm phys}+m_{\rm res}$($m_h^{\rm sim}$).
}
\begin{ruledtabular}
\begin{tabular}{cccccccccccc}
label & $\beta$ & $L^3\times T\times L_s$ & $a^{-1}$ & $a$ &
$aL$ & $m_{ud}^{\rm phys}$/$m_s^{\rm phys}$ & $m_{\rm res}$($m_h^{\rm phys}$) &
$m_{\rm res}$($m_h^{\rm sim}$)& $m_l/m_h$ & $m_{\pi}$($m_h^{\rm phys}$) &
$m_{\pi}L$ \\
     &          &                         & [GeV]    & [fm] &
[fm] &                                 &                                 &
                              &           & [MeV]                       &
            \\ \hline
24c1 & $2.13$ & $24^3\times64\times16$ & $1.729(25)$ & $0.114$ &
$2.74$ & $0.00134(4)$ & $0.003076(58)$ &
$0.003152(43)$ & $0.005/0.04$ & $327$ & $4.54$ \\
24c2 &        &                        &             & & &
\;\;\;\;\;\;\;\;/$0.0379(11)$ &        &                &  $0.01/0.04$ & $418$ & $4.79$ \\ \hline
32c1 & $2.25$ & $32^3\times64\times16$ & $2.280(28)$ & $0.0864$ &
$2.76$ & $0.00100(3)$ & $0.006643(82)$ & $0.0006664(76)$ &
$0.004/0.03$ & $289$ & $4.06$ \\
32c2 &        &                        &             & & &
\;\;\;\;\;\;/$0.0280(7)$ &        &                 & $0.006/0.03$ & $344$ & $4.83$ \\
32c3 &        &                        &             & & &
         &        &                 & $0.008/0.03$ & $393$ & $5.52$
\end{tabular}
\end{ruledtabular}
\end{table*}
Two lattice spacings $a\sim 0.114$ [fm] and $0.0864$ [fm] are used
to take a continuum limit.
We label the coarser and finer lattices as ``24c'' and ``32c'', respectively,
representing their lattice sizes.
The physical box size is set to be modest, which is around $2.75$ [fm].
The size of the fifth dimension is $L_s=16$ making
the chiral symmetry breaking quite small
with residual masses $m_{\rm res}\simeq0.003$ and $0.0007$
for 24c and 32c, respectively.
Degenerate $u$ and $d$ quark mass parameters are chosen so that
the simulation covers the pion mass range of $290$--$420$ [MeV].
The smallest value of $m_{\pi}L$ is $4.06$, which implies finite
volume effect would be small at simulation points.
Only one sea $s$ quark mass parameter is taken in our lattice ensemble
for both lattice spacings, which is larger than
the physical $s$ quark mass by a small amount.
As we will explain in Sec.~\ref{SEC:chiral_extrapolation},
we basically use SU(2) chiral perturbation theory fit functions
assuming sea $s$ quark mass sits on physical point,
while the actual sea $s$ quark mass in this simulation is not physical one.
The uncertainty from this inconsistency is estimated by
the partially quenched SU(3) chiral perturbation theory
as explained in Sec.~\ref{SEC:Systematic_errors} and turns out to be
less than $1\%$.
For a valence $s$ quark, we make measurements with two $s$ quark mass parameters
sandwiching the physical $s$ quark mass and make a linear interpolation.

\section{Matching}
\label{SEC:matching}

In this work, we adopt a two step matching: the first is a matching between
QCD full theory and HQET in the continuum, the second is a matching between
the continuum and the lattice in HQET.
The matching is carried out by one-loop perturbative calculation.
Here we summarize key points of the matching.
\begin{itemize}
\item The full theory operators in the continuum are renormalized in
$\overline{\rm MS}$(NDR) scheme at $\mu_b=m_b$, $b$ quark mass scale.
Fierz transformations in arbitrary dimensions are specified in the
naive dimensional regularization (NDR)
scheme by Buras and Weisz evanescent operators~\cite{Buras:1989xd}.
\item The HQET operators in the continuum are also renormalized in
$\overline{\rm MS}$(NDR) scheme at some scale $\mu$.
\item Matching operators between full theory and HQET in the continuum
is carried out by perturbatively calculating matrix elements
of the operators in both theories and comparing them.
\item The matching above is performed at scale $\mu=m_b$ to avoid 
a large logarithm of $\mu/m_b$.
We then use renormalization group running in the HQET to go down to a
lower scale.
\item The HQET operators with the lattice regularization are calculated
using a DWF formalism for light quarks to maintain good chiral
symmetry, which is important to control the operator mixing.
\item Matching HQET operators between continuum and lattice is perturbatively
carried out at a lattice cut-off scale $\mu=a^{-1}$,
where $a$ denotes a lattice spacing.
\item In the perturbative matching, we introduce a fictitious gluon mass
to regulate IR divergences.
The structure of the IR divergences should be the same between the continuum
and the lattice theories, otherwise they cannot be matched to each other.
\item In the matching of HQET operators between continuum and lattice,
$O(a)$ discretization errors are taken into account.
We employ on-shell $O(a)$ improvement program, in which we impose the
equation of motion on the external heavy and light quark lines.
In the improvement, we include both $O(pa)$ and $O(ma)$ contributions,
where $p$ and $m$ denote light quark momentum and mass, respectively.
\item The theory with static approximation of the heavy quark is renormalizable
and perturbative renormalization is justified;
though this is impossible once $O(1/m_Q)$ correction is included,
in which case non-perturbative subtraction of the $1/a$ power divergence
is necessary~\cite{Maiani:1991az, Heitger:2003nj}.
Inclusion of the $O(a)$ improvement operators does not alter the justification
of the perturbative treatment:
the $O(a)$ operators just bring $O(\alpha_s^{k+1})$ uncertainty at
$k$th-loop perturbation by mixing with $O(a^0)$ operators,
not causing destruction in taking a continuum limit.
\end{itemize}
In the following, the details are presented.

\subsection{Continuum matching}

In the continuum, the QCD full theory and HQET are renormalized
at a scale $\mu$, which we specify as a matching point.
The operator relation of heavy-light quark bilinear $J_{\Gamma}$
and $\Delta B=2$ four quark operator $O_L$ between the two theories
is written as
\begin{eqnarray}
J_{\Gamma}^{\rm full}(\mu)&=&
C_{\Gamma}(\mu)J_{\Gamma}^{\rm HQET}(\mu)+O(\Lambda_{\rm QCD}/m_b),\\
O_L^{\rm full}(\mu)&=&
Z_1(\mu)O_L^{\rm HQET}(\mu)+Z_2(\mu)O_S^{\rm HQET}(\mu)\nonumber\\
&&+O(\Lambda_{\rm QCD}/m_b).
\end{eqnarray}
The one-loop perturbative matching factor for heavy-light axial-vector
current is~\cite{Eichten:1989zv}
\begin{eqnarray}
C_{\gamma_0\gamma_5}(\mu)&=&
1+\left(\frac{g}{4\pi}\right)^2\frac{4}{3}
\left[-\frac{3}{2}\ln\left(\frac{\mu^2}{m_b^2}\right)-2\right].
\end{eqnarray}
For the four quark operator, the one-loop perturbative matching factors are
~\cite{Flynn:1990qz, Buchalla:1996ys}
\begin{eqnarray}
Z_1(\mu)&=&
1+\left(\frac{g}{4\pi}\right)^2
\left[-6\ln\left(\frac{\mu^2}{m_b^2}\right)-14\right],\\
Z_2(\mu)&=&
-8\left(\frac{g}{4\pi}\right)^2.
\end{eqnarray}
The numerical values of the matching factors at $\mu=m_b$ are presented
in Tab.~\ref{TAB:numerical_continuum_matching}.

\subsection{RG running in HQET}

To avoid large logarithm of $\mu/m_b$,
we match the theories at $\mu=m_b$ in the continuum matching
and use RG running to reach a smaller
energy scale $\mu$ in the HQET side.
The running is governed by the RG equation:
\begin{eqnarray}
\mu^2\frac{d}{d\mu^2}C_{\Gamma}(\mu)&=&
\frac{1}{2}C_{\Gamma}(\mu)\gamma_{\Gamma},
\label{EQ:RG_eq_bilinear}\\
\mu^2\frac{d}{d\mu^2}
\begin{bmatrix} Z_1(\mu) & Z_2(\mu) \end{bmatrix}&=&
\frac{1}{2}\begin{bmatrix} Z_1(\mu) & Z_2(\mu) \end{bmatrix}
\begin{bmatrix}
\gamma_{11} & \gamma_{12} \\ \gamma_{21} & \gamma_{22}
\end{bmatrix},\nonumber\\
\label{EQ:RG_eq_four_quark}
\end{eqnarray}
where $\gamma$'s are anomalous dimensions.
Solutions of the RG equations (\ref{EQ:RG_eq_bilinear}) and
(\ref{EQ:RG_eq_four_quark}) are generally written as:
\begin{eqnarray}
C_{\Gamma}(\mu)&=&C_{\Gamma}(\mu')U_{\Gamma}(\mu', \mu),\\
\begin{bmatrix} Z_1(\mu) & Z_2(\mu) \end{bmatrix}&=&
\begin{bmatrix} Z_1(\mu') & Z_2(\mu') \end{bmatrix}
{\bf U}_L(\mu', \mu),
\end{eqnarray}
where
\begin{eqnarray}
{\bf U}_L(\mu', \mu)=
\begin{bmatrix}
U_L^{(11)}(\mu', \mu) & U_L^{(12)}(\mu', \mu) \\
U_L^{(21)}(\mu', \mu) & U_L^{(22)}(\mu', \mu)
\end{bmatrix}.
\end{eqnarray}
Note that heavy-quark spin symmetry gives constraints on $\gamma$'s:
\begin{eqnarray}
\gamma_{12}=0, \;\;\gamma_{22}=\gamma_{11}+4\gamma_{21},
\end{eqnarray}
which turn into
\begin{eqnarray}
U_L^{(12)}(\mu', \mu)&=&0,\\
U_L^{(22)}(\mu', \mu)&=& U_L^{(11)}(\mu', \mu)+4U_L^{(21)}(\mu', \mu).
\end{eqnarray}
Each $U$'s are expressed as
\begin{eqnarray}
U_{\Gamma}(\mu', \mu)&=&
\left(1+\frac{\alpha_s(\mu)-\alpha_s(\mu')}{4\pi}J_{\Gamma}\right)
\left[\frac{\alpha_s(\mu')}{\alpha_s(\mu)}\right]^{d_{\Gamma}}\nonumber\\
&&+O(\alpha_s^2),
\label{EQ:RG-U_Gamma}\\
U_L^{(11)}(\mu', \mu)&=&
\left(1+\frac{\alpha_s(\mu)-\alpha_s(\mu')}{4\pi}J_1\right)
\left[\frac{\alpha_s(\mu')}{\alpha_s(\mu)}\right]^{d_1}\nonumber\\
&&+O(\alpha_s^2),
\label{EQ:RG-U_11}\\
U_L^{(21)}(\mu', \mu)&=&
-\frac{1}{4}\left(
\left[\frac{\alpha_s(\mu')}{\alpha_s(\mu)}\right]^{d_1}
-\left[\frac{\alpha_s(\mu')}{\alpha_s(\mu)}\right]^{d_2}\right)\nonumber\\
&&+O(\alpha_s),
\label{EQ:RG-U_21}\\
U_L^{(22)}(\mu', \mu)&=&
\left[\frac{\alpha_s(\mu')}{\alpha_s(\mu)}\right]^{d_2}+O(\alpha_s),
\label{EQ:RG-U_22}
\end{eqnarray}
where $\alpha_s=g^2/(4\pi)$.
In the one-loop matching, two-loop calculations of the anomalous dimensions
and beta-function are required for obtaining $J_{\Gamma}$, $d_{\Gamma}$,
$J_1$, $d_1$ and $d_2$ in Eqs.~(\ref{EQ:RG-U_Gamma})--(\ref{EQ:RG-U_22}).
The two-loop anomalous dimensions were calculated in
Refs.~\cite{Ji:1991pr, Broadhurst:1991fz} for quark bilinears and
in Refs.~\cite{Gimenez:1992is, Ciuchini:1996sr, Buchalla:1996ys} for
four-quark operators.

Because we include sea quarks only for $u$, $d$ and $s$
in our simulations ($N_f=2+1$) and our lattice cutoff scale is higher than
$c$ quark mass, we employ a two-step RG running to reach a scale $\mu=a^{-1}$:
making running from $\mu=m_b$ to $m_c$ scale using $N_f=4$ theory and
running back to $a^{-1}$ scale using $N_f=3$ theory, such as
\begin{eqnarray}
U_{\Gamma}(m_b, a^{-1})&=&
U_{\Gamma}^{N_f=4}(m_b, m_c)U_{\Gamma}^{N_f=3}(m_c, a^{-1}),\;\;\\
{\bf U}_L(m_b, a^{-1})&=&
{\bf U}_L^{N_f=4}(m_b, m_c){\bf U}_L^{N_f=3}(m_c, a^{-1}),\;\;
\end{eqnarray}
in which
\begin{eqnarray}
d_{\Gamma}^{N_f=4}=-\frac{6}{25},\;\; d_1^{N_f=4}&=&-\frac{12}{25},\;\;
d_2^{N_f=4}=-\frac{4}{25},\\
J_{\Gamma}^{N_f=4}=0.910,&&J_1^{N_f=4}=1.864,\\
d_{\Gamma}^{N_f=3}=-\frac{2}{9},\;\; d_1^{N_f=3}&=&-\frac{4}{9},\;\;
d_2^{N_f=3}=-\frac{4}{27},\\
J_{\Gamma}^{N_f=3}=0.755,&&J_1^{N_f=3}=1.698.
\end{eqnarray}
The RG-running coefficients are summarized
in Tab.~\ref{TAB:numerical_continuum_matching}.
\begin{table}[t]
\caption{
\label{TAB:numerical_continuum_matching}
Numerical values of the one-loop continuum matching factors and
RG-running coefficients.~\cite{Ishikawa:2011dd}
}
\begin{ruledtabular}
\begin{tabular}{c|cc}
& $24$c ($1.73$ GeV) & $32$c ($2.28$ GeV)\\ \hline
$\alpha_s(m_b=4.18~{\rm GeV}\mbox{\cite{Beringer:1900zz}})$ &
\multicolumn{2}{c}{$0.2261$} \\
$\alpha_s(m_c=1.275~{\rm GeV}\mbox{\cite{Beringer:1900zz}})$ &
\multicolumn{2}{c}{$0.3908$} \\
$\alpha_s(a^{-1})$ & $0.3204$ & $0.2773$ \\ \hline
$C_{\gamma_0\gamma_5}(m_b)$ & \multicolumn{2}{c}{$0.9520$} \\
$U_{\Gamma}^{N_f=4}(m_b, m_c)$ & \multicolumn{2}{c}{$1.1550$} \\
$U_{\Gamma}^{N_f=3}(m_c, a^{-1})$ & $0.9521$ & $0.9196$ \\ \hline 
$Z_1(m_b)$ & \multicolumn{2}{c}{$0.7483$} \\
$Z_2(m_b)$ & \multicolumn{2}{c}{$-0.1439$} \\
$U_L^{(11)N_f=4}(m_b, m_c)$ & \multicolumn{2}{c}{$1.3345$} \\
$U_L^{(21)N_f=4}(m_b, m_c)$ & \multicolumn{2}{c}{$-0.0526$} \\
$U_L^{(22)N_f=4}(m_b, m_c)$ & \multicolumn{2}{c}{$1.0921$} \\
$U_L^{(11)N_f=3}(m_c, a^{-1})$ & $0.9055$ & $0.8442$ \\
$U_L^{(21)N_f=3}(m_c, a^{-1})$ & $0.0141$ & $0.0231$ \\ 
$U_L^{(22)N_f=3}(m_c, a^{-1})$ & $0.9706$ & $0.9500$
\end{tabular}
\end{ruledtabular}
\end{table}

\subsection{Static effective theory matching}

The matching of the static effective theory between continuum and lattice
is carried out at a scale $\mu=a^{-1}$ using one-loop perturbation.
In the matching, lattice discretization errors are taken into account
up to $O(pa)$ and $O(m_qa)$, where $p$ and $m_q$ are
typical light quark momentum
and light quark mass, respectively.
To include these discretization errors, higher dimensional operators
need to be added in the matching.
The operator mixing pattern is constrained by symmetries, typically,
chiral symmetry, heavy quark spin symmetry,
and discrete symmetries such as ${\cal P}$, ${\cal T}$ and ${\cal C}$.

The operator relation for quark bilinear is written as:
\begin{eqnarray}
J_{\Gamma}^{\rm cont}&=&Z_{\Gamma}J_{\Gamma}^{\rm imp},
\label{EQ:static_matching_general_OGamma}
\end{eqnarray}
where $J_{\Gamma}^{\rm imp}$ is $O(a)$ improved lattice bilinear:
\begin{eqnarray}
J_{\Gamma}^{\rm imp}=
J_{\Gamma}+ac_{\Gamma}^{(pa)}GJ_{\Gamma D}
+ac_{\Gamma}^{(ma)}GJ_{\Gamma M},
\label{EQ:improved_bilinear}
\end{eqnarray}
in which
\begin{eqnarray}
J_{\Gamma D}=
\overline{h}\Gamma
(\mbox{\boldmath$\gamma$}\cdot\overrightarrow{\mbox{\boldmath$D$}})q,
\;
J_{\Gamma M}=m_q\overline{h}\Gamma q,
\end{eqnarray}
and $G$ is defined by $\gamma_0\Gamma\gamma_0=G\Gamma$.
For four-quark operators:
\begin{eqnarray}
O_L^{\rm cont}&=&Z_L O_L^{\rm imp},
\label{EQ:static_matching_general_OL}\\
O_S^{\rm cont}&=&Z_S O_S^{\rm imp},
\label{EQ:static_matching_general_OS}
\end{eqnarray}
where $O_L^{\rm imp}$ and $O_S^{\rm imp}$ are $O(a)$ improved lattice
operators:
\begin{eqnarray}
O_L^{\rm imp}&=&O_L+ac_L^{(pa)}(O_{ND}+2O_{ND}')\nonumber\\
&&+ac_L^{(ma)}(O_{NM}+2O_{NM}'),
\label{EQ:improved_OL}\\
O_S^{\rm imp}&=&O_S+ac_S^{(pa)}(O_{ND}-2O_{ND}')\nonumber\\
&&+ac_S^{(ma)}(O_{NM}-2O_{NM}'),
\label{EQ:improved_OS}
\end{eqnarray}
with
\begin{eqnarray}
O_{ND}&=&
2[\overline{h}\gamma_{\mu}^R(\bm{\gamma}\cdot\overrightarrow{\bm{D}})q]
[\overline{h}\gamma_{\mu}^Lq],\\
O_{ND}'&=&
2[\overline{h}P_R(\bm{\gamma}\cdot\overrightarrow{\bm{D}})q]
[\overline{h}P_Lq],\\
O_{NM}&=&
2m_q[\overline{h}\gamma_{\mu}^Rq][\overline{h}\gamma_{\mu}^Lq],\\
O_{NM}'&=&
2m_q[\overline{h}P_Rq][\overline{h}P_Lq].
\end{eqnarray}
We note that the coefficients for the quark bilinear operator do not
depend on $\Gamma$, which is a consequence of chiral and heavy quark spin
symmetry~\cite{Ishikawa:2011dd, Becirevic:2003hd, Blossier:2007hg} and
this fact holds nonperturbatively.
(For the four-quark operators, it is claimed that more higher order
operators are required in Eqs.~(\ref{EQ:improved_OL}) and (\ref{EQ:improved_OS})
for the $O(a)$ improvement at higher loop or non-perturbative level
\cite{Papinutto:2013cra}.)

For the one-loop calculation of coefficients in
Eqs.~(\ref{EQ:static_matching_general_OGamma}),
(\ref{EQ:static_matching_general_OL}) and (\ref{EQ:static_matching_general_OS}),
we use mean-field (MF) improvement to remove huge tad-pole
contribution on the lattice perturbation~\cite{Lepage:1992xa}.
Measured plaquette value $P$ or $u_0=P^{1/4}$ comes into
the matching for the MF improvement.

We employ DWF as light quarks, thus the physical light quark propagator is
written as
\begin{eqnarray}
S_q(p)&=&\langle q(-p)\overline{q}(p)\rangle\nonumber\\
&=&\frac{1-w_0^2}{i\not{\!p}+(1-w_0^2)m_f}\left(1+O(p^2, pm_f, m_f^2)\right),
\;\;
\end{eqnarray}
where $w_0=1-M_5$.
The physical quark propagator suggests that the quark wave function has
a domain-wall specific factor $(1-w_0^2)^{1/2}$ and the quark mass
should be identified by $m_q=(1-w_0^2)m_f$, which would appear in the matching
coefficients.

The matching coefficients at one-loop level are
\begin{eqnarray}
Z_{\Gamma}&=&
{\cal Z}_w^{-1/2}
\left\{1+\left(\frac{g_{\overline{\rm MS}}}{4\pi}\right)^2\frac{4}{3}
\hat{z}_{\Gamma}^{\rm MF}\right\}+O(g^4),\\
c_{\Gamma}^{(pa)}&=&
\frac{1}{u_0}\left(\frac{g_{\overline{\rm MS}}}{4\pi}\right)^2\frac{4}{3}
\hat{z}_{\Gamma}^{(pa)\rm MF}+O(g^4),\\
c_{\Gamma}^{(ma)}&=&
\frac{1}{u_0}\left(\frac{g_{\overline{\rm MS}}}{4\pi}\right)^2\frac{4}{3}
\hat{z}_{\Gamma}^{(ma)\rm MF}+O(g^4),
\end{eqnarray}
\begin{eqnarray}
Z_L&=&
{\cal Z}_w^{-1}
\left\{1+\left(\frac{g_{\overline{\rm MS}}}{4\pi}\right)^2\frac{4}{3}
\hat{z}_L^{\rm MF}\right\}+O(g^4),\\
c_L^{(pa)}&=&
\frac{1}{u_0}
\left(\frac{g_{\overline{\rm MS}}}{4\pi}\right)^2\frac{4}{3}
\hat{z}_L^{(pa)\rm MF}+O(g^4),\\
c_L^{(ma)}&=&
\frac{1}{u_0}
\left(\frac{g_{\overline{\rm MS}}}{4\pi}\right)^2\frac{4}{3}
\hat{z}_L^{(ma)\rm MF}+O(g^4),\\
Z_S&=&
{\cal Z}_w^{-1}+O(g^2),
\label{EQ:Z_S}\\
c_S^{(pa)}&=&O(g^2),\\
c_S^{(ma)}&=&O(g^2),
\end{eqnarray}
where
\begin{eqnarray}
{\cal Z}_w=
\frac{1-(w_0^{\rm MF})^2}{u_0}
\left(1+\left(\frac{g_{\overline{\rm MS}}}{4\pi}\right)^2\frac{4}{3}
\hat{z}_w^{\rm MF}\right)+O(g^4),\;\;\;
\end{eqnarray}
and the renormalized coupling in the continuum $\overline{\rm MS}$ scheme
$g_{\overline{\rm MS}}$ at scale $\mu=a^{-1}$ is related to the bare
lattice coupling $g_0$ as:
\begin{eqnarray}
\frac{1}{g_{\overline{\rm MS}}^2}=\frac{P}{g_0^2}+d_g+c_p+N_fd_f,
\label{EQ:MF-coupling}
\end{eqnarray}
in which $d_g$ and $c_p$ are dependent on the gluon action and $d_f$ is
dependent on the fermion action.
Note that the continuum matching coefficient for $O_S$ is already $O(g^2)$,
therefore only tree-level static matching coefficient for this operator
is needed in the one-loop matching procedure.
Nevertheless, we include partly the $O(g^2)$ in Eq.~(\ref{EQ:Z_S}) to keep
the same form of ${\cal Z}_w$ as that for $Z_L$, which does not matter
at the one-loop level.
The coefficients for this simulation are summarized
in Tab.~\ref{TAB:numerical_HQET_matching}.
\begin{table}[t]
\caption{
\label{TAB:numerical_HQET_matching}
Numerical values of the one-loop static effective theory matching factors.
~\cite{Ishikawa:2011dd}
}
\begin{ruledtabular}
\begin{tabular}{c|cccc}
\multirow{2}{5mm}{} &
\multicolumn{2}{c}{$24$c} & \multicolumn{2}{c}{$32$c} \\
 & HYP1 & HYP2 & HYP1 & HYP2 \\
\hline
$P$ \mbox{(chiral limit)}&
\multicolumn{2}{c}{$0.5883$} & \multicolumn{2}{c}{$0.6156$} \\
$M_5^{\rm MF}$ &
\multicolumn{2}{c}{$1.3032$} & \multicolumn{2}{c}{$1.3432$} \\
$g_{\rm MS}^2/4\pi$ &
\multicolumn{2}{c}{$0.1769$} & \multicolumn{2}{c}{$0.1683$} \\
$Z_{\Gamma=\gamma_0\gamma_5}$ &
$0.9105$ & $0.9383$ & $0.9256$ & $0.9526$ \\
$c_{\Gamma=\gamma_0\gamma_5}^{(pa)}$ &
$0.0790$ & $0.1374$ & $0.0744$ & $0.1294$ \\
$c_{\Gamma=\gamma_0\gamma_5}^{(ma)}$ &
$0.0864$ & $0.1660$ & $0.0739$ & $0.1482$ \\
$Z_L$ &
$0.8260$ & $0.8911$ & $0.8546$ & $0.9187$ \\
$c_L^{(pa)}$ &
$0.1185$ & $0.2061$ & $0.1117$ & $0.1942$ \\
$c_L^{(ma)}$ &
$0.1296$ & $0.2489$ & $0.1108$ & $0.2222$ \\
$Z_S$ &
\multicolumn{2}{c}{$0.9645$} & \multicolumn{2}{c}{$1.0040$} \\
\end{tabular}
\end{ruledtabular}
\end{table}

\section{Measurement and data extraction}
\label{SEC:measurement}

In this section, we present details of measurements on the gluon
configurations introduced in Sec.~\ref{SEC:Lattice_actions}.

\subsection{Correlators}

In the static limit, energies of states do not depend on their momentum.
This fact requires special treatment of correlators,
because even in the large separation of source and sink positions
in time $t$,
unique ground state cannot be obtained~\cite{Christ:2007cn, Albertus:2010nm}.
Especially, the Gaussian source and sink smearing used in this work
requires taking into account this feature.
In this subsection, we follow discussions
in Refs.~\cite{Christ:2007cn, Albertus:2010nm} and explicitly show
an extension to any form of source and sink smearing function.

We start with defining our state convention.
Static action (\ref{EQ:static_action}) is invariant under spatial local
phase rotation of heavy quark fields:
\begin{eqnarray}
h(\vec{x}, t)\longrightarrow e^{i\theta(\vec{x})}h(\vec{x}, t),\\
\overline{h}(\vec{x}, t)\longrightarrow
e^{-i\theta(\vec{x})}\overline{h}(\vec{x}, t),
\end{eqnarray}
which leads to Noether's current:
\begin{eqnarray}
J_h(\vec{x}, t)=\overline{h}(\vec{x}, t)h(\vec{x}, t),
\end{eqnarray}
with a conservation law:
\begin{eqnarray}
\partial_0J_h(\vec{x}, t)=0,
\end{eqnarray}
indicating time-independent charge (heavy quark number density operator)
at each spatial point:
\begin{eqnarray}
N_h(\vec{x})=J_h(\vec{x}, t),
\end{eqnarray}
which commutes with the Hamiltonian.
We can define $B$ meson states in the PDG notation,
$B=(\bar{b}q)$ and $\bar{B}=(b\bar{q})$,
as eigenstates of $N_h(\vec{x})$,
\begin{eqnarray}
N_h(\vec{y})|\tilde{B}(\vec{x})\rangle_L
&=&-\delta_{\vec{x}, \vec{y}}^{(3)}|\tilde{B}(\vec{x})\rangle_L,
\label{EQ:new_eigenstates}\\
\langle\tilde{B}(\vec{x})|\tilde{B}(\vec{y})\rangle_L
&=&\delta_{\vec{x}, \vec{y}}^{(3)},
\label{EQ:new_eigenstates_norm}
\end{eqnarray}
where ``$L$'' indicates states in the static limit
with finite spatial size $L$.
Using these, $B$ meson states with spacial momentum $\vec{p}$
are defined as
\begin{eqnarray}
|B(\vec{p})\rangle_L&=&
\sqrt{2a^3}\sum_{\vec{x}} e^{-i\vec{p}\cdot\vec{x}}|\tilde{B}(\vec{x})\rangle_L,
\label{EQ:momentum_states}
\end{eqnarray}
where momentum $\vec{p}$ takes discrete values:
\begin{eqnarray}
\vec{p}=\frac{2\pi}{La}(n_1, n_2, n_3),\;\;\;
0 < n_1, n_2, n_3\leq L.
\end{eqnarray}
This state convention gives a normalization
\begin{eqnarray}
\langle B(\vec{p})|B(\vec{q})\rangle_L
&=&
2(La)^3\delta_{\vec{p}, \vec{q}}^{(3)}\nonumber\\
&\xrightarrow[La\rightarrow\infty]{}&
2(2\pi)^3\delta^{(3)}(\vec{p}-\vec{q}),
\end{eqnarray}
which leads to a relation between finite and infinite volume momentum
eigenstates
\begin{eqnarray}
|B(\vec{p})\rangle_L&\xrightarrow[La\rightarrow\infty]{}&
|B(\vec{p})\rangle,\\
\langle B(\vec{p})|B(\vec{q})\rangle
&=&2(2\pi)^3\delta^{(3)}(\vec{p}-\vec{q}),
\label{EQ:state_normalization_infinite_voluum}
\end{eqnarray}
so that infinite volume static states $|B(\vec{p})\rangle$ give
a conventional normalization (\ref{EQ:state_normalization_infinite_voluum}).
Thus what we need to calculate in the finite volume are
\begin{eqnarray}
\langle 0|A_0(\vec{0}, 0)|B(\vec{p}=0)\rangle_L
&\xrightarrow[La\rightarrow\infty]{}&
\Phi_B,\\
\langle B(\vec{p}=0)|O_{\rm 4q}(\vec{0}, 0)|B(\vec{p}=0)\rangle_L
&\xrightarrow[La\rightarrow\infty]{}&
M_B,
\end{eqnarray}
where $A_0(\vec{x}, t)$ and $O_{\rm 4q}(\vec{x}, t)$ are
local heavy-light axial-vector current (in time direction)
and four-quark operators defined in Eqs.~(\ref{EQ:def_Phi_B_HQET}),
(\ref{EQ:M_L}) and (\ref{EQ:M_S}).
The statement mentioned earlier in this subsection
that the $B$ meson energy does not depend on its momentum
is understandable as the $B$ meson states defined
in Eq.~(\ref{EQ:new_eigenstates})
are also energy eigenstates, where the energy does not depend on
spatial coordinates due to translational invariance of the system,
and the energy, as a consequence, is independent of momentum
by Eq.~(\ref{EQ:momentum_states}).
This property requires unfamiliar treatment on correlators.
A typical example is an operator which includes spatially smeared quark field:
\begin{eqnarray}
A_0^S(\vec{x}, t)&=&
\left(\sum_{\vec{y}}f(\vec{y})\overline{h}(\vec{x}+\vec{y}, t)\right)
\gamma_0\gamma_5\nonumber\\
&&\cdot\left(\sum_{\vec{z}}g(\vec{z})q(\vec{x}+\vec{z}, t)\right),
\label{EQ:smeared_operator}
\end{eqnarray}
where $f(\vec{y})$ and $g(\vec{z})$ are smearing functions,
such as Gaussian and wall-type.
Consider $B$ meson decay amplitude with the smeared operator and take a large
$t$ limit:
\begin{eqnarray}
&&\langle 0|A_0^S(\vec{x}, t)|B(\vec{p})\rangle_L\nonumber\\
&\xrightarrow[t\gg 0]&
e^{i\vec{p}\cdot\vec{x}}e^{-E_0t}
\langle 0|A_0^S(\vec{0}, 0)|B(\vec{p})\rangle_L\nonumber\\
&\not=&
\delta_{\vec{p}, \vec{0}}^{(3)}~e^{-E_0t}
\langle 0|A_0^S(\vec{0}, 0)|B(\vec{p}=0)\rangle_L,
\end{eqnarray}
where $E_0$ represents an energy of $B$ meson ground state.
Thus we cannot obtain unique zero-momentum state
even in the large $t$ limit,
because $B$ meson energy does not depend on spatial momentum $\vec{p}$
any more.
This fact causes unusual derivation of matrix elements.
Let us demonstrate it here.
We consider three-point function with smeared quark fields
\begin{eqnarray}
{\cal C}_{\rm 4q}^{SS}(t_f, t, 0)&=&a^3\sum_{\vec{x}}
\langle A_0^S(\vec{0}, t_f)O_{\rm 4q}(\vec{x}, t)
A_0^{S\dagger}(\vec{0}, 0)\rangle.\;\;\;\;\;\;\;
\end{eqnarray}
Using completeness of states:
\begin{eqnarray}
{\bm 1}&=&\frac{1}{2(La)^3}\sum_{\vec{p}}
|B(\vec{p})\rangle_L\langle B(\vec{p})|
+(\mbox{higher states}),\;\;\;\;
\label{EQ:completeness_of_states}
\end{eqnarray}
the three-point function becomes
\begin{eqnarray}
&&{\cal C}_{\rm 4q}^{SS}(t_f, t, 0)
\xrightarrow[t_f\gg t\gg 0]\nonumber\\
&&\frac{1}{4(La)^3}\sum_{\vec{p}}e^{-E_0t_f}
\langle 0|A_0^S(\vec{0}, 0)|B(\vec{p})\rangle_L\nonumber\\
&&\cdot\langle B(\vec{p})|O_{\rm 4q}(\vec{0}, 0)
|B(\vec{p})\rangle_L
\langle B(\vec{p})|A_0^{S\dagger}(\vec{0}, 0)|0\rangle_L\nonumber\\
&=&
\frac{1}{2}{\cal C}^{SS}(t_f, 0)M_B,
\label{EQ:derivation_of_matrix_elements}
\end{eqnarray}
where
\begin{eqnarray}
{\cal C}^{SS}(t, 0)
&=&
\langle A_0^S(\vec{0}, t)A_0^{S\dagger}(\vec{0}, 0)\rangle\nonumber\\
&\xrightarrow[t\gg 0]{}&
\frac{1}{2(La)^3}e^{-E_0t}\sum_{\vec{p}}
|\langle 0|A_0^S(\vec{0}, 0)|B(\vec{p})\rangle_L|^2\nonumber\\
&=&
{\cal A}^{SS}e^{-E_0t},
\end{eqnarray}
and we used
\begin{eqnarray}
&&
\langle B(\vec{p})|O_{\rm 4q}(\vec{0}, 0)|B(\vec{p})\rangle_L\nonumber\\
&=&
\langle B(\vec{p}=0)|O_{\rm 4q}(\vec{0}, 0)|B(\vec{p}=0)\rangle_L,
\end{eqnarray}
following Eq.~(\ref{EQ:momentum_states}).
As seen in Eq.~(\ref{EQ:derivation_of_matrix_elements}),
we inevitably have to use ${\cal C}^{SS}(t, 0)$, in which sink position is not
spatially volume summed, which results in large statistical noise.
The matrix element $M_B$ is then obtained as:
\begin{eqnarray}
{\cal C}_{\rm 4q}^{SS}(t_f, t, 0)&\xrightarrow[t_f\gg t\gg 0]{}&
{\cal A}_{\rm 4q},\\
M_B&=&
\frac{2{\cal A}_{\rm 4q}}{{\cal A}^{SS}e^{-E_0t_f}}.
\label{EQ:M_B}
\end{eqnarray}
To obtain zero-momentum state in two-point functions,
we have to use a projection by spatial volume summation
of sink operator.
What we need to measure for $\Phi_B $ are two-point correlation functions:
\begin{eqnarray}
{\cal C}^{\tilde{L}S}(t, 0)
&=&
a^3\sum_{\vec{x}}\langle A_0(\vec{x}, t)A_0^{S\dagger}(\vec{0}, 0)\rangle,\\
{\cal C}^{\tilde{S}S}(t, 0)
&=&
a^3\sum_{\vec{x}}\langle A_0^S(\vec{x}, t)A_0^{S\dagger}(\vec{0}, 0)\rangle,
\end{eqnarray}
in which sink operators are volume summed to project into the
zero-momentum state, otherwise we cannot obtain the unique state by just
taking the large $t$ limit.
By inserting completeness of states (\ref{EQ:completeness_of_states}),
these two point correlation functions in $t\gg0$ can be easily written as
\begin{eqnarray}
{\cal C}^{\tilde{L}S}(t, 0)
&\xrightarrow[t\gg 0]{}&
\frac{1}{2}\langle 0|A_0(\vec{0}, 0)
|B(\vec{p}=0)\rangle_L\nonumber\\
&&\times\langle B(\vec{p}=0)|A_0^S(\vec{0}, 0)|0\rangle_L
e^{-E_0t}\nonumber\\
&=&
{\cal A}^{\tilde{L}S}e^{-E_0t},\\
{\cal C}^{\tilde{S}S}(t, 0)
&\xrightarrow[t\gg 0]{}&
\frac{1}{2}|\langle 0|A_0^S(\vec{0}, 0)
|B(\vec{p}=0)\rangle_L|^2 e^{-E_0t}\nonumber\\
&=&
{\cal A}^{\tilde{S}S}e^{-E_0t}.
\end{eqnarray}
$\Phi_B$ is then obtained through
\begin{eqnarray}
\Phi_B
\xrightarrow[t\gg 0]{}
\sqrt{2}\frac{{\cal C}^{\tilde{L}S}(t, 0)}
{\sqrt{{\cal C}^{\tilde{S}S}(t, 0)e^{-E_0t}}}
=\sqrt{2}\frac{{\cal A}^{\tilde{L}S}}{\sqrt{{\cal A}^{\tilde{S}S}}},
\label{EQ:Phi_B}
\end{eqnarray}
in which noisy correlator ${\cal C}^{SS}(t, 0)$ is not needed
in contrast to $M_B$.
In the actual simulation, we use $O(a)$ improved operators
to remove $O(a)$ lattice artifact, as indicated
in Eqs.~(\ref{EQ:improved_bilinear}), (\ref{EQ:improved_OL})
and (\ref{EQ:improved_OS}) in Sec.~\ref{SEC:matching}.

\subsection{Source and sink smearing}

In an attempt to obtain a better overlap with ground state,
we use gauge-invariant Gaussian smearing for source and sink operators.
We follow the smearing procedure
in Refs.~\cite{Alexandrou:1992ti, Berruto:2005hg}.
We choose a Gaussian function with width $\omega$ as a smearing function
in Eq.~(\ref{EQ:smeared_operator}) for both static and light quarks:
\begin{eqnarray}
f(\vec{x})=g(\vec{x})=\exp(-x^2/\omega^2).
\label{EQ:gaussian_finction}
\end{eqnarray}
To achieve this smearing in gauge-invariant way,
we use an implementation:
\begin{eqnarray}
\sum_{\vec{y}}f(\vec{y})\psi(\vec{x}+\vec{y}, t)
=\left(1+\frac{\omega^2}{4N_G}\nabla^2\right)^{N_G}\psi(\vec{x}, t),\;\;\;\;
\end{eqnarray}
with the hopping matrix
\begin{eqnarray}
\left[\nabla^2\right]_{xy}\equiv\sum_{i=1}^3
\left[U_i(x+\hat{i})\delta_{x+\hat{i}, y}
+U_i^{\dagger}(x-\hat{i})\delta_{x-\hat{i}, y}\right],\;\;\;\;\;\;\;\;
\end{eqnarray}
where $N_G$ is the number of times the smearing kernel acts on fermion field
$\psi(\vec{x}, t)$, which leads Gaussian function (\ref{EQ:gaussian_finction})
in $N_G\rightarrow\infty$ limit.
The choice of parameters $\omega$ and $N_G$ is summarized
in Tab.~\ref{TAB:measurement}, which gives physical Gaussian width
around $0.45$~fm.

\subsection{Measurement parameters}

Measurement parameters are summarized in Tab.~\ref{TAB:measurement}.
The valence $d$ quark mass parameter is the same as the degenerate
sea $u$ and $d$ quark's.
To interpolate to a physical $s$ quark mass, we take two values of $s$
valence quark mass parameters sandwiching the physical point and one of them
is set to be the same as sea $s$ quark's.
The physical $s$ quark mass is slightly different from the sea $s$ quark mass,
so we estimate the uncertainty from this inconsistency by using the partially
quenched SU(3) chiral perturbation theory as we describe later.
\begin{table*}
\caption{\label{TAB:measurement}
Measurement parameters.
$N_G$ and $\omega$ are source and sink Gaussian smearing parameters.
$\Delta t_{\rm src-sink}$ represents source-sink separation in three-point
functions.
}
\begin{ruledtabular}
\begin{tabular}{cccccccc}
label & $m_q$ &
Measured MD traj & \# of data & \# of src & $N_G$ &
$\omega$ & $\Delta t_{\rm src-sink}$\\ \hline
24c1 & $0.005$, $0.034$, $0.040$ &
        $900$--$8980$ every $40$ & $203$ & $4$ & $32$ & $4$ & $20$\\
24c2 & $0.010$, $0.034$, $0.040$ &
       $1460$--$8540$ every $40$ & $178$ & $2$ &      &     \\ \hline
32c1 & $0.004$, $0.027$, $0.030$ &
        $520$--$6800$ every $20$ & $304$ & $1$ & $40$ & $5$ & $24$\\
32c2 & $0.006$, $0.027$, $0.030$ &
        $1000$--$7220$ every $20$ & $312$ & $1$ &      &     \\
32c2 & $0.008$, $0.027$, $0.030$ &
        $520$--$5540$ every $20$ & $252$ & $1$ &      &     \\
\end{tabular}
\end{ruledtabular}
\end{table*}

\subsection{Autocorrelations}

The autocorrelation time of the ensemble is investigated
using the integrated autocorrelation time for both static heavy-light
two-point and three-point functions.
The integrated autocorrelation time of two-point functions is
measured at $t=12$ for ${\cal C}^{\tilde{L}S}(t, 0)$ and
${\cal C}^{\tilde{S}S}(t, 0)$,
but at $t=15$ for ${\cal C}^{SS}(t, 0)$ in both 24c and 32c ensemble.
We measure it at mid-point between source and sink location for
three-point functions.
Based on this analysis, we choose to make blocking,
so that the blocking size is $80$ MD trajectories for 32c1 ensemble
(lightest quark mass parameter),
whereas $40$ MD trajectories for other ensembles.
Note that in the study of light hadron spectrum on these ensembles,
the blocking size was $20$ MD trajectories~\cite{Aoki:2010dy}.

\subsection{Correlator fits}
\label{SEC:Correlator_fits}
In figures in Appendix~\ref{APP:Effective_mass},
we show effective masses of two-point functions and
amplitudes of three-point functions.
We perform simultaneous fits of three types of two-point correlators
${\cal C}^{\tilde{L}S}(t, 0)$, ${\cal C}^{\tilde{S}S}(t, 0)$ and
${\cal C}^{SS}(t, 0)$, assuming $E_0$ is common in these correlators.
To take into account the periodicity in the lattice box, a $\cosh$ function
is assumed in the fit:
\begin{eqnarray}
{\cal C}^{\tilde{L}S}(t, 0)&=&
{\cal A}^{\tilde{L}S}(e^{-E_0t}+e^{-E_0(T-t)}),\\
{\cal C}^{\tilde{S}S}(t, 0)&=&
{\cal A}^{\tilde{S}S}(e^{-E_0t}+e^{-E_0(T-t)}),\\
{\cal C}^{SS}(t, 0)&=&
{\cal A}^{SS}(e^{-E_0t}+e^{-E_0(T-t)}).
\end{eqnarray}
For the three-point correlators ${\cal C}_L^{SS}(t_f, t, 0)$ and
${\cal C}_S^{SS}(t_f, t, 0)$, constant fits are made:
\begin{eqnarray}
{\cal C}_L^{SS}(t_f, t, 0)={\cal A}_L^{SS},\\
{\cal C}_S^{SS}(t_f, t, 0)={\cal A}_S^{SS},
\end{eqnarray}
where $t_f$ is fixed to be source-sink separation shown in
Tab.~\ref{TAB:measurement}.
Fit ranges are shown in the effective mass and amplitude plots in
Appendix~\ref{APP:Effective_mass} and
the fit results are presented in Tabs.~\ref{TAB:Correlator_fit_results}
and \ref{TAB:Correlator_fit_results_imp}.
Note that the $O(a)$-improved ${\cal C}_S^{SS}(t_f, t, 0)$ is not
calculated, as the one-loop level matching does not require it.
\begin{table*}
\begin{center}
\caption{Correlator fit results ($O(a)$-unimproved).}
\label{TAB:Correlator_fit_results}
\begin{tabular}{c|c|ccccc|cc|cc}
\hline\hline
\multicolumn{11}{c}{24c1, $m_h=0.040$, $m_l=0.005$} \\
\hline
\rule{0mm}{4mm} smear & $m_q$ & 
$E_0$ & ${\cal A}^{\widetilde{L}S}$ & ${\cal A}^{\widetilde{S}S}$ &
${\cal A}^{SS}$ & $\chi^2$/dof &
${\cal A}_L^{SS}$ & $\chi^2$/dof &
${\cal A}_S^{SS}$ & $\chi^2$/dof \\ \hline
HYP1 & $0.005$ &0.5107(28)&0.1291(33)e+5&0.1386(33)e+10&0.2663(66)e+7&1.3&
0.294(13)e+2&0.5&-0.1741(61)e+2&0.5\\
     & $0.034$ &0.5440(13)&0.1542(18)e+5&0.1512(16)e+10&0.2984(40)e+7&1.5&
0.2230(43)e+2&0.6&-0.1357(24)e+2&0.5\\
     & $0.04$ &0.5510(12)&0.1589(17)e+5&0.1531(15)e+10&0.3038(38)e+7&1.4&
0.2064(37)e+2&0.6&-0.1262(21)e+2&0.4\\
\hline
HYP2 & $0.005$ &0.4656(22)&0.1124(23)e+5&0.1407(28)e+10&0.2670(56)e+7&1.2&
0.509(15)e+2&0.4&-0.3258(89)e+2&0.2\\
     & $0.034$ &0.4998(11)&0.1330(13)e+5&0.1543(14)e+10&0.3041(33)e+7&2.1&
0.3789(64)e+2&0.4&-0.2412(40)e+2&0.6\\
     & $0.04$ &0.5073(10)&0.1370(12)e+5&0.1565(13)e+10&0.3099(32)e+7&2.1&
0.3487(55)e+2&0.5&-0.2221(35)e+2&0.7\\
\hline
\multicolumn{11}{c}{24c2, $m_h=0.040$, $m_l=0.01$} \\
\hline
\rule{0mm}{4mm} smear & $m_q$ &
$E_0$ & ${\cal A}^{\widetilde{L}S}$ & ${\cal A}^{\widetilde{S}S}$ &
${\cal A}^{SS}$ & $\chi^2$/dof &
${\cal A}_L^{SS}$ & $\chi^2$/dof &
${\cal A}_S^{SS}$ & $\chi^2$/dof \\ \hline 
HYP1 & $0.01$ &0.5117(36)&0.1291(42)e+5&0.1368(42)e+10&0.276(10)e+7&1.8&
0.299(16)e+2&1.0&-0.1850(81)e+2&1.0\\
     & $0.034$ &0.5408(22)&0.1493(30)e+5&0.1475(27)e+10&0.3043(69)e+7&1.8&
0.2288(73)e+2&1.1&-0.1424(39)e+2&0.5\\
     & $0.04$ &0.5480(20)&0.1540(28)e+5&0.1494(25)e+10&0.3095(65)e+7&1.8&
0.2115(63)e+2&1.2&-0.1322(34)e+2&0.6\\
\hline
HYP2 & $0.01$ &0.4645(30)&0.1094(30)e+5&0.1351(36)e+10&0.2706(83)e+7&0.9&
0.547(20)e+2&1.2&-0.344(11)e+2&0.7\\
     & $0.034$ &0.4955(17)&0.1269(21)e+5&0.1477(23)e+10&0.3002(54)e+7&1.1&
0.3986(94)e+2&1.0&-0.2543(62)e+2&0.7\\
     & $0.04$ &0.5033(16)&0.1309(20)e+5&0.1501(22)e+10&0.3062(52)e+7&1.2&
0.3648(82)e+2&1.0&-0.2333(54)e+2&0.7\\
\hline
\multicolumn{11}{c}{32c1, $m_h=0.030$, $m_l=0.004$} \\
\hline
\rule{0mm}{4mm} smear & $m_q$ &
$E_0$ & ${\cal A}^{\widetilde{L}S}$ & ${\cal A}^{\widetilde{S}S}$ &
${\cal A}^{SS}$ & $\chi^2$/dof &
${\cal A}_L^{SS}$ & $\chi^2$/dof &
${\cal A}_S^{SS}$ & $\chi^2$/dof \\ \hline 
HYP1 & $0.004$ &0.4231(29)&0.753(19)e+4&0.1195(30)e+10&0.1105(32)e+7&0.4&
0.481(44)e+1&0.5&-0.284(21)e+1&0.5\\
     & $0.027$ &0.4519(14)&0.925(12)e+4&0.1343(16)e+10&0.1262(21)e+7&0.6&
0.379(15)e+1&0.7&-0.2264(83)e+1&1.0\\
     & $0.03$ &0.4557(14)&0.945(12)e+4&0.1355(15)e+10&0.1278(20)e+7&0.7&
0.363(14)e+1&0.7&-0.2170(77)e+1&1.1\\
\hline
HYP2 & $0.004$ &0.3816(28)&0.674(16)e+4&0.1198(28)e+10&0.1096(28)e+7&1.0&
0.1041(73)e+2&2.0&-0.661(33)e+1&0.1\\
     & $0.027$ &0.4118(14)&0.832(10)e+4&0.1365(16)e+10&0.1280(19)e+7&1.4&
0.802(23)e+1&2.2&-0.496(15)e+1&0.8\\
     & $0.03$ &0.4157(14)&0.849(10)e+4&0.1379(15)e+10&0.1296(19)e+7&1.5&
0.764(21)e+1&2.2&-0.473(13)e+1&0.9\\
\hline
\multicolumn{11}{c}{32c2, $m_h=0.030$, $m_l=0.006$} \\
\hline
\rule{0mm}{4mm} smear & $m_q$ & 
$E_0$ & ${\cal A}^{\widetilde{L}S}$ & ${\cal A}^{\widetilde{S}S}$ &
${\cal A}^{SS}$ & $\chi^2$/dof &
${\cal A}_L^{SS}$ & $\chi^2$/dof &
${\cal A}_S^{SS}$ & $\chi^2$/dof \\ \hline 
HYP1 & $0.006$ &0.4293(22)&0.809(17)e+4&0.1280(24)e+10&0.1168(26)e+7&0.8&
0.480(34)e+1&0.6&-0.297(15)e+1&0.8\\
     & $0.027$ &0.4530(15)&0.943(13)e+4&0.1381(17)e+10&0.1290(18)e+7&0.9&
0.387(12)e+1&0.9&-0.2379(71)e+1&0.4\\
     & $0.03$ &0.4557(15)&0.957(13)e+4&0.1390(17)e+10&0.1301(18)e+7&1.0&
0.374(11)e+1&1.0&-0.2301(67)e+1&0.4\\
\hline
HYP2 & $0.006$ &0.3855(19)&0.708(12)e+4&0.1256(21)e+10&0.1148(22)e+7&0.8&
0.1019(48)e+2&0.3&-0.648(26)e+1&0.2\\
     & $0.027$ &0.4114(14)&0.834(11)e+4&0.1378(17)e+10&0.1288(17)e+7&1.7&
0.792(20)e+1&0.5&-0.500(12)e+1&0.3\\
     & $0.03$ &0.4143(14)&0.846(11)e+4&0.1388(16)e+10&0.1301(16)e+7&1.7&
0.764(19)e+1&0.5&-0.483(12)e+1&0.3\\
\hline
\multicolumn{11}{c}{32c3, $m_h=0.030$, $m_l=0.008$} \\
\hline
\rule{0mm}{4mm} smear & $m_q$ &
$E_0$ & ${\cal A}^{\widetilde{L}S}$ & ${\cal A}^{\widetilde{S}S}$ &
${\cal A}^{SS}$ & $\chi^2$/dof &
${\cal A}_L^{SS}$ & $\chi^2$/dof &
${\cal A}_S^{SS}$ & $\chi^2$/dof \\ \hline 
HYP1 & $0.008$ &0.4296(24)&0.795(17)e+4&0.1232(24)e+10&0.1135(25)e+7&0.7&
0.491(37)e+1&2.1&-0.285(18)e+1&2.6\\
     & $0.027$ &0.4529(17)&0.924(15)e+4&0.1337(20)e+10&0.1258(21)e+7&0.8&
0.378(18)e+1&1.8&-0.2261(95)e+1&1.6\\
     & $0.03$ &0.4567(17)&0.943(14)e+4&0.1349(19)e+10&0.1275(21)e+7&0.8&
0.361(16)e+1&1.8&-0.2164(88)e+1&1.5\\
\hline
HYP2 & $0.008$ &0.3895(23)&0.717(14)e+4&0.1247(22)e+10&0.1141(24)e+7&1.8&
0.958(51)e+1&1.2&-0.605(31)e+1&1.6\\
     & $0.027$ &0.4126(16)&0.827(11)e+4&0.1351(17)e+10&0.1268(19)e+7&1.8&
0.739(28)e+1&1.5&-0.465(17)e+1&2.0\\
     & $0.03$ &0.4164(15)&0.843(11)e+4&0.1363(16)e+10&0.1284(18)e+7&1.8&
0.704(25)e+1&1.5&-0.444(16)e+1&1.9\\
\hline\hline
\end{tabular}
\end{center}
\end{table*}

\begin{table*}
\begin{center}
\caption{Correlator fit results ($O(a)$-improved).}
\label{TAB:Correlator_fit_results_imp}
\begin{tabular}{c|c|ccccc|cc}
\hline\hline
\multicolumn{9}{c}{24c1, $m_h=0.040$, $m_l=0.005$} \\
\hline
\rule{0mm}{4mm} smear & $m_q$ &
$E_0$ & ${\cal A}^{\widetilde{L}S}$ & ${\cal A}^{\widetilde{S}S}$ &
${\cal A}^{SS}$ & $\chi^2$/dof &
${\cal A}_L^{SS}$ & $\chi^2$/dof \\ \hline 
HYP1 & $0.005$ &0.5107(27)&0.1337(34)e+5&0.1387(32)e+10&0.2664(65)e+7&1.3&
0.311(13)e+2&0.4\\
     & $0.034$ &0.5440(13)&0.1600(18)e+5&0.1512(15)e+10&0.2984(40)e+7&1.4&
0.2372(45)e+2&0.6\\
     & $0.04$ &0.5510(12)&0.1650(17)e+5&0.1531(14)e+10&0.3037(38)e+7&1.4&
0.2198(38)e+2&0.6\\
\hline
HYP2 & $0.005$ &0.4654(22)&0.1183(24)e+5&0.1405(27)e+10&0.2668(55)e+7&1.3&
0.559(16)e+2&0.4\\
     & $0.034$ &0.4997(10)&0.1406(13)e+5&0.1542(14)e+10&0.3041(33)e+7&2.1&
0.4187(68)e+2&0.3\\
     & $0.04$ &0.5072(10)&0.1450(13)e+5&0.1564(13)e+10&0.3098(31)e+7&2.2&
0.3858(59)e+2&0.3\\
\hline
\multicolumn{9}{c}{24c2, $m_h=0.040$, $m_l=0.01$} \\
\hline
\rule{0mm}{4mm} smear & $m_q$ & 
$E_0$ & ${\cal A}^{\widetilde{L}S}$ & ${\cal A}^{\widetilde{S}S}$ &
${\cal A}^{SS}$ & $\chi^2$/dof &
${\cal A}_L^{SS}$ & $\chi^2$/dof \\ \hline 
HYP1 & $0.01$ &0.5118(35)&0.1339(43)e+5&0.1371(41)e+10&0.276(10)e+7&1.8&
0.318(16)e+2&0.9\\
     & $0.034$ &0.5409(21)&0.1552(30)e+5&0.1477(27)e+10&0.3046(68)e+7&1.8&
0.2438(76)e+2&1.1\\
     & $0.04$ &0.5481(19)&0.1601(28)e+5&0.1496(25)e+10&0.3098(64)e+7&1.8&
0.2255(66)e+2&1.1\\
\hline
HYP2 & $0.01$ &0.4647(29)&0.1155(31)e+5&0.1354(35)e+10&0.2710(81)e+7&0.9&
0.600(21)e+2&1.1\\
     & $0.034$ &0.4956(17)&0.1344(21)e+5&0.1479(22)e+10&0.3003(53)e+7&1.1&
0.439(10)e+2&0.9\\
     & $0.04$ &0.5034(15)&0.1388(21)e+5&0.1503(21)e+10&0.3063(50)e+7&1.1&
0.4028(90)e+2&0.9\\
\hline
\multicolumn{9}{c}{32c1, $m_h=0.030$, $m_l=0.004$} \\
\hline
\rule{0mm}{4mm} smear & $m_q$ & 
$E_0$ & ${\cal A}^{\widetilde{L}S}$ & ${\cal A}^{\widetilde{S}S}$ &
${\cal A}^{SS}$ & $\chi^2$/dof &
${\cal A}_L^{SS}$ & $\chi^2$/dof \\ \hline 
HYP1 & $0.004$ &0.4232(28)&0.775(19)e+4&0.1197(29)e+10&0.1107(31)e+7&0.4&
0.505(45)e+1&0.5\\
     & $0.027$ &0.4519(14)&0.953(12)e+4&0.1342(16)e+10&0.1263(21)e+7&0.6&
0.399(16)e+1&0.9\\
     & $0.03$ &0.4557(14)&0.973(12)e+4&0.1355(15)e+10&0.1279(21)e+7&0.7&
0.382(14)e+1&1.0\\
\hline
HYP2 & $0.004$ &0.3818(28)&0.704(17)e+4&0.1200(27)e+10&0.1098(28)e+7&1.0&
0.1131(76)e+2&2.0\\
     & $0.027$ &0.4116(14)&0.869(10)e+4&0.1363(15)e+10&0.1278(19)e+7&1.4&
0.867(23)e+1&2.3\\
     & $0.03$ &0.4155(13)&0.887(10)e+4&0.1377(15)e+10&0.1295(19)e+7&1.4&
0.826(21)e+1&2.3\\
\hline
\multicolumn{9}{c}{32c2, $m_h=0.030$, $m_l=0.006$} \\
\hline
\rule{0mm}{4mm} smear & $m_q$ & 
$E_0$ & ${\cal A}^{\widetilde{L}S}$ & ${\cal A}^{\widetilde{S}S}$ &
${\cal A}^{SS}$ & $\chi^2$/dof &
${\cal A}_L^{SS}$ & $\chi^2$/dof \\ \hline 
HYP1 & $0.006$ &0.4291(21)&0.830(17)e+4&0.1277(24)e+10&0.1166(26)e+7&0.8&
0.506(35)e+1&0.7\\
     & $0.027$ &0.4528(15)&0.969(14)e+4&0.1379(17)e+10&0.1288(18)e+7&0.9&
0.409(13)e+1&1.2\\
     & $0.03$ &0.4556(14)&0.984(13)e+4&0.1387(16)e+10&0.1299(17)e+7&1.0&
0.396(12)e+1&1.3\\
\hline
HYP2 & $0.006$ &0.3853(18)&0.737(13)e+4&0.1253(20)e+10&0.1147(22)e+7&0.8&
0.1114(50)e+2&0.4\\
     & $0.027$ &0.4113(14)&0.871(12)e+4&0.1376(16)e+10&0.1288(16)e+7&1.8&
0.860(22)e+1&0.5\\
     & $0.03$ &0.4142(14)&0.884(12)e+4&0.1386(16)e+10&0.1301(16)e+7&1.8&
0.830(20)e+1&0.5\\
\hline
\multicolumn{9}{c}{32c3, $m_h=0.030$, $m_l=0.008$} \\
\hline
\rule{0mm}{4mm} smear & $m_q$ & 
$E_0$ & ${\cal A}^{\widetilde{L}S}$ & ${\cal A}^{\widetilde{S}S}$ &
${\cal A}^{SS}$ & $\chi^2$/dof &
${\cal A}_L^{SS}$ & $\chi^2$/dof \\ \hline
HYP1 & $0.008$ &0.4296(23)&0.818(17)e+4&0.1233(24)e+10&0.1135(25)e+7&0.7&
0.513(37)e+1&1.9\\
     & $0.027$ &0.4529(17)&0.952(15)e+4&0.1338(19)e+10&0.1259(21)e+7&0.8&
0.399(18)e+1&1.6\\
     & $0.03$ &0.4567(16)&0.972(15)e+4&0.1350(19)e+10&0.1275(20)e+7&0.8&
0.381(17)e+1&1.6\\
\hline
HYP2 & $0.008$ &0.3895(22)&0.748(14)e+4&0.1247(21)e+10&0.1142(23)e+7&1.7&
0.1035(53)e+2&1.3\\
     & $0.027$ &0.4127(15)&0.865(12)e+4&0.1351(16)e+10&0.1269(19)e+7&1.7&
0.804(29)e+1&1.5\\
     & $0.03$ &0.4164(15)&0.882(11)e+4&0.1364(16)e+10&0.1285(18)e+7&1.7&
0.766(27)e+1&1.5\\
\hline\hline
\end{tabular}
\end{center}
\end{table*}
For some quark mass parameters, $\chi^2$/d.o.f. exceeds $2$.
We, however, keep fit ranges unaltered throughout all quark mass
parameters, to avoid human bias.
Then our correlator fit results have non-negligible fit-range dependence.
As we will explain in Sec.~\ref{SEC:fit-range_dependence},
the fit-range dependence are taken into account as an uncertainty of
our calculation.

\subsection{Decay constants, matrix elements and $B$-parameters}

The $B$ meson decay constants $f_B$ and mixing matrix elements ${\cal M}_B$
are obtained by Eq.~(\ref{EQ:Phi_M_HQET}) through
Eqs.~(\ref{EQ:Phi_B}) and (\ref{EQ:M_B}).
The results obtained are presented in Tab.~\ref{TAB:quantities_at_simu-pt}.
The statistical error at each simulation point is less than $2\%$ for
decay constants while sometimes reaching $5\%$ for matrix elements and
$B$-parameters.
\begin{table*}
\begin{center}
\caption{Decay constants, matrix elements and $B$-parameters
in lattice unit at simulation points.
$\Phi_{B_s}$, $\Phi_{B_s}/\Phi_B$, $M_{B_s}$, $(M_{B_s}/M_B)^{1/2}$,
$B_{B_s}$ and $(B_{B_s}/B_B)^{1/2}$ are interpolated to physical $s$ quark mass.
Matching factors are multiplied.}
\label{TAB:quantities_at_simu-pt}
\begin{tabular}{cc|ccc|ccc|ccc}
\hline\hline
\multicolumn{11}{c}{HYP1, $O(a)$-unimproved}\\
\hline
vol & $m_l$ &
$\Phi_B$ & $\Phi_{B_s}$ & $\Phi_{B_s}/\Phi_B$ &
$M_B$ & $M_{B_s}$ & $(M_{B_s}/M_B)^{1/2}$ &
$B_B$ & $B_{B_s}$ & $(B_{B_s}/B_B)^{1/2}$
\\
\hline
24c & $0.005$ & $0.2613(38)$ & $0.2998(21)$ & $1.147(12)$ & $0.1580(86)$ & $0.2098(56)$ & $1.152(22)$ & $0.867(38)$ & $0.875(17)$ & $1.004(15)$ \\
24c & $0.01$ & $0.2630(48)$ & $0.2940(33)$ & $1.118(11)$ & $0.158(10)$ & $0.1986(74)$ & $1.118(20)$ & $0.861(42)$ & $0.861(23)$ & $1.000(15)$ \\
32c & $0.004$ & $0.1611(22)$ & $0.1872(15)$ & $1.162(12)$ & $0.568(49)$ & $0.788(29)$ & $1.178(42)$ & $0.820(67)$ & $0.843(28)$ & $1.014(34)$ \\
32c & $0.006$ & $0.1674(20)$ & $0.1880(16)$ & $1.1230(71)$ & $0.625(44)$ & $0.808(30)$ & $1.136(28)$ & $0.837(56)$ & $0.857(26)$ & $1.012(24)$ \\
32c & $0.008$ & $0.1676(20)$ & $0.1873(17)$ & $1.1179(73)$ & $0.658(49)$ & $0.806(38)$ & $1.107(25)$ & $0.879(59)$ & $0.861(35)$ & $0.990(19)$ \\
\hline
\multicolumn{11}{c}{HYP2, $O(a)$-unimproved}\\
\hline
vol & $m_l$ &
$\Phi_B$ & $\Phi_{B_s}$ & $\Phi_{B_s}/\Phi_B$ &
$M_B$ & $M_{B_s}$ & $(M_{B_s}/M_B)^{1/2}$ &
$B_B$ & $B_{B_s}$ & $(B_{B_s}/B_B)^{1/2}$
\\
\hline
24c & $0.005$ & $0.2327(27)$ & $0.2638(15)$ & $1.134(10)$ & $0.1193(49)$ & $0.1555(34)$ & $1.142(17)$ & $0.825(26)$ & $0.837(14)$ & $1.007(11)$ \\
24c & $0.01$ & $0.2312(35)$ & $0.2573(23)$ & $1.113(10)$ & $0.1237(57)$ & $0.1521(44)$ & $1.109(14)$ & $0.867(30)$ & $0.861(18)$ & $0.996(10)$ \\
32c & $0.004$ & $0.1483(20)$ & $0.1718(12)$ & $1.158(11)$ & $0.491(33)$ & $0.670(23)$ & $1.168(32)$ & $0.837(51)$ & $0.851(24)$ & $1.008(25)$ \\
32c & $0.006$ & $0.1522(15)$ & $0.1713(14)$ & $1.1256(69)$ & $0.505(24)$ & $0.652(20)$ & $1.136(18)$ & $0.817(35)$ & $0.833(20)$ & $1.010(15)$ \\
32c & $0.008$ & $0.1547(18)$ & $0.1716(14)$ & $1.1093(72)$ & $0.525(25)$ & $0.637(21)$ & $1.102(16)$ & $0.822(35)$ & $0.811(23)$ & $0.993(11)$ \\
\hline
\multicolumn{11}{c}{HYP1, $O(a)$-improved}\\
\hline
vol & $m_l$ &
$\Phi_B$ & $\Phi_{B_s}$ & $\Phi_{B_s}/\Phi_B$ &
$M_B$ & $M_{B_s}$ & $(M_{B_s}/M_B)^{1/2}$ &
$B_B$ & $B_{B_s}$ & $(B_{B_s}/B_B)^{1/2}$
\\
\hline
24c & $0.005$ & $0.2706(38)$ & $0.3112(21)$ & $1.150(12)$ & $0.1661(89)$ & $0.2217(58)$ & $1.156(22)$ & $0.850(36)$ & $0.858(17)$ & $1.005(15)$ \\
24c & $0.01$ & $0.2726(49)$ & $0.3053(34)$ & $1.120(11)$ & $0.167(10)$ & $0.2105(77)$ & $1.120(20)$ & $0.846(40)$ & $0.846(22)$ & $1.000(14)$ \\
32c & $0.004$ & $0.1658(22)$ & $0.1928(15)$ & $1.163(12)$ & $0.593(50)$ & $0.824(30)$ & $1.178(41)$ & $0.809(65)$ & $0.831(28)$ & $1.013(33)$ \\
32c & $0.006$ & $0.1720(20)$ & $0.1935(17)$ & $1.1252(71)$ & $0.654(44)$ & $0.847(31)$ & $1.138(28)$ & $0.829(54)$ & $0.848(25)$ & $1.011(23)$ \\
32c & $0.008$ & $0.1724(21)$ & $0.1930(17)$ & $1.1196(72)$ & $0.686(49)$ & $0.845(39)$ & $1.110(24)$ & $0.865(57)$ & $0.850(34)$ & $0.991(18)$ \\
\hline
\multicolumn{11}{c}{HYP2, $O(a)$-improved}\\
\hline
vol & $m_l$ &
$\Phi_B$ & $\Phi_{B_s}$ & $\Phi_{B_s}/\Phi_B$ &
$M_B$ & $M_{B_s}$ & $(M_{B_s}/M_B)^{1/2}$ &
$B_B$ & $B_{B_s}$ & $(B_{B_s}/B_B)^{1/2}$
\\
\hline
24c & $0.005$ & $0.2450(28)$ & $0.2791(15)$ & $1.139(10)$ & $0.1295(52)$ & $0.1699(37)$ & $1.145(17)$ & $0.809(25)$ & $0.818(13)$ & $1.006(10)$ \\
24c & $0.01$ & $0.2439(36)$ & $0.2724(24)$ & $1.1166(98)$ & $0.1348(61)$ & $0.1664(47)$ & $1.111(13)$ & $0.849(28)$ & $0.840(17)$ & $0.9950(96)$ \\
32c & $0.004$ & $0.1547(20)$ & $0.1795(13)$ & $1.160(11)$ & $0.530(34)$ & $0.717(24)$ & $1.163(31)$ & $0.830(48)$ & $0.834(23)$ & $1.002(23)$ \\
32c & $0.006$ & $0.1585(16)$ & $0.1790(15)$ & $1.1292(69)$ & $0.545(25)$ & $0.701(21)$ & $1.134(17)$ & $0.813(33)$ & $0.820(19)$ & $1.004(14)$ \\
32c & $0.008$ & $0.1614(18)$ & $0.1796(15)$ & $1.1124(72)$ & $0.562(27)$ & $0.687(23)$ & $1.106(16)$ & $0.809(34)$ & $0.799(22)$ & $0.994(11)$ \\
\hline\hline
\end{tabular}
\end{center}
\end{table*}

\section{Chiral/Continuum extrapolation}
\label{SEC:chiral_extrapolation}

\subsection{NLO SU(2)HM$\chi$PT formula}

Physical quantities at simulated light ($u$ and $d$)
quark mass points are extrapolated to physical degenerate light quark value.
In this work, we use next-to-leading order SU(2) heavy-light meson
chiral perturbation theory (NLO SU(2)HM$\chi$PT)
depicted in Ref.~\cite{Albertus:2010nm}.
(See also Ref.~\cite{Allton:2008pn} for SU(2)$\chi$PT.)
In SU(2)$\chi$PT, $s$ quark is integrated out of the theory;
effects from $s$ quark are included in low-energy constants (LEC's).
The SU(2)$\chi$PT formula is obtained from SU(3)$\chi$PT assuming
$u$ and $d$ quark masses are much smaller than $s$ quark mass.
The formula does not depend on $s$ quark mass in an explicit way.
The convergence of the chiral fit is improved by using the SU(2)$\chi$PT
as long as $u$ and $d$ quark masses
are sufficiently small~\cite{Allton:2008pn}.
In Ref.~\cite{Allton:2008pn}, it is argued that the RBC/UKQCD DWF
ensemble does not show convergence of NLO SU(2)$\chi$PT above
the pion mass of $420$ MeV for the light hadron masses
and decay constants.
The ensembles we use in this work stay below that border.

The NLO SU(2)$\chi$PT formula for $B_d$ and $B_s$ quantity
(${\cal Q}_{B_d}$ and ${\cal Q}_{B_s}$, respectively)
with unitary $d$ quark is generally written as
\begin{eqnarray}
{\cal Q}_{B_d}&=&
{\cal Q}_0^{\rm SU(2)}\biggl(
1+X_{\cal Q}
\frac{Y_{\cal Q}+Z_{\cal Q}(g_{B^{\ast}B_{\pi}}^{\rm SU(2)})^2}
{(4\pi f^{\rm SU(2)})^2}
\ell(m_{LL}^2)
\nonumber\\
&&+C_{{\cal Q}l}^{\rm SU(2)}m_{LL}^2
+C_{{\cal Q}h}^{\rm SU(2)}(m_{HH}^2-m_{HH,~\rm phys}^2)
\nonumber\\
&&+C_{{\cal Q}a}^{\rm SU(2)}a^2\biggr),
\label{EQ:SU2ChPT_QBd}\\
{\cal Q}_{B_s}&=&
{\cal Q}_0^{(s)}
\left(1+C_{{\cal Q}l}^{\rm (s)}m_{LL}^2
+C_{{\cal Q}h}^{\rm (s)}(m_{HH}^2-m_{HH,~\rm phys}^2)\right.
\nonumber\\
&&\left.+C_{{\cal Q}s}^{\rm (s)}(m_{SS}^2-m_{HH,~\rm phys}^2)
+C_{{\cal Q}a}^{\rm (s)}a^2\right),
\label{EQ:SU2ChPT_QBs}
\end{eqnarray}
where
\begin{eqnarray}
\ell(m_{LL}^2)&=&m_{LL}^2\ln\left(\frac{m_{LL}^2}{\Lambda_{\chi}^2}\right),
\label{EQ:chral_log}\\
m_{LL}^2&=&2B_0^{\rm SU(2)}(m_l+m_{\rm res}),\\
m_{HH}^2&=&2B_0^{\rm SU(2)}(m_h+m_{\rm res}),\\
m_{SS}^2&=&2B_0^{\rm SU(2)}(m_s+m_{\rm res}),\\
m_{HH,~\rm phys}^2&=&2B_0^{\rm SU(2)}(m_h^{\rm phys}+m_{\rm res}),
\end{eqnarray}
with $m_l$, $m_h$ and $m_s$ depicting unitary degenerate $u$ and $d$,
sea $s$ and valence $s$ quark mass, respectively.
$X_{\cal Q}$, $Y_{\cal Q}$ and $Z_{\cal Q}$ are constants specific to
each physical quantity, given in Tab.~\ref{TAB:constants_XYZ}.
\begin{table}
\caption{\label{TAB:constants_XYZ}
Constants $X_{\cal Q}$, $Y_{\cal Q}$ and $Z_{\cal Q}$
in Eqs.~(\ref{EQ:SU2ChPT_QBd}) and (\ref{EQ:SU2ChPT_QBr}).
}
\begin{ruledtabular}
\begin{tabular}{cccc}
${\cal Q}$   & $X_{\cal Q}$ & $Y_{\cal Q}$ & $Z_{\cal Q}$ \\ \hline
$\Phi$ & -3/4         & 1            & 3            \\
$M$    & -1           & 2            & 3            \\ 
$B$    & -1/2         & 1            & -3         
\end{tabular}
\end{ruledtabular}
\end{table}
$f^{\rm SU(2)}$, $B_0^{\rm SU(2)}$, $g_{B^{\ast}B_{\pi}}^{\rm SU(2)}$,
${\cal Q}_0^{\rm SU(2)}$, $C_{{\cal Q}l}^{\rm SU(2)}$,
$C_{{\cal Q}h}^{\rm SU(2)}$, $C_{{\cal Q}a}^{\rm SU(2)}$,
${\cal Q}_0^{\rm (s)}$, $C_{{\cal Q}l}^{\rm (s)}$,
$C_{{\cal Q}h}^{\rm (s)}$, $C_{{\cal Q}s}^{\rm (s)}$ and
$C_{{\cal Q}a}^{\rm (s)}$ are LEC's.
Note that these LEC's are specific to the SU(2)$\chi$PT,
in which the effects of $s$ quark are integrated out at a physical $s$
quark mass $m_h^{\rm phys}$.
The $s$ quark mass dependence needs to be included,
unless the $s$ quark mass has a physical value.
It can be implemented by Taylor expansion
of LEC's around the physical $s$ quark mass as shown
in Eqs.~(\ref{EQ:SU2ChPT_QBd}) and (\ref{EQ:SU2ChPT_QBs}).
In this work, we use two kinds of link smearing in the static quark action.
Only coefficients in front of $a^2$ are dependent on the smearing.
We here mention that because the $B$-parameters express how the VSA holds well,
its quark mass dependence is expected to be mild.
In fact, the logarithm in the $\chi$PT formula for $B_B$
is suppressed for
$g_{B^{\ast}B\pi}=0.449$~\cite{Detmold:2011bp}
used in this study,
which leads to smaller coefficient of the logarithm term compared to that of
the decay constant and matrix element.
For the SU(3) breaking ratios, the expression up to NLO becomes
\begin{eqnarray}
\frac{{\cal Q}_{B_s}}{{\cal Q}_{B_d}}&=&
\widetilde{\cal Q}_0^{\rm SU(2)}\biggl(
1-X_{\cal Q}\frac{Y_{\cal Q}+Z_{\cal Q}(g_{B^{\ast}B_{\pi}}^{\rm SU(2)})^2}
{(4\pi f^{\rm SU(2)})^2}\ell(m_{LL}^2)
\nonumber\\
&&+\widetilde{C}_{{\cal Q}l}^{\rm SU(2)}m_{LL}^2
+\widetilde{C}_{{\cal Q}h}^{\rm SU(2)}(m_{HH}^2-m_{HH,~\rm phys}^2)
\nonumber\\
&&
+C_{{\cal Q}s}^{\rm (s)}(m_{SS}^2-m_{HH,~\rm phys}^2)
+\widetilde{C}_{{\cal Q}a}^{\rm SU(2)}a^2\biggr).
\nonumber\\
\label{EQ:SU2ChPT_QBr}
\end{eqnarray}
Note that these expressions do not give unity even at
$m_l=m_s$ point, because SU(3) flavor symmetry is lost,
and SU(2)$\chi$PT formula can be applied only for the region of $m_l\ll m_s$.

\subsection{Details of the chiral fitting}

For the chiral fit, we use the values of $f^{\rm SU(2)}$ and $B_0^{\rm SU(2)}$
from Ref.~\cite{Aoki:2010dy}, of $g_{B^{\ast}B_{\pi}}^{\rm SU(2)}$
from Ref.~\cite{Detmold:2011bp}, which are summarized in Tab.~\ref{TAB:LECs}.
\begin{table}
\caption{\label{TAB:LECs}
Low-energy constants used in this work.
}
\begin{ruledtabular}
\begin{tabular}{ccc}
LEC's & NLO SU(2)$\chi$PT & NLO SU(2)$\chi$PT(FV) \\ \hline
$B_0^{\rm SU(2)}$ [GeV]~\cite{Aoki:2010dy} & $4.12(7)$   & $4.03(7)$  \\
$f^{\rm SU(2)}$ [GeV]~\cite{Aoki:2010dy}   & $0.110(2)$  & $0.112(2)$ \\
$g_{B^{\ast}B_{\pi}}^{\rm SU(2)}$~\cite{Detmold:2011bp} &
\multicolumn{2}{c}{$0.449(47)_{\rm stat}(19)_{\rm sys}$} \\
$\Lambda_{\chi}$ [GeV] & \multicolumn{2}{c}{$1.0$} \\
\end{tabular}
\end{ruledtabular}
\end{table}

We carry out combined fits using HYP1 and HYP2 link smearing data assuming
that terms unrelated to the lattice spacing are common among the smearings.
Their correlation is taken into account.
As mentioned in the previous subsection, we introduce
$s$ quark mass dependence up to linear term.
To fully track the sea $s$ quark dependence, however,
at least three independent data in the $(m_h, a)$ parameter space
are required.
Our simulation setting has only one sea $s$ quark mass parameter
for each lattice spacings and the parameter is not tuned to be physical one,
therefore the data cannot be fitted using the formula
(\ref{EQ:SU2ChPT_QBd}), (\ref{EQ:SU2ChPT_QBs}) and (\ref{EQ:SU2ChPT_QBr}).
Nevertheless, we use those formula assuming sea $s$ quark mass parameter is on
physical point, leading to sea $s$ quark terms being vanished.
Later on we estimate the uncertainty from this inconsistency
using partially quenched SU(3)$\chi$PT.
On the other hand, we have two valence $s$ quark mass data.
In our analysis we first linearly interpolate the data to physical $s$
quark mass point using the two valence data,
then the fit functions are applied setting $m_s=m_h^{\rm phys}$.

In order to take into account ambiguity of the chiral fit function ansatz,
we also use a linear fit function form:
\begin{eqnarray}
{\cal G}(m_L, a)={\cal G}_0\left(1+C'_lm_{LL}^2+C'_aa^2\right),
\end{eqnarray}
for $B_d$ quantities and SU(3) breaking ratios,
which has the same form as that for $B_s$ sector in SU(2)$\chi$PT
framework.
We also investigate the uncertainty from chiral fits
by eliminating the heaviest quark mass data in both of 24c and 32c ensembles.

\subsection{Scaling check and $O(a)$ improvement}

\begin{figure*}
\begin{center}
\includegraphics[scale=1.00, viewport = 0 0 500 180, clip]
{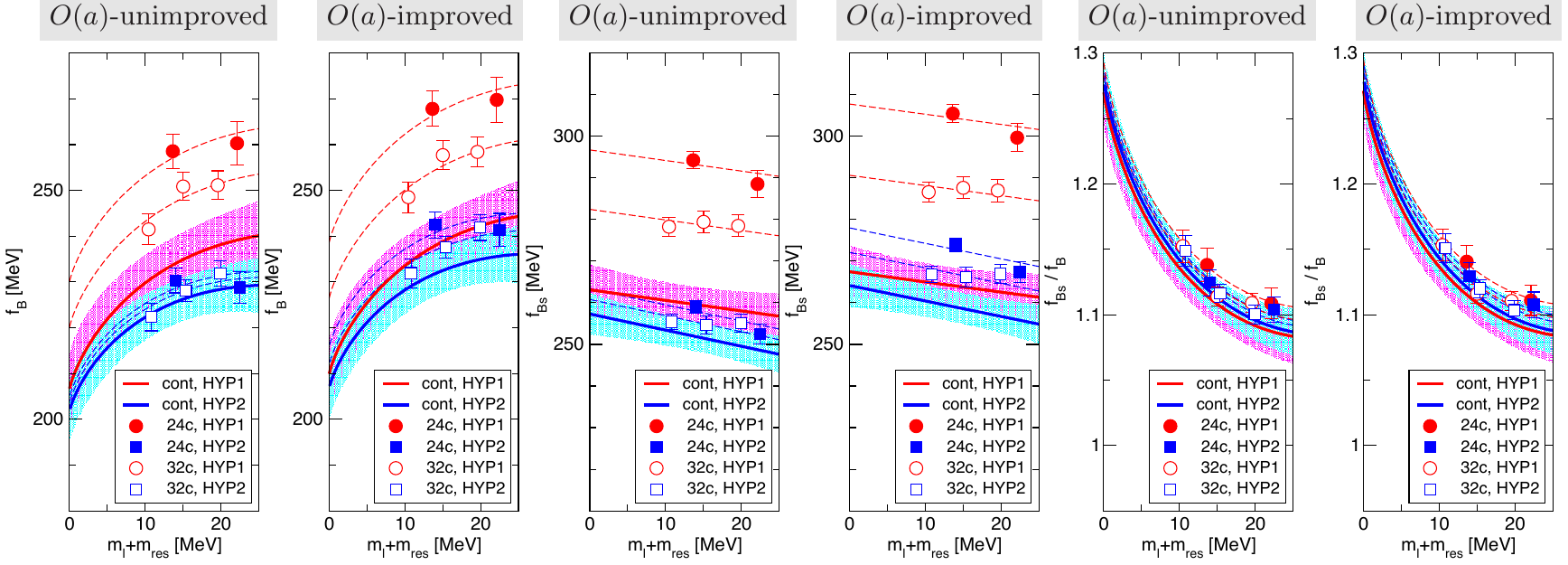}
\caption{SU(2)$\chi$PT fit of $f_B$, $f_{B_s}$ and
$f_{B_s}/f_B$ comparing $O(a)$-unimproved and -improved data.
HYP1 and HYP2 data are fit independently.
Thick lines with band represent continuum limit.}
\label{FIG:Chiral_fit_FB_SU(2)ChPT}
\end{center}
\end{figure*}

\begin{figure*}
\begin{center}
\includegraphics[scale=1.00, viewport = 0 0 500 180, clip]
{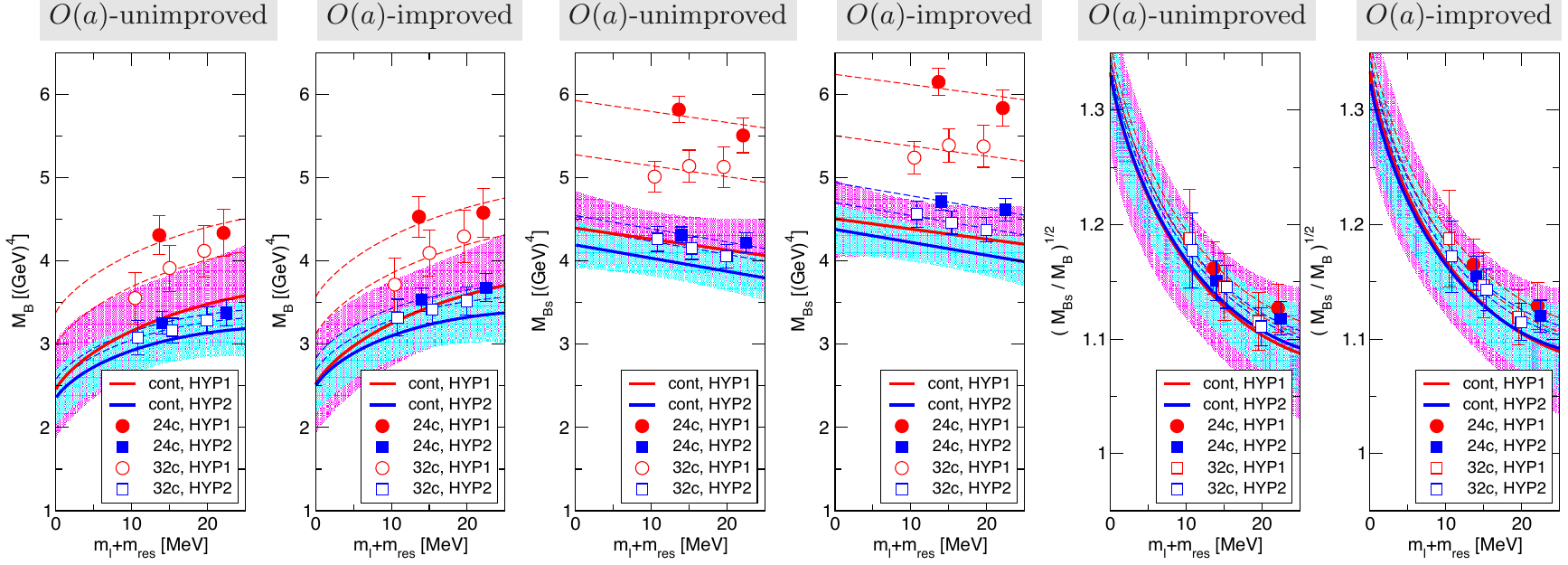}
\caption{
SU(2)$\chi$PT fit of ${\cal M}_B$, ${\cal M}_{B_s}$ and
$\sqrt{{\cal M}_{B_s}/{\cal M}_B}$ comparing $O(a)$-unimproved
and -improved data.
HYP1 and HYP2 data are fit independently.
Thick lines with band represent continuum limit.}
\label{FIG:Chiral_fit_MB_SU(2)ChPT}
\end{center}
\end{figure*}

We present fit results using SU(2)$\chi$PT formula
in Tab.~\ref{TAB:fit_results_SU2}.
\begin{table*}
\caption{\label{TAB:fit_results_SU2}
Chiral fit results in lattice unit using SU(2)$\chi$PT formula.
The values show physical point and continuum limit results.
Matching factors are multiplied.
}
\begin{ruledtabular}
\begin{tabular}{c|cccc|cccc|cccc}
 &
\multicolumn{4}{c|}{HYP1} & \multicolumn{4}{c|}{HYP2} &
\multicolumn{4}{c}{combined} \\
 &
\multicolumn{2}{c}{$O(a)$-unimp} & \multicolumn{2}{c|}{$O(a)$-imp} &
\multicolumn{2}{c}{$O(a)$-unimp} & \multicolumn{2}{c|}{$O(a)$-imp} &
\multicolumn{2}{c}{$O(a)$-unimp} & \multicolumn{2}{c}{$O(a)$-imp} \\
 &
value & $\chi^2$/dof & value & $\chi^2$/dof &
value & $\chi^2$/dof & value & $\chi^2$/dof &
value & $\chi^2$/dof & value & $\chi^2$/dof \\ \hline
$\Phi_{B}$ & $0.1437(50)$ & $0.78$ & $0.1460(50)$ & $0.67$ & $0.1400(41)$ & $1.17$ & $0.1436(42)$ & $1.06$ & $0.1392(41)$ & $2.00$ & $0.1428(42)$ & $1.51$ \\
$\Phi_{B_s}$ & $0.1766(37)$ & $0.77$ & $0.1795(38)$ & $0.75$ & $0.1725(30)$ & $1.27$ & $0.1771(31)$ & $1.36$ & $0.1726(31)$ & $2.18$ & $0.1772(32)$ & $1.30$ \\
$\Phi_{B_s}/\Phi_{B}$ & $1.228(23)$ & $0.74$ & $1.229(22)$ & $0.66$ & $1.236(20)$ & $0.33$ & $1.238(20)$ & $0.22$ & $1.233(20)$ & $0.74$ & $1.235(20)$ & $0.69$ \\
\hline
$M_{B}$ & $0.432(91)$ & $0.37$ & $0.443(93)$ & $0.34$ & $0.410(50)$ & $0.03$ & $0.435(52)$ & $0.01$ & $0.402(50)$ & $0.36$ & $0.430(54)$ & $0.33$ \\
$M_{B_s}$ & $0.686(64)$ & $0.68$ & $0.704(67)$ & $0.70$ & $0.653(40)$ & $0.09$ & $0.683(43)$ & $0.04$ & $0.636(39)$ & $1.03$ & $0.669(41)$ & $0.86$ \\
$\sqrt{M_{B_s}/M_{B}}$ & $1.261(62)$ & $0.07$ & $1.261(61)$ & $0.05$ & $1.262(42)$ & $0.19$ & $1.255(41)$ & $0.06$ & $1.262(43)$ & $0.08$ & $1.255(42)$ & $0.04$ \\
\hline
$B_{B}$ & $0.79(11)$ & $0.20$ & $0.79(11)$ & $0.17$ & $0.753(74)$ & $0.34$ & $0.763(70)$ & $0.42$ & $0.757(78)$ & $0.57$ & $0.766(75)$ & $0.62$ \\
$B_{B_s}$ & $0.833(53)$ & $0.20$ & $0.829(52)$ & $0.22$ & $0.807(40)$ & $1.23$ & $0.802(38)$ & $1.09$ & $0.804(41)$ & $1.56$ & $0.802(39)$ & $1.45$ \\
$\sqrt{B_{B_s}/B_{B}}$ & $1.019(45)$ & $0.19$ & $1.020(44)$ & $0.15$ & $1.025(30)$ & $0.13$ & $1.018(29)$ & $0.05$ & $1.023(31)$ & $0.09$ & $1.016(29)$ & $0.09$ \\
\end{tabular}
\end{ruledtabular}
\end{table*}
\begin{table*}
\caption{\label{TAB:fit_results_linear}
Chiral fit results in lattice unit using linear fit function.
The values show physical point and continuum limit results.
Matching factors are multiplied.
}
\begin{ruledtabular}
\begin{tabular}{c|cccc|cccc|cccc}
 &
\multicolumn{4}{c|}{HYP1} & \multicolumn{4}{c|}{HYP2} &
\multicolumn{4}{c}{combined} \\
 &
\multicolumn{2}{c}{$O(a)$-unimp} & \multicolumn{2}{c|}{$O(a)$-imp} &
\multicolumn{2}{c}{$O(a)$-unimp} & \multicolumn{2}{c|}{$O(a)$-imp} &
\multicolumn{2}{c}{$O(a)$-unimp} & \multicolumn{2}{c}{$O(a)$-imp} \\
 &
value & $\chi^2$/dof & value & $\chi^2$/dof &
value & $\chi^2$/dof & value & $\chi^2$/dof &
value & $\chi^2$/dof & value & $\chi^2$/dof \\ \hline
$\Phi_{B}$ & $0.1500(53)$ & $1.17$ & $0.1523(54)$ & $1.03$ & $0.1463(44)$ & $1.63$ & $0.1501(46)$ & $1.48$ & $0.1455(44)$ & $2.14$ & $0.1492(45)$ & $1.63$ \\
$\Phi_{B_s}$ & $0.1766(37)$ & $0.77$ & $0.1795(38)$ & $0.75$ & $0.1725(30)$ & $1.27$ & $0.1771(31)$ & $1.36$ & $0.1726(31)$ & $2.18$ & $0.1772(32)$ & $1.30$ \\
$\Phi_{B_s}/\Phi_{B}$ & $1.164(22)$ & $1.37$ & $1.165(21)$ & $1.27$ & $1.172(20)$ & $1.01$ & $1.174(19)$ & $0.82$ & $1.169(20)$ & $1.00$ & $1.171(19)$ & $0.94$ \\
\hline
$M_{B}$ & $0.47(10)$ & $0.44$ & $0.48(10)$ & $0.41$ & $0.450(55)$ & $0.03$ & $0.477(58)$ & $0.01$ & $0.442(56)$ & $0.36$ & $0.472(60)$ & $0.32$ \\
$M_{B_s}$ & $0.686(64)$ & $0.68$ & $0.704(67)$ & $0.70$ & $0.653(40)$ & $0.09$ & $0.683(43)$ & $0.04$ & $0.636(39)$ & $1.03$ & $0.669(41)$ & $0.86$ \\
$\sqrt{M_{B_s}/M_{B}}$ & $1.186(58)$ & $0.16$ & $1.187(57)$ & $0.13$ & $1.186(40)$ & $0.29$ & $1.180(38)$ & $0.12$ & $1.187(40)$ & $0.12$ & $1.180(39)$ & $0.06$ \\
\hline
$B_{B}$ & $0.80(11)$ & $0.20$ & $0.80(11)$ & $0.17$ & $0.760(75)$ & $0.33$ & $0.769(71)$ & $0.41$ & $0.763(79)$ & $0.57$ & $0.773(75)$ & $0.61$ \\
$B_{B_s}$ & $0.833(53)$ & $0.20$ & $0.829(52)$ & $0.22$ & $0.807(40)$ & $1.23$ & $0.802(38)$ & $1.09$ & $0.804(41)$ & $1.56$ & $0.802(39)$ & $1.45$ \\
$\sqrt{B_{B_s}/B_{B}}$ & $1.015(45)$ & $0.19$ & $1.015(44)$ & $0.15$ & $1.020(30)$ & $0.13$ & $1.013(28)$ & $0.05$ & $1.019(30)$ & $0.09$ & $1.012(29)$ & $0.09$ \\
\end{tabular}
\end{ruledtabular}
\end{table*}
We also show chiral fit using the SU(2)$\chi$PT formula
in Figs.~\ref{FIG:Chiral_fit_FB_SU(2)ChPT} and
\ref{FIG:Chiral_fit_MB_SU(2)ChPT},
in which both $O(a)$-unimproved and -improved results are presented.
The features of the data are as follows:
\begin{itemize}
\item
The data shows HYP1 smearing gives larger scaling violations than
HYP2.
\item
HYP1 and HYP2 results are almost consistent with each other
in the continuum limit.
This consistency is seen even in the $O(a)$-unimproved case within
large statistical errors.
While the $O(a)$-improved data shows slightly better consistency than
unimproved one, we cannot see clear effectiveness of the $O(a)$-improvement
at current statistics.
\item
The $O(a)$-improvement slightly pushes data up
for decay constants and matrix elements at each simulation point.
\item
Being a ratio, the scaling violation for $\xi$ and $f_{B_s}/f_B$ is tiny.
HYP1 and HYP2, $O(a)$-improved and unimproved results are consistent
at each simulation points.
\end{itemize}
When $O(a)$-improvement is successfully accomplished and $a^2$ scaling
is used in the continuum extrapolation
(assuming $O(\alpha_s^2a)$ and $O(a^3)$ contributions are small),
HYP1 and HYP2 results must give the same value in the continuum limit,
and our data is actually consistent with this observation.
Therefore we use combined fit of HYP1 and HYP2 assuming chiral fit parameter
for each smearing is different only for the coefficients of $a^2$ term.

\subsection{Fit results and criteria for final results}

\begin{figure*}
\begin{center}
\includegraphics[scale=1.0, viewport = 0 0 500 220, clip]
{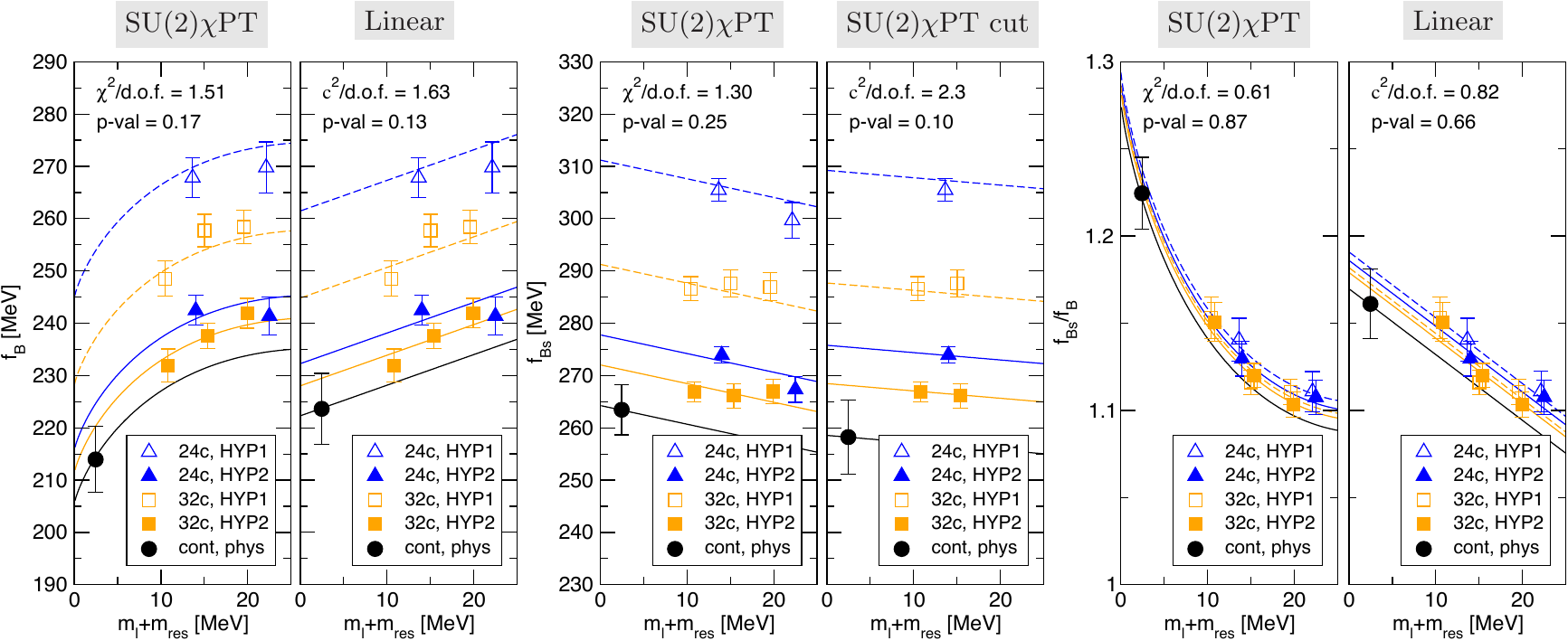}
\caption{Chiral fit of $f_B$, $f_{B_s}$ and $f_{B_s}/f_B$
using SU(2)$\chi$PT and linear.
''cut'' indicates heaviest quark mass points
at each lattice spacing are removed in the fitting.}
\label{FIG:Chiral_fit_FB}
\end{center}
\end{figure*}

\begin{figure*}
\begin{center}
\includegraphics[scale=1.0, viewport = 0 0 500 220, clip]
{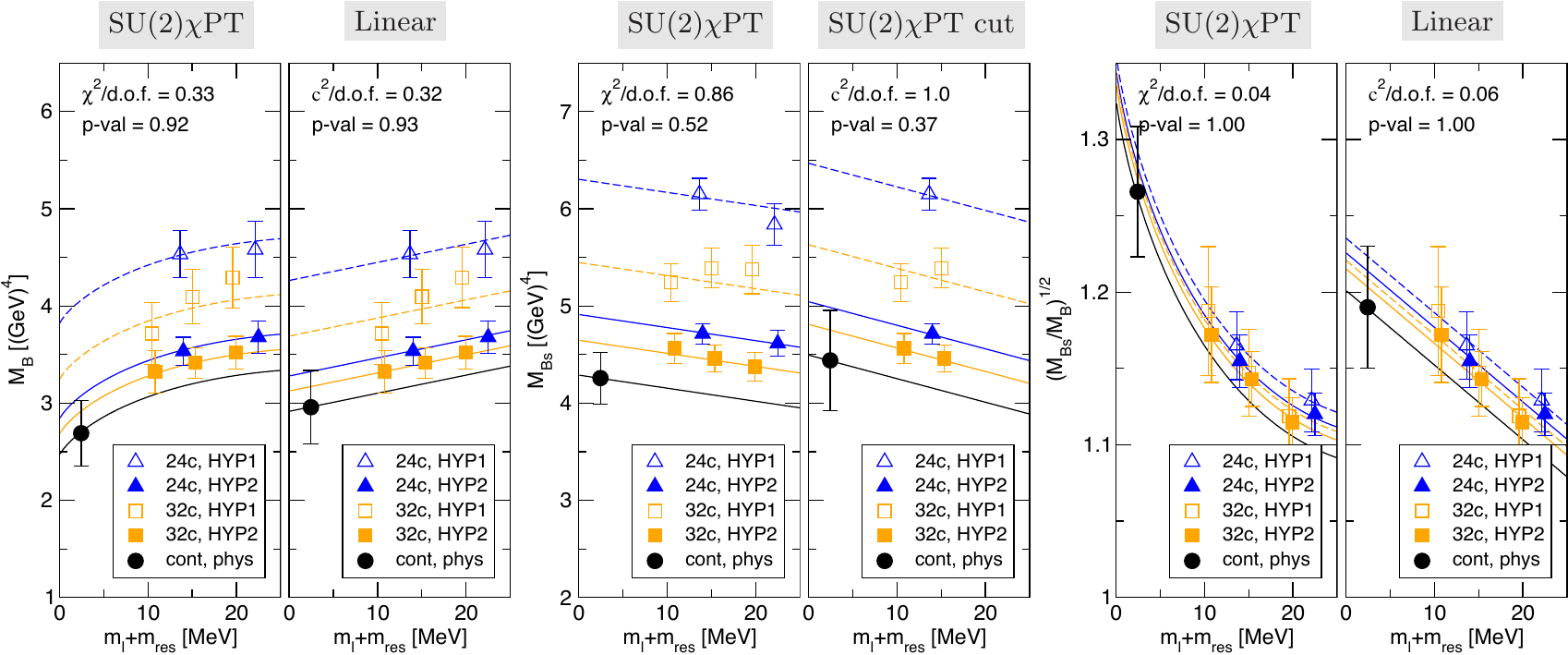}
\caption{Chiral fit of ${\cal M}_B$, ${\cal M}_{B_s}$ and
$({\cal M}_{B_s}/{\cal M}_B)^{1/2}$
using SU(2)$\chi$PT and linear.
''cut'' indicates heaviest quark mass points
at each lattice spacing are removed in the fitting.}
\label{FIG:Chiral_fit_MB}
\end{center}
\end{figure*}

\begin{figure*}
\begin{center}
\includegraphics[scale=1.0, viewport = 0 0 500 210, clip]
{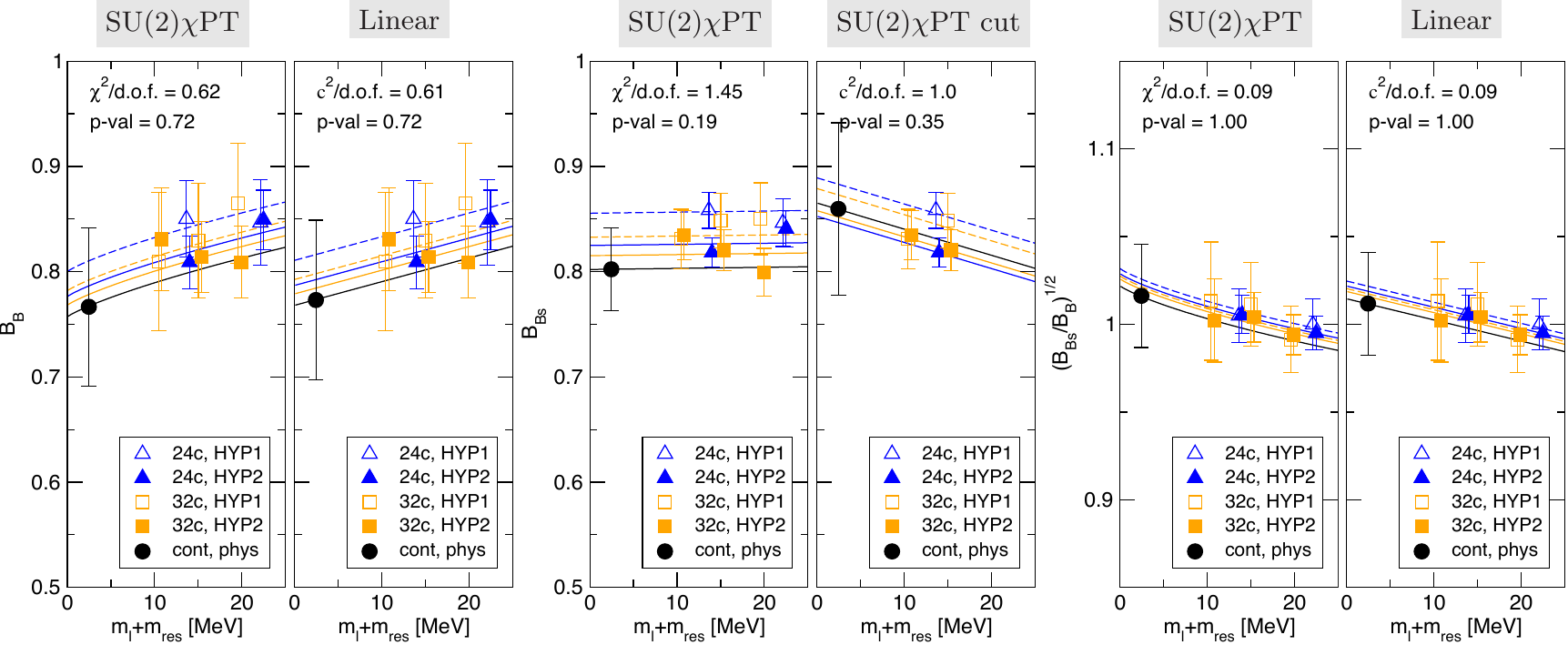}
\caption{Chiral fit of $B_B$, $B_{B_s}$ and $(B_{B_s}/B_B)^{1/2}$
using SU(2)$\chi$PT and linear.
''cut'' indicates heaviest quark mass points
at each lattice spacing are removed in the fitting.}
\label{FIG:Chiral_fit_BB}
\end{center}
\end{figure*}

In this work, $O(a)$-improved data are taken for the final results.
Hereafter, numerical data and figures indicate the $O(a)$-improved one.
We present chiral fit results 
in Figs.~\ref{FIG:Chiral_fit_FB}, \ref{FIG:Chiral_fit_MB} and
\ref{FIG:Chiral_fit_BB}.
Correlations between two kinds of link smearing in the static action is
included in the fitting.
$\chi^2$/d.o.f.'s and $p$-values in the fits are presented in the figures.
$\chi^2$/d.o.f. in each fit are all acceptable level,
thus it is hard to exclude any of the fit at the ansatz.
We thus take following criteria for the chiral and continuum extrapolations:
\begin{itemize}
\item
For $B_d$ quantities and SU(3) breaking ratios,
an average of results from SU(2)$\chi$PT and the linear fit,
whose physical point values are presented in Table XI, is taken.
We then take half of the full difference between the SU(2)$\chi$PT and
the linear results as an uncertainty from chiral fit function ansatz.
\item
For $B_s$ quantities, SU(2)$\chi$PT fit (linear fit) results
are taken as central values.
To investigate the chiral fit form ambiguity,
data in region of $m_{\pi}>350$~MeV are removed
and we see its effect to the extrapolated value.
We take difference between the full data and cut data,
where the heaviest quark mass points at each lattice spacing are removed
(``SU(2)$\chi$PT cut'' in Figs.~\ref{FIG:Chiral_fit_FB},
\ref{FIG:Chiral_fit_MB} and \ref{FIG:Chiral_fit_BB}),
as a chiral fit ambiguity.
\end{itemize}

Combining with the ratio of the decay constants, $\xi$ can be obtained through
Eq.~(\ref{EQ:xi_from_b-para}).
While the ratio of the $B$-parameters is well determined,
current data for the decay constants has a large uncertainty from chiral
extrapolation, which also leads to a poor determination of $\xi$ from
Eq.~(\ref{EQ:xi_from_b-para}), not giving any gain.
We hence simply use Eq.~(\ref{EQ:xi}) to calculate $\xi$ in this work.

\subsection{Finite volume effect}

Our lattice has modest physical volume around $2.75$ fm and the lowest
$m_{\pi}L$ is about $4$, thus we may estimate
finite volume (FV) uncertainty using FV NLO$\chi$PT.
The FV correction can be included in the $\chi$PT formula
by replacing chiral logarithms~(\ref{EQ:chral_log})
with~\cite{Bernard:2001yj, Colangelo:2005gd}
\begin{eqnarray}
\ell(m_{LL}^2)&=&m_{LL}^2\ln\left(\frac{m_{LL}^2}{\Lambda_{\chi}^2}
+\delta_1(m_{LL}L)\right),\\
\delta_1(m_{LL}L)&=&
\frac{4}{m_{LL}L}\sum_{\vec{r}\not=0}\frac{K_1(|r|m_{LL}L)}{|r|},
\label{EQ:fv_delta1}
\end{eqnarray}
where $K_1$ is modified Bessel functions of the 2nd kind.
For the numerical implementation of Eq.~(\ref{EQ:fv_delta1}),
we use the multiplicities depicted
in Refs.~\cite{Bernard:2001yj, Allton:2008pn}.
With SU(2)$\chi$PT for the chiral extrapolation,
we cannot evaluate the FV effect for $B_s$ sector in this procedure.
The effect is, however, expected to be quite small in this sector,
we estimate this uncertainty is negligible.
In the simulated quark mass region, the FV correction slightly pushes
the data up for $B_d$ quantities, and hence down for the SU(3) breaking ratios,
$f_{B_s}/f_B$ and $\xi$.

\section{Systematic errors}
\label{SEC:Systematic_errors}

\begin{figure*}
\begin{center}
\includegraphics[scale=0.4, viewport = 0 0 800 240, clip]
{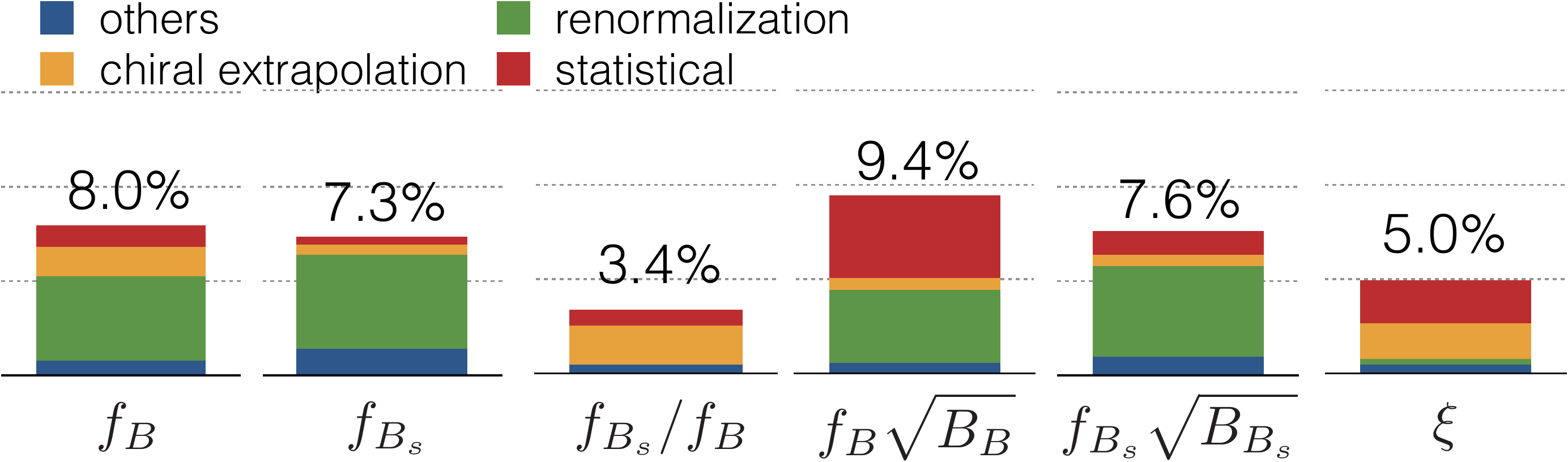}
\caption{Error budget for final quantities.
The height of the bars denotes total error,
while relative size of the colors is determined by squared errors.}
\label{FIG:error_budget}
\end{center}
\end{figure*}

In this section we clarify the systematic errors we take into account.
A summary of the systematic errors is shown in Tab.~\ref{TAB:error_budget}
and also in Fig.~\ref{FIG:error_budget}.
\begin{table*}
\begin{center}
\caption{Error budget [$\%$] for final quantities.}
\label{TAB:error_budget}
\begin{tabular}{c|ccccccccc}
\hline\hline
&
$f_B$ & $f_{B_s}$ & $f_{B_s}/f_B$ &
$f_B\sqrt{\hat{B}_B}$ & $f_{B_s}\sqrt{\hat{B}_{B_s}}$ & $\xi$ &
$\hat{B}_B$ & $\hat{B}_{B_s}$ & $B_{B_s}/B_B$ \\
\hline
Statistics &
$2.99$ & $1.81$ & $1.65$ & $6.34$ & $3.12$ & $3.36$ & $9.80$ & $4.93$ & $5.80$
\\
Chiral/continuum extrapolation &
$3.54$ & $1.98$ & $2.66$ & $2.55$ & $2.13$ & $3.08$ & $14.84$ & $7.15$ & $3.66$
\\
Finite volume effect &
$0.82$ & $0.0$ & $1.00$ & $0.76$ & $0.00$ & $1.07$ & $0.15$ & $0.0$ & $0.16$
\\
Discretization &
$1.0$ & $1.0$ & $0.2$ & $1.0$ & $1.0$ & $0.2$ & $1.0$ & $1.0$ & $0.2$
\\
One-loop renormalization &
$6.0$ & $6.0$ & $0.0$ & $6.0$ & $6.0$ & $1.2$ & $6.0$ & $6.0$ & $1.2$
\\
$g_{B^{\ast}B\pi}$ &
$0.24$ & $0.00$ & $0.35$ & $0.14$ & $0.00$ & $0.25$ & $0.20$ & $0.00$ & $0.22$
\\
Scale &
$1.82$ & $1.85$ & $0.04$ & $1.84$ & $1.86$ & $0.05$ & $0.04$ & $0.05$ & $0.02$
\\
Physical quark mass &
$0.05$ & $0.01$ & $0.06$ & $0.06$ & $0.19$ & $0.20$ & $0.03$ & $0.00$ & $0.02$
\\
Off-physical sea s quark mass &
$0.84$ & $0.69$ & $0.79$ & $0.20$ & $0.39$ & $0.91$ & $0.28$ & $0.19$ & $0.42$
\\
Fit-range &
$0.44$ & $2.31$ & $0.26$ & $0.10$ & $1.74$ & $0.58$ & $3.14$ & $0.00$ & $1.54$
\\
\hline
Total systematic error &
$7.38$ & $7.09$ & $3.00$ & $6.90$ & $6.94$ & $3.66$ & $16.34$ & $9.39$ & $4.18$
\\
Total error (incl. statistical) &
$7.96$ & $7.32$ & $3.42$ & $9.37$ & $7.61$ & $4.97$ & $19.05$ & $10.61$ & $7.15$
\\
\hline\hline
\end{tabular}
\end{center}
\end{table*}

\subsection{Chiral extrapolation}

As described in Sec.~\ref{SEC:chiral_extrapolation},
we use SU(2)$\chi$PT formula for the chiral and continuum extrapolations.
The linear fit function ansatz cannot be excluded in the current statistics,
thus we take their average.
The method to estimate the associated systematic errors has been described
in detail in Sec.\ref{SEC:chiral_extrapolation}.

\subsection{$g_{B^{\ast}B\pi}$}

In the chiral fit, we use $g_{B^{\ast}B\pi}=0.449(47)(19)$,
where the first uncertainty is statistical and the second is
systematic~\cite{Detmold:2011bp}.
This value was obtained using the $2+1$ flavor dynamical DWF configurations,
which is the same as that we use in this simulation.
The systematic errors are fully evaluated in Ref.~\cite{Detmold:2011bp},
thus we quote this value as a reliable one.
We use $0.449$ as a central value and change it by $\pm0.051$ in
the chiral fit for the uncertainty of this coupling.

\subsection{Discretization}

The static heavy and light quark system has $O(a)$ lattice discretization
errors even if chiral fermions are employed for the light quarks,
in which case the $O(a)$ discretization errors start with $O(\alpha_s a)$.
In this simulation, $O(a)$-improvement is made using one-loop
perturbation theory~\cite{Ishikawa:2011dd}.
Thus, the remaining $O(a)$ lattice artifact is supposed to be
$O(\alpha_s^2a)$ at each simulated lattice spacing $a$.
For the lattice artifact, the coupling should be the lattice one, i.e.,
defined by Eq.~(\ref{EQ:MF-coupling}), whose actual value is shown in
Tab.~\ref{TAB:numerical_HQET_matching}.
Conservatively assuming $\Lambda_{\rm QCD}\sim500$ MeV, the order of magnitude
for each discretization error is summarized in Tab.~\ref{TAB:disc_order}.
\begin{table}
\begin{center}
\caption{Power counting for perturbation and discretization error estimation.
We here define $\hat{a}=a\Lambda_{\rm QCD}$.}
\label{TAB:disc_order}
\begin{tabular}{c|c|c}
\hline\hline
$\alpha_s(m_b)$ & \multicolumn{2}{c}{$0.2261$} \\
$(\alpha_s(m_b))^2$ & \multicolumn{2}{c}{$0.0511$} \\
\hline\hline
  & 24c     & 32c \\
\hline
$\alpha_s^{\rm LAT}$ & $0.1769$ & $0.1683$ \\
$(\alpha_s^{\rm LAT})^2$ & $0.0313$ & $0.0283$ \\
$\hat{a}$ & $0.29$ & $0.22$ \\
$\hat{a}^2$ & $0.084$ & $0.048$ \\
$\hat{a}^3$ & $0.024$ & $0.011$ \\
$\alpha_s^{\rm LAT}\hat{a}$ & $0.051$ & $0.037$ \\
$(\alpha_s^{\rm LAT})^2\hat{a}$ & $0.0091$ & $0.0062$ \\
$\alpha_s^{\rm LAT}\hat{a}^3$ & $0.0042$ & $0.0019$ \\
$\hat{a}^2$~:~$\alpha_s^{\rm LAT}\hat{a}$~:~$\alpha_s^{\rm LAT}\hat{a}^3$ &
$1$~:~$0.61$~:~$0.05$ & $1$~:~$0.77$~:~$0.04$ \\
$\hat{a}^2$~:~$(\alpha_s^{\rm LAT})^2\hat{a}$~:~$\alpha_s^{\rm LAT}\hat{a}^3$ &
$1$~:~$0.11$~:~$0.05$ & $1$~:~$0.13$~:~$0.04$ \\
\hline\hline
\end{tabular}
\end{center}
\end{table}
While without one-loop perturbative $O(a)$-improvement,
the magnitude of $O(\alpha_s a)$ term is more than half of $O(a^2)$ term,
the improvement makes a substantial reduction of it.
The uncertainty from $O(\alpha_s^2 a)$ is
$\sim0.9\%$ (24c) and $\sim0.6\%$ (32c) level.
The uncertainty from $O(a^3)$ contribution, which starts at one-loop level,
is even smaller than that.
Thus we take $1\%$ as an uncertainty from remaining
$O(a)$ and $O(a^3)$ contribution in the continuum.
For SU(3) breaking ratios, the lattice artifact comes with a factor of
$(m_s-m_d)/\Lambda_{\rm QCD}\sim0.2$, which leads to reduced uncertainty
down to $0.2\%$.

\subsection{Renormalization}

In this work, renormalization is carried out in one-loop perturbation
framework.
We here use power counting for the estimation of higher order
uncertainty of the perturbation.
We use two-step matching procedure; first, QCD full theory and HQET are matched
in the continuum at a scale $\mu=m_b$, second, continuum and lattice HQET
are matched at a scale $\mu=a^{-1}$.
Values of $\alpha_s$ in these matchings are presented
in Tab.~\ref{TAB:disc_order}.
Assuming coefficients of the power expansion to be one,
the counting estimation shows two-loop uncertainty of $5.1\%$
in the first matching and of $3.1\%$ in the second.
We add them in quadrature leading to $6\%$.
For the ratio of decay constants, the renormalization factor is completely
canceled out, thus the perturbation ambiguity is negligible.
For $\xi$, however, non-vanishing contribution $Z_2/Z_1$ remains in the
ratio, which causes an uncertainty.
Nevertheless, because this uncertainty is suppressed by a factor of
$(m_s-m_l)/\Lambda_{\rm QCD}\sim0.2$, the one-loop ambiguity is reduced
to $1.2\%$.
We note that one-loop perturbation ambiguity exists also in the
$O(a)$ improvement coefficients, which is counted as
the discretization error as discussed the previous subsection.

\subsection{Scale}

As shown in Tab.~\ref{TAB:ensembles}, lattice scales used in this study
have $1\%$ level uncertainty.
We investigate systematic error from this by varying
the value of lattice spacing within the uncertainty.
While matching factors and $O(a)$ improved coefficients need to be
implicitly varied for this search, the effect is negligible.
Thus we take into account the error only when the lattice units are converted
into physical units and chiral/continuum extrapolations are carried out.

\subsection{Light quark mass}

Light quark masses at physical point also have $3\%$ level uncertainty
as seen in Tab.~\ref{TAB:measurement}.
It affects values of physical observables.
We check the effects by varying the physical quark mass values within
the uncertainty.

\subsection{Off physical sea $s$ quark mass}
\label{SEC:quenching}

Our gluon ensemble has only one dynamical $s$ quark mass parameter,
which is slightly off from the physical $s$ quark mass.
In spite of this, we use SU(2)$\chi$PT fit functions assuming
the sea $s$ quark is on physical mass.
The uncertainty from this inconsistency must be investigated.
To deal with it, we make an estimation using SU(3)$\chi$PT as a model.
We use partially quenched
SU(3)HM$\chi$PT~\cite{Arndt:2004bg, Aubin:2005aq},
whose explicit formula are also presented in Ref.~\cite{Albertus:2010nm}.
The ambiguity from off-physical $s$ quark mass effect is investigated by
taking the difference between correct treatment of our
simulation setup and fake treatment where the $s$ quark mass is
assumed to be on physical point.

\subsection{Finite volume}

FV effect is estimated using FV$\chi$PT as mentioned
in Sec.~\ref{SEC:chiral_extrapolation}.
Uncertainty from FV effect is estimated from the difference
between SU(2)$\chi$PT and FV~SU(2)$\chi$PT.
The effect for $B_s$ quantities is expected to be significantly small,
thus is neglected in our analysis.

\subsection{Fit-range dependence}
\label{SEC:fit-range_dependence}

As mentioned in Sec.~\ref{SEC:Correlator_fits},
our correlator fit results have non-negligible fit-range dependence.
Although this uncertainty is rather statistical than systematic,
we count it as a systematic error here.
To take into account the uncertainty of the fit range choices,
we shift the minimal value of $t$ in the fit range toward larger value by $2$
for two-point functions and shorten the range by $2$ for
three-point functions.
In Appendix ~\ref{APP:Fit-range-dependence}, the physical quantities
with the original and the shifted fit range at each simulation parameters
are shown in Figs.~\ref{FIG:fit_range_dependence_PB},
\ref{FIG:fit_range_dependence_MB} and \ref{FIG:fit_range_dependence_BB}.
We find non-negligible fit range dependences remain in some cases,
where the cases that the difference between fit range choices is
beyond $1$-$\sigma$ statistical error are listed in the caption in each figure.
We define the uncertainty of the fit range dependences as:
\begin{enumerate}
 \item When physical quantities at some quark mass parameter move beyond
       $1\sigma$ statistical error by changing the fit range,
       the data at the mass parameter for both HYP1 and HYP2 are
       replaced to see the effect of the move.
 \item Chiral/continuum fits are performed to investigate
       the shift caused by the replacement of the data.
 \item We repeat this procedure for each data which has large move
       beyond $1$-$\sigma$ statistical error by changing the fit range.
 \item The final uncertainty is obtained by adding each shift of the
       chiral/continuum extrapolated value in quadrature.
\end{enumerate}
The resulting uncertainty is taken as a systematic error in our calculation.

\section{Conclusions}
\label{SEC:conclusions}

\subsection{Results of physical quantities}

We present final results for $B$ meson quantities
in the static limit of $b$ quark:
\begin{eqnarray}
\left[~f_B~\right]^{\rm static}
&=&218.8(6.5)_{\rm stat}(16.1)_{\rm sys}~{\rm MeV},~~~~~ \\
\left[~f_{B_s}~\right]^{\rm static}
&=&263.5(4.8)_{\rm stat}(18.7)_{\rm sys}~{\rm MeV}, \\
\left[~f_{B_s}/f_B~\right]^{\rm static}
&=&1.193(20)_{\rm stat}(36)_{\rm sys}, \\
\left[~f_B\sqrt{\hat{B}_B}~\right]^{\rm static}
&=&240(15)_{\rm stat}(17)_{\rm sys}~{\rm MeV}, \\
\left[~f_{B_s}\sqrt{\hat{B}_{B_s}}~\right]^{\rm static}
&=&290(09)_{\rm stat}(20)_{\rm sys}~{\rm MeV}, \\
\left[~\xi~\right]^{\rm static}
&=&1.208(41)_{\rm stat}(44)_{\rm sys}, \\
\left[~\hat{B}_B~\right]^{\rm static}
&=&1.17(11)_{\rm stat}(19)_{\rm sys}, \\
\left[~\hat{B}_{B_s}~\right]^{\rm static}
&=&1.22(06)_{\rm stat}(11)_{\rm sys}, \\
\left[~B_{B_s}/B_B~\right]^{\rm static}
&=&1.028(60)_{\rm stat}(43)_{\rm sys},
\end{eqnarray}
where first errors indicate statistical while second ones are systematic.
Note that $O(1/m_b)$ uncertainty, which is mentioned in the next subsection,
is not included in the systematic errors above.
We also show final results including $O(1/m_b)$ uncertainty in the
systematic error:
\begin{eqnarray}
f_B
&=&218.8(6.5)_{\rm stat}(30.8)_{\rm sys}~{\rm MeV}, \\
f_{B_s}
&=&263.5(4.8)_{\rm stat}(36.7)_{\rm sys}~{\rm MeV}, \\
f_{B_s}/f_B
&=&1.193(20)_{\rm stat}(44)_{\rm sys}, \\
f_B\sqrt{\hat{B}_B}
&=&240(15)_{\rm stat}(33)_{\rm sys}~{\rm MeV}, \\
f_{B_s}\sqrt{\hat{B}_{B_s}}
&=&290(09)_{\rm stat}(40)_{\rm sys}~{\rm MeV}, \\
\xi
&=&1.208(41)_{\rm stat}(52)_{\rm sys}, \\
\hat{B}_B
&=&1.17(11)_{\rm stat}(24)_{\rm sys}, \\
\hat{B}_{B_s}
&=&1.22(06)_{\rm stat}(19)_{\rm sys}, \\
B_{B_s}/B_B
&=&1.028(60)_{\rm stat}(49)_{\rm sys}.
\end{eqnarray}

We here present the constraint on a ratio of CKM matrix element
(\ref{EQ:ratio_of_CKM})
obtained through Eq.~(\ref{EQ:ratio_of_CKM_xi}):
\begin{eqnarray}
\left|\frac{V_{td}}{V_{ts}}\right|=0.206(13),
\end{eqnarray}
where statistical and systematic errors
including $O(1/m_b)$ uncertainty are all added in quadrature.

\subsection{Comparison with other approaches and $1/m_b$ ambiguity}

Since we use static approximation for $b$ quark, there exists
$O(\Lambda_{\rm QCD}/m_b)$ uncertainty for the physical quantities.
Here, we take PDG value of $b$ quark mass in $\overline{\rm MS}$ scheme
$m_b=4.18(03)$~GeV~\cite{Beringer:1900zz}
and assume $\Lambda_{\rm QCD}=0.5$~GeV.
The uncertainty from static approximation becomes $12\%$.
For the SU(3) breaking ratios, however, there would be suppression factor
coming from SU(3) light flavor symmetry, which leads:
\begin{eqnarray}
\frac{\Lambda_{\rm QCD}}{m_b}\times\frac{m_s-m_d}{\Lambda_{\rm QCD}}
\sim 2.2\%.
\label{EQ:one_ov_mb_for_ratio}
\end{eqnarray}

\begin{figure*}
\begin{center}
\includegraphics[scale=1.00, viewport = 0 0 500 450, clip]
{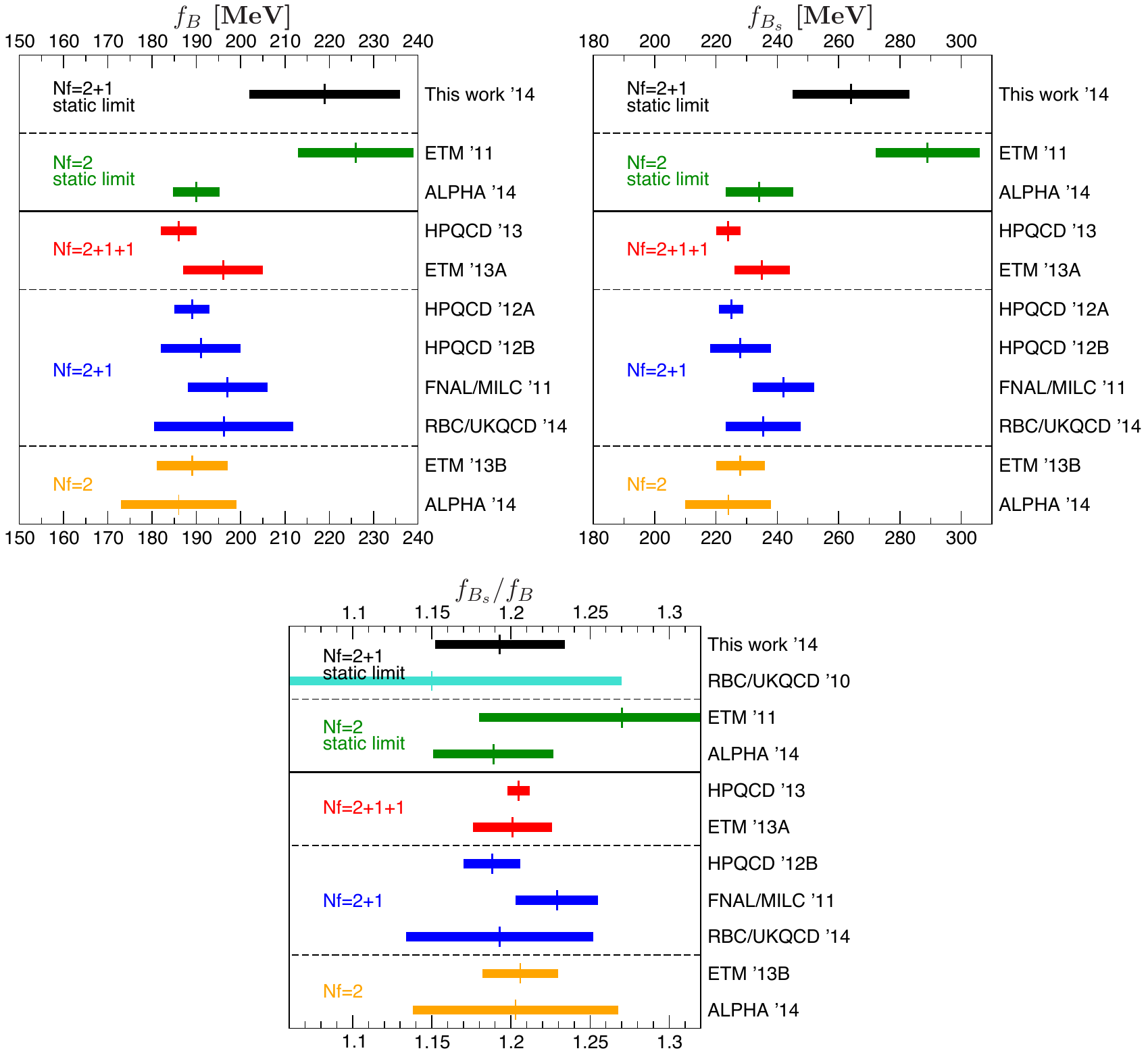}
\end{center}
\caption{
Comparison of $f_B$, $f_{B_s}$ and $f_{B_s}/f_B$ with other works.
The data is cited from Refs.~\cite{Dimopoulos:2011gx} (ETM '11),
\cite{Bernardoni:2014fva} (ALPHA '14),
\cite{Dowdall:2013tga} (HPQCD '13),
\cite{Carrasco:2013naa} (ETM '13A),
\cite{McNeile:2011ng, Na:2012kp} (HPQCD '12A, HPQCD '12B),
\cite{Bazavov:2011aa} (FNAL/MILC '11),
\cite{Christ:2014uea} (RBC/UKQCD '14),
\cite{Carrasco:2013zta} (ETM '13B)
and
\cite{Albertus:2010nm} (RBC/UKQCD '10).
The values of $f_B$ and $f_{B_s}$ in ETM '11 are obtained from $\Phi_B$ and
$\Phi_{B_s}$ divided by $\sqrt{m_B}$ and $\sqrt{m_{B_s}}$, respectively.
Errors for the static limit results do not contain $1/m_b$ uncertainty.}
\label{FIG:comparison_decay_constants}
\end{figure*}

\begin{figure*}
\begin{center}
\includegraphics[scale=1.00, viewport = 0 0 500 290, clip]
{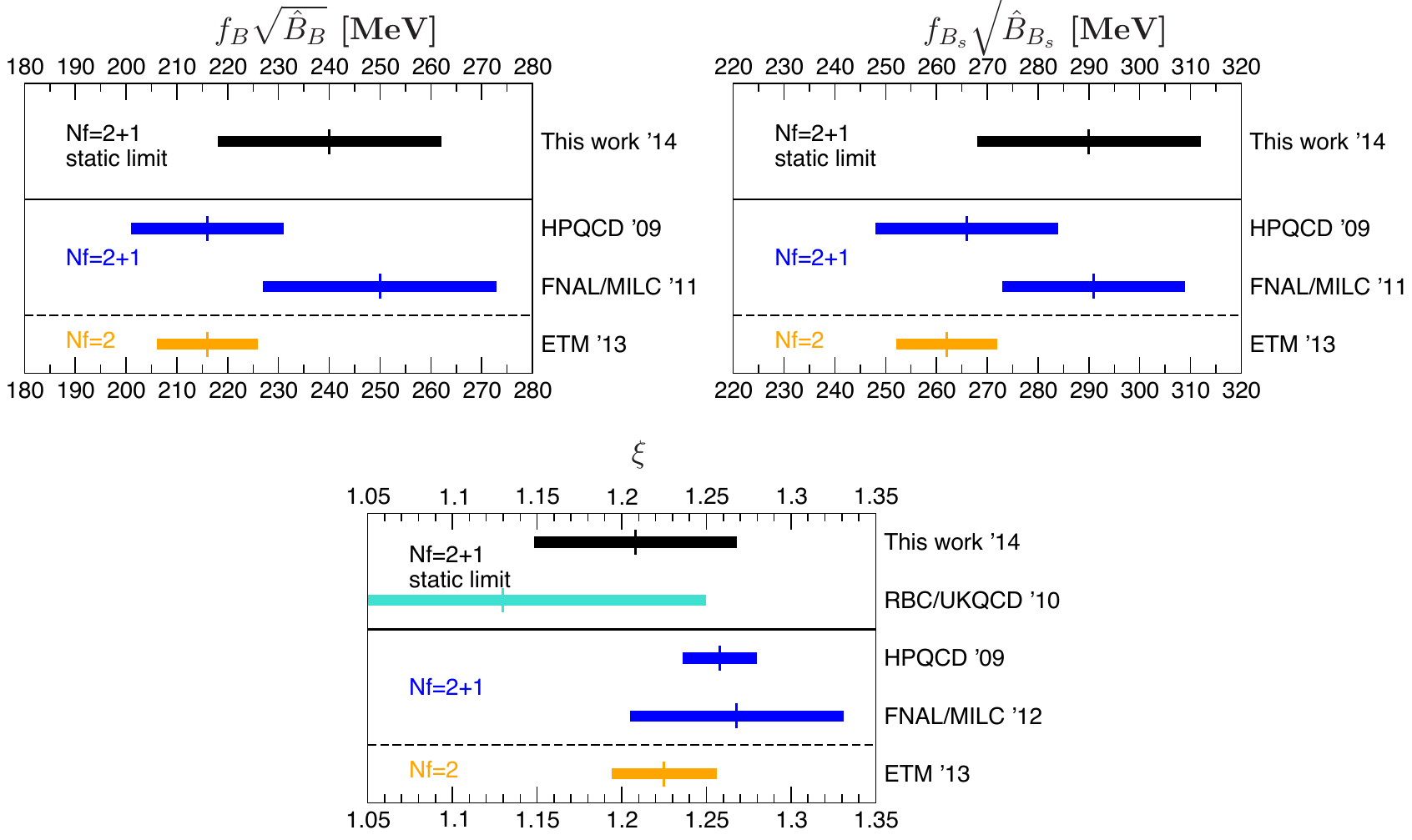}
\end{center}
\caption{
Comparison of $f_B\sqrt{\hat{B}_B}$, $f_{B_s}\sqrt{\hat{B}_{B_s}}$
and $\xi$ with other works.
The data is cited from Refs.~\cite{Gamiz:2009ku} (HPQCD '09),
\cite{Bouchard:2011xj} (FNAL/MILC '11),
\cite{Bazavov:2012zs} (FNAL/MILC '12),
\cite{Carrasco:2013zta} (ETM '13) and
\cite{Albertus:2010nm} (RBC/UKQCD '10).
The RGI value of $f_B\sqrt{\hat{B}_B}$ and $f_{B_s}\sqrt{\hat{B}_{B_s}}$
in FNAL/MILC '11 are obtained by converting
$f_B\sqrt{B_B}$ and $f_{B_s}\sqrt{B_{B_s}}$ at $\mu=m_b$ in
Ref.~\cite{Bouchard:2011xj} with the two-loop multiplicative factor $1.516$.
Errors for the static limit results do not contain $1/m_b$ uncertainty.}
\label{FIG:comparison_matrix_elements}
\end{figure*}

\begin{figure*}
\begin{center}
\includegraphics[scale=1.00, viewport = 0 -5 500 270, clip]
{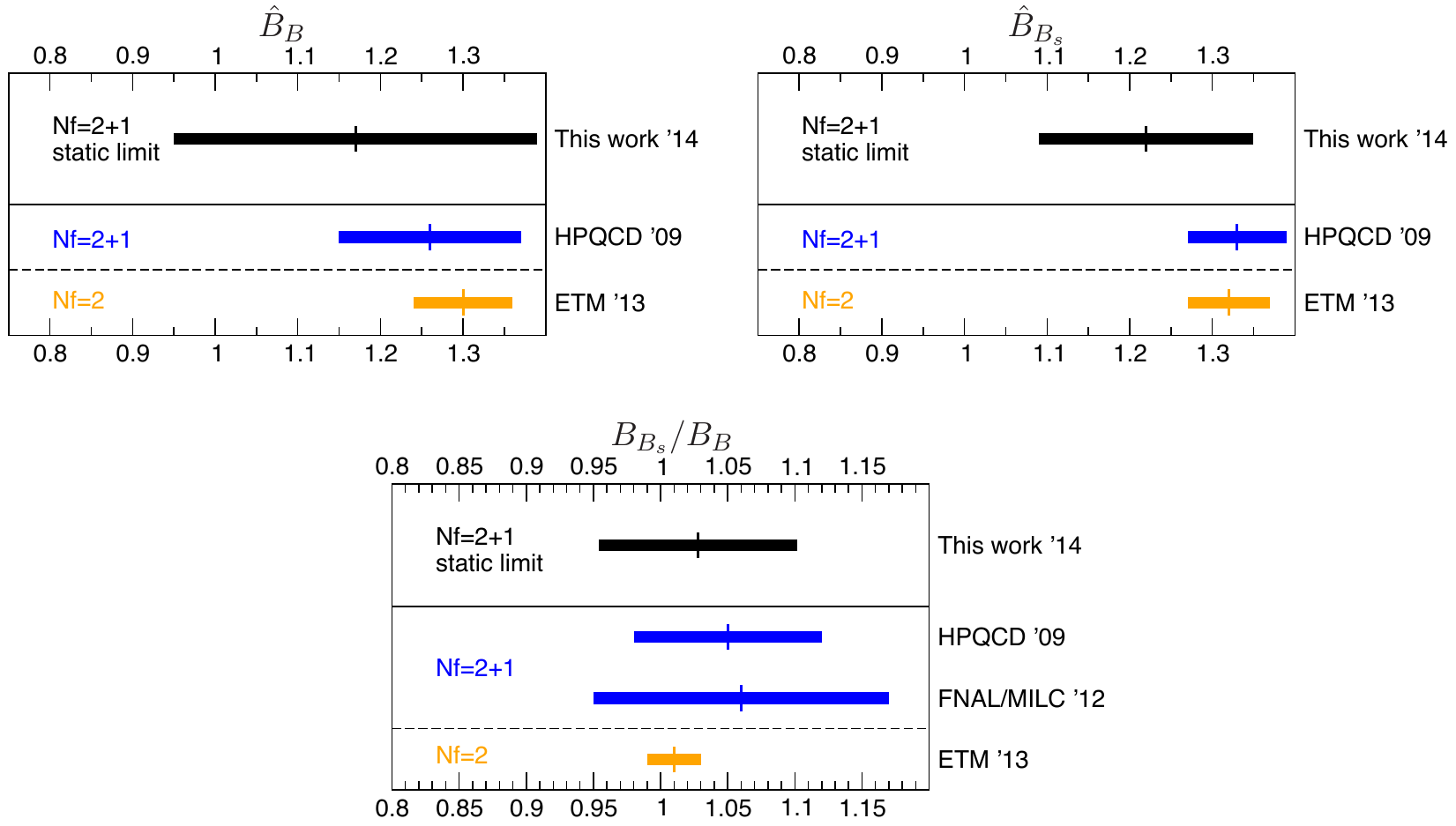}
\end{center}
\caption{Comparison of $\hat{B}_B$, $\hat{B}_{B_s}$ and $B_{B_s}/B_B$
with other works.
The data is cited from
Refs.~\cite{Gamiz:2009ku} (HPQCD '09),
\cite{Carrasco:2013zta} (ETM '13) and
\cite{Bazavov:2012zs} (FNAL/MILC '12).
Errors for the static limit results do not contain $1/m_b$ uncertainty.}
\label{FIG:comparison_B_parameters}
\end{figure*}

We show comparison with other works for our obtained quantities
in Figs.~\ref{FIG:comparison_decay_constants},
\ref{FIG:comparison_matrix_elements} and \ref{FIG:comparison_B_parameters}.
(See also Review of lattice results by Flavor Lattice Averaging Group (FLAG)
~\cite{Aoki:2013ldr}.)
Our results have $\sim10\%$ larger value for decay constants $f_B$ and $f_{B_s}$
from other works at physical $b$ quark mass point,
which would be plausibly understood by the static approximation ambiguity.
The ETM Collaboration's results at the static limit
in Ref.~\cite{Dimopoulos:2011gx} also shows this tendency.
However, ALPHA Collaboration's results on $f_B$ and $f_{B_s}$
in the static limit indicate much smaller deviation from those at physical
$b$ quark mass point~\cite{Bernardoni:2014fva}.
We cannot conclude the reason of this difference from us,
because our current uncertainty is still large.
On the other hand, there is no clear difference from
the physical $b$ quark point in $f_B\sqrt{\hat{B}_B}$,
$f_{B_s}\sqrt{\hat{B}_{B_s}}$, $\hat{B}_B$ and $\hat{B}_{B_s}$,
because of the large error.
For the SU(3) breaking ratios, significant deviation from others is not seen,
since the static approximation uncertainty is largely reduced
by SU(3) light flavor symmetry factor as described
in Eq.~(\ref{EQ:one_ov_mb_for_ratio}).

\begin{figure*}
\begin{center}
\includegraphics[scale=1.00, viewport = 0 -5 250 140, clip]
{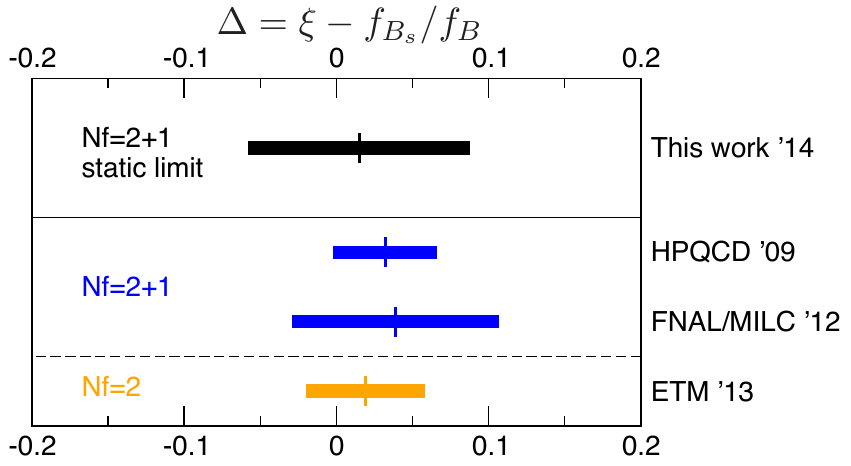}
\end{center}
\caption{
Comparison of $\Delta=\xi-f_{B_s}/f_B$ with other works.
The data is cited from Refs.~\cite{Gamiz:2009ku} (HPQCD '09),
\cite{Bazavov:2012zs} (FNAL/MILC '12) and \cite{Carrasco:2013zta} (ETM '13).
In calculating $\Delta$, correlations between $\xi$ and $f_{B_s}/f_B$
are not taken into account.}
\label{FIG:comparison_Delta}
\end{figure*}

Finally, it would be interesting to see a correspondence
between $\xi$ and $f_{B_s}/f_B$.
In this study we obtained the difference:
\begin{eqnarray}
\Delta=\xi-\frac{f_{B_s}}{f_B}=0.015(73),
\end{eqnarray}
where correlation between $\xi$ and $f_{B_s}/f_B$ is omitted.
As mentioned in Sec.~\ref{SEC:observables_QCD_full},
naive factorization suggests $\xi$ is close to $f_{B_s}/f_B$, and
our result supports this observation in the static limit of $b$ quark.
In Fig.~\ref{FIG:comparison_Delta} we show $\Delta$ in other works
together with our results.
No discrepancy between $\xi$ and $f_{B_s}/f_B$ beyond $1\sigma$ error
has yet been seen.

\subsection{Further improvements for next step}

Although the obtained results in this work is encouraging,
there exist limitations due to insufficient statistics and various
systematic errors.
As the current error budget in Fig.~\ref{FIG:error_budget} shows,
dominant uncertainties are statistical error, chiral extrapolation
and uncertainty from renormalization.
To overcome the current situation, possible options are following.

\begin{list}{}{\itemindent=9mm \leftmargin=5mm}
\item[\bf All-Mode-Averaging (AMA):]
Currently, our results have large statistical error and
chiral extrapolation is suffering from the lack of statistics.
Gluon link smearings in the static action help to improve signal qualities
to some extent, the statistical error is, however, not enough small.
All-mode-averaging (AMA) technique~\cite{Blum:2012uh} provides
a substantial computational cost reduction, which leads to 
the improved statistics.
In the AMA, a bunch of source points are put to increase statistics keeping
computational cost small by using a conjugate gradient (CG) solver
with relaxed convergence conditions.
\item[\bf Physical light quark mass point simulation:]
The lightest pion mass in this paper is $\sim 290$ [MeV],
which leaves large uncertainty from the chiral extrapolation.
This error would be significantly reduced by the physical point
simulation, where the simulated pion mass is $\sim 135$ [MeV].
The $2+1$ flavor dynamical ensembles are being generated by RBC/UKQCD
Collaborations using M\"{o}bius DWF~\cite{Brower:2004xi},
keeping almost the same lattice spacings as those in this work,
but with doubled physical volume~\cite{Blum:2014wsa}.
It would increase computational cost by a large amount,
hence the AMA technique mentioned above is crucial.
\item[\bf Non-perturbative renormalization:]
While one-loop renormalization uncertainty is $0\%$ or quite small for
SU(3) breaking ratios, it is estimated to be, at the most, $6\%$
for non-ratio quantities.
Non-perturbative renormalization is, hence, required for the non-ratio
quantities to reduce the large uncertainty.
The renormalization would be accomplished using
the momentum-subtraction (RI/MOM) scheme
\cite{Martinelli:1993dq, Martinelli:1994ty},
in which an additional renormalization condition is required to manage
the $1/a$ power divergence.
\end{list}

These programs are non-trivial but promising directions.
Part of them are currently on-going~\cite{Ishikawa:2013faa}
and we plan to present more definite results
on this project in near future.

\begin{acknowledgements}
We thank members of RBC/UKQCD Collaborations, especially Oliver Witzel 
for useful discussions. 
The calculations reported here were performed on the QCDOC computers of
RIKEN-BNL Research Center and the USQCD Collaboration at
Brookhaven National Laboratory (BNL),
RIKEN Integrated Cluster of Clusters (RICC) at RIKEN, Wako,
KMI computer $\varphi$ at Nagoya University
and resources provided by the USQCD Collaboration
funded by the U.S. Department of Energy.
The software used includes the CPS QCD codes
(http://qcdoc.phys.columbia.edu/cps.html),
supported in part by the U.S. DOE SciDAC program.
This work is supported in past by JSPS Kakenhi grant
Nos.~21540289 and 22224003 (Y.~A.).
T.~Izubuchi, C.~L. and A.~S. were supported in part by
U.S. DOE contract DE-AC02-98CH10886 and
T.~Izubuchi also by JSPS Grants 22540301, 23105715 and 26400261.
\end{acknowledgements}

\appendix
\section{Effective mass and correlator plots}
\label{APP:Effective_mass}

\begin{figure*}
\begin{center}
\includegraphics[scale=0.93, viewport = 0 0 510 670, clip]
{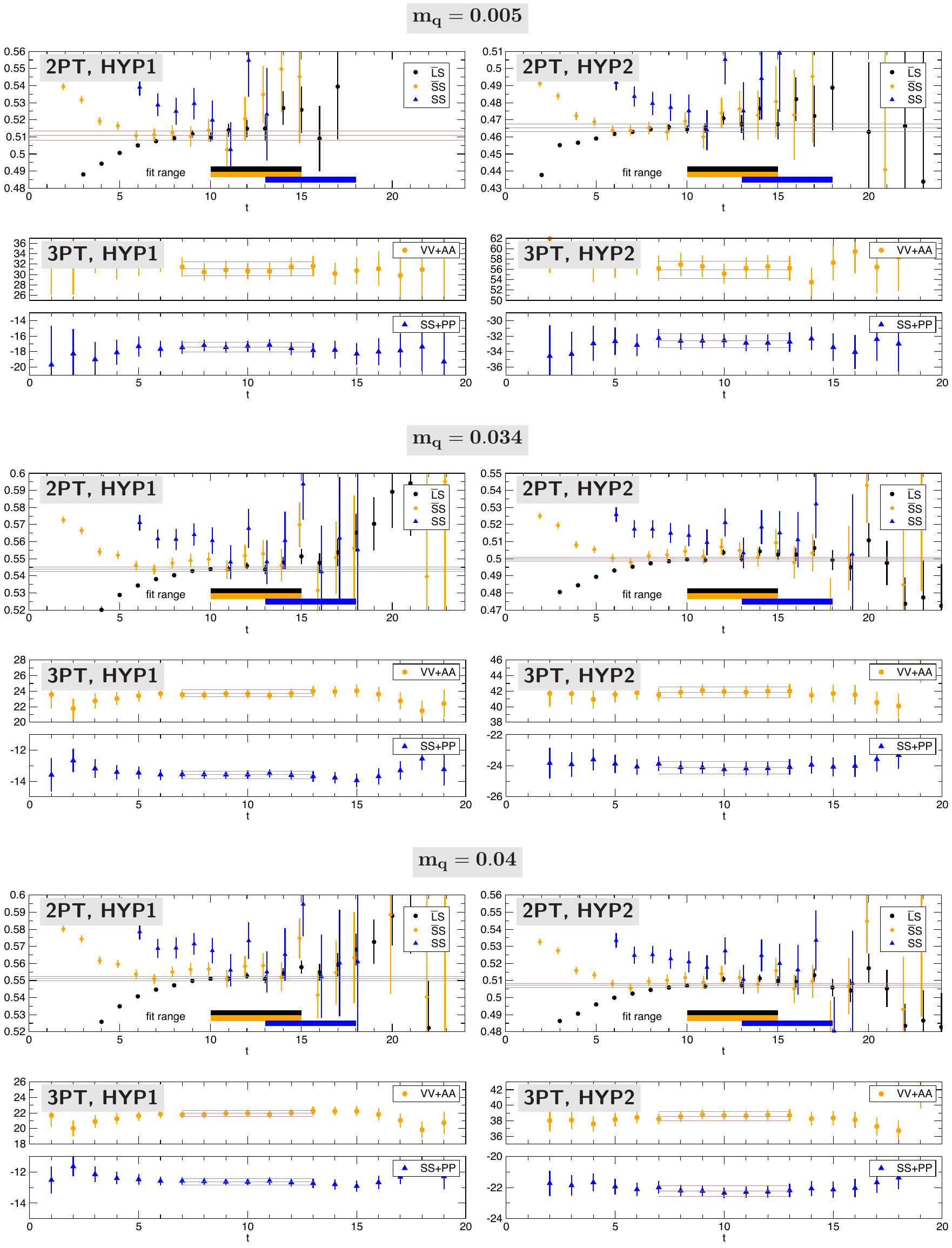}
\caption{Effective mass (two-point function) and three-point function
plot for 24c1.
The figures show
$E_{\rm eff}=-\ln(C^{\rm XX}(t+1, 0)/C^{\rm XX}(t, 0))$
with ${\rm XX}=(\tilde{L}S, \tilde{S}S, SS)$ for 2PT,
$C_L^{SS}(t_f, t, 0)$ for 3PT VV+AA and
$C_S^{SS}(t_f, t, 0)$ for 3PT SS+PP.
Fit ranges and fit results are shown in the figures.
For three-point functions $t_f$ is fixed to be $20$.}
\label{FIG:effective_mass_plot_24c1}
\end{center}
\end{figure*}

\begin{figure*}
\begin{center}
\includegraphics[scale=0.93, viewport = 0 0 510 670, clip]
{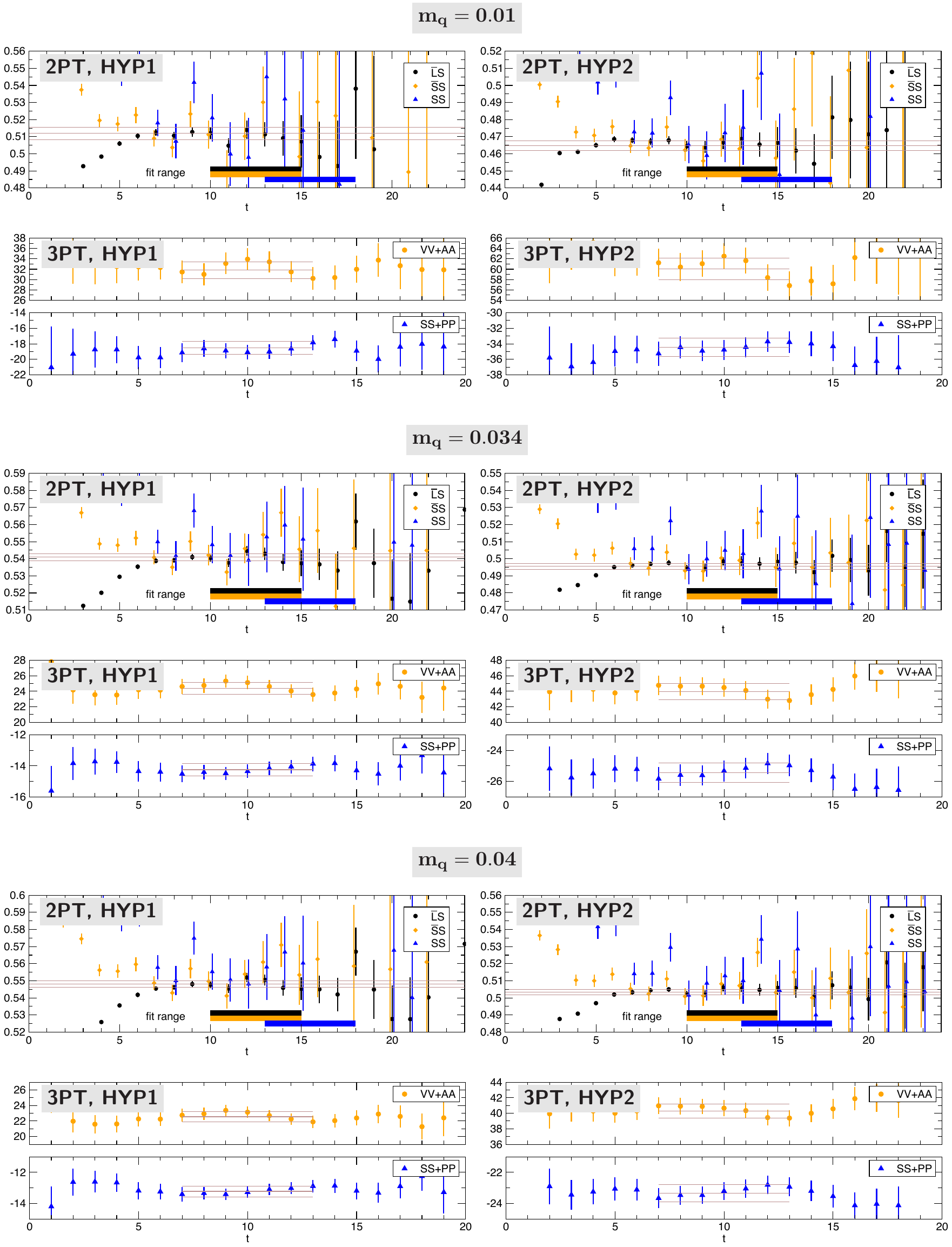}
\caption{Effective mass (two-point function) and three-point function
 plot for 24c2.
The figures show
$E_{\rm eff}=-\ln(C^{\rm XX}(t+1, 0)/C^{\rm XX}(t, 0))$
with ${\rm XX}=(\tilde{L}S, \tilde{S}S, SS)$ for 2PT,
$C_L^{SS}(t_f, t, 0)$ for 3PT VV+AA and
$C_S^{SS}(t_f, t, 0)$ for 3PT SS+PP.
Fit ranges and fit results are shown in the figures.
For three-point functions $t_f$ is fixed to be $20$.}
\label{FIG:effective_mass_plot_24c2}
\end{center}
\end{figure*}

\begin{figure*}
\begin{center}
\includegraphics[scale=0.93, viewport = 0 0 510 670, clip]
{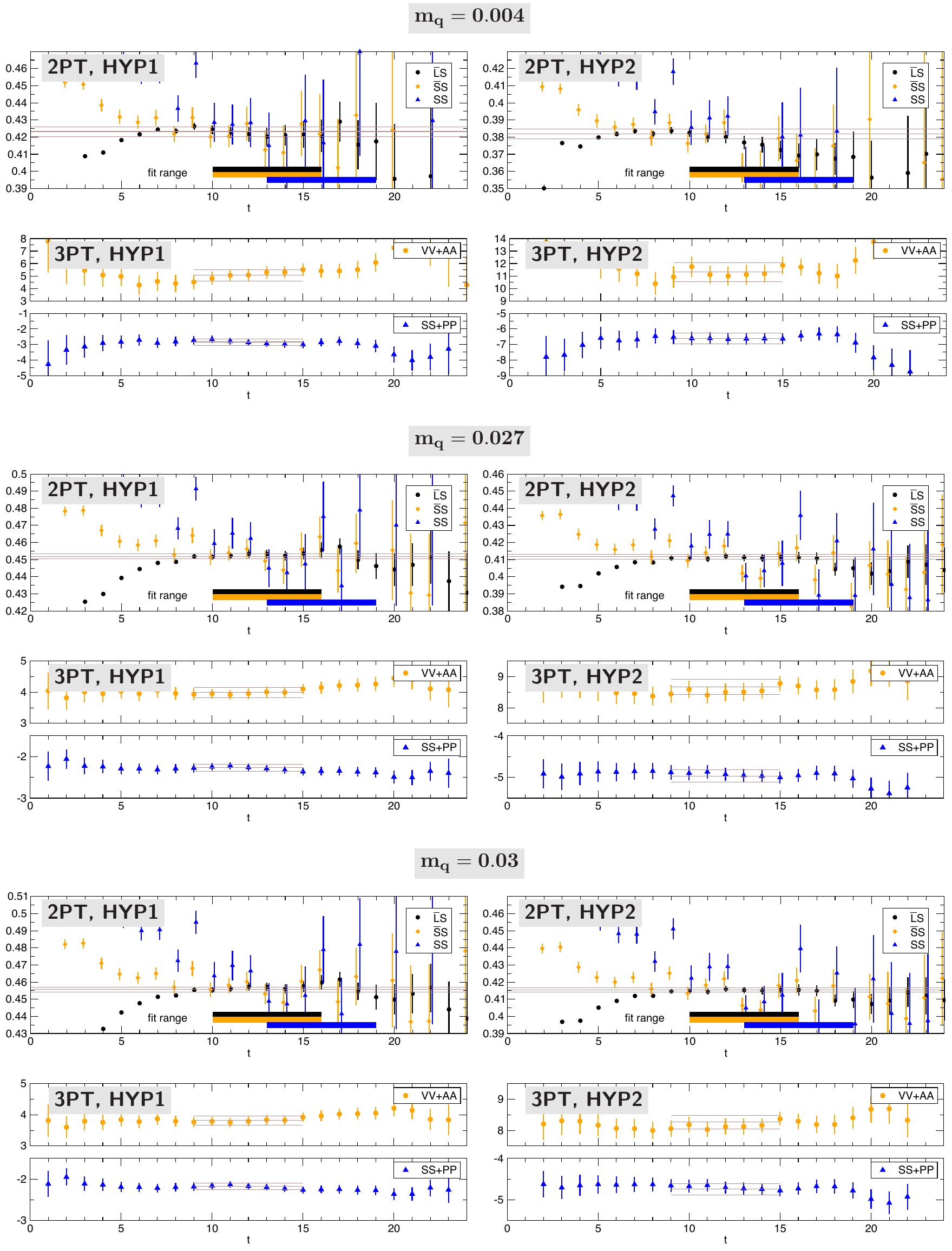}
\caption{Effective mass (two-point function) and three-point function
 plot for 32c1.
The figures show
$E_{\rm eff}=-\ln(C^{\rm XX}(t+1, 0)/C^{\rm XX}(t, 0))$
with ${\rm XX}=(\tilde{L}S, \tilde{S}S, SS)$ for 2PT,
$C_L^{SS}(t_f, t, 0)$ for 3PT VV+AA and
$C_S^{SS}(t_f, t, 0)$ for 3PT SS+PP.
Fit ranges and fit results are shown in the figures.
For three-point functions $t_f$ is fixed to be $24$.}
\label{FIG:effective_mass_plot_32c1}
\end{center}
\end{figure*}

\begin{figure*}
\begin{center}
\includegraphics[scale=0.93, viewport = 0 0 510 670, clip]
{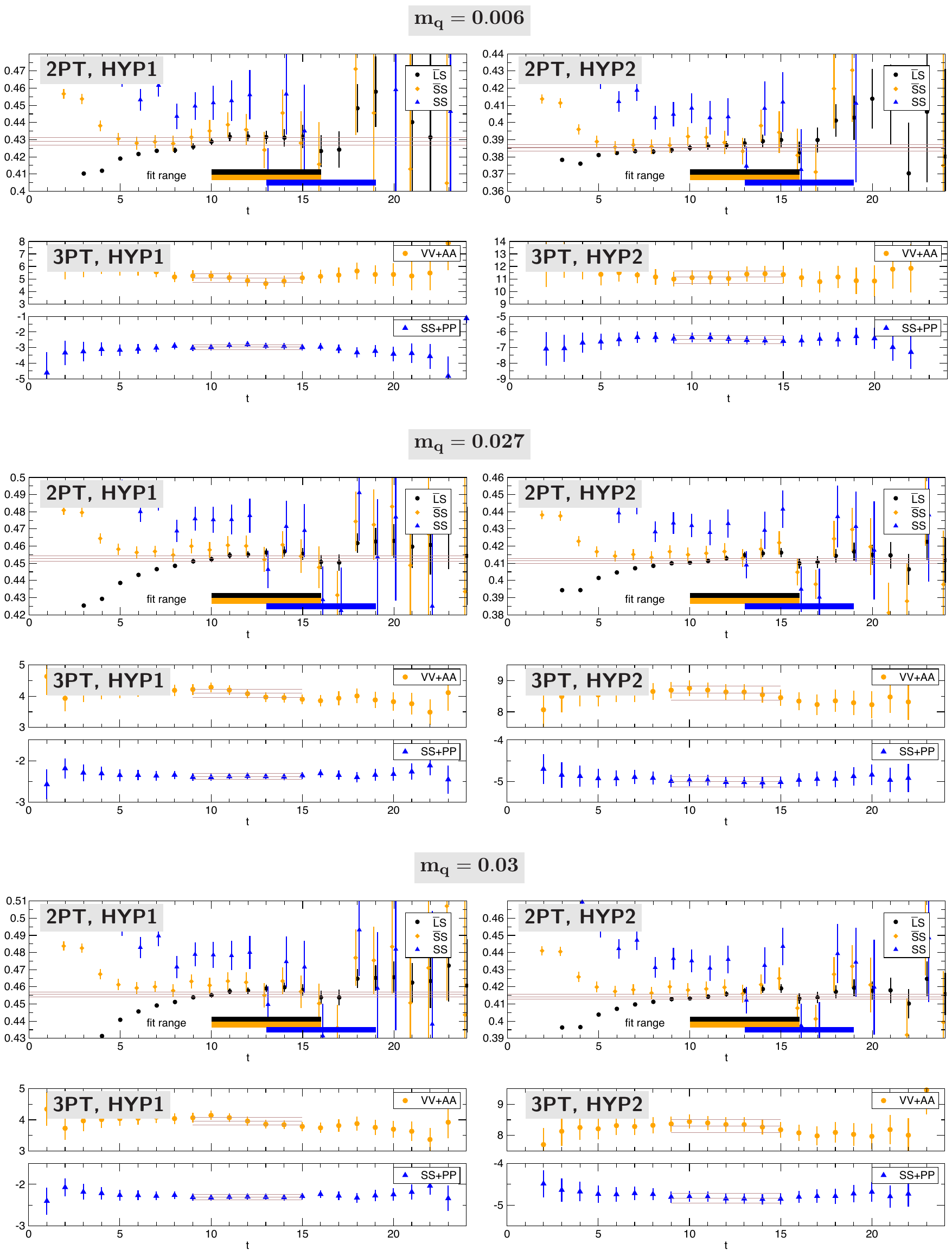}
\caption{Effective mass (two-point function) and three-point function
 plot for 32c2.
The figures show
$E_{\rm eff}=-\ln(C^{\rm XX}(t+1, 0)/C^{\rm XX}(t, 0))$
with ${\rm XX}=(\tilde{L}S, \tilde{S}S, SS)$ for 2PT,
$C_L^{SS}(t_f, t, 0)$ for 3PT VV+AA and
$C_S^{SS}(t_f, t, 0)$ for 3PT SS+PP.
Fit ranges and fit results are shown in the figures.
For three-point functions $t_f$ is fixed to be $24$.}
\label{FIG:effective_mass_plot_32c2}
\end{center}
\end{figure*}

\begin{figure*}
\begin{center}
\includegraphics[scale=0.93, viewport = 0 0 510 670, clip]
{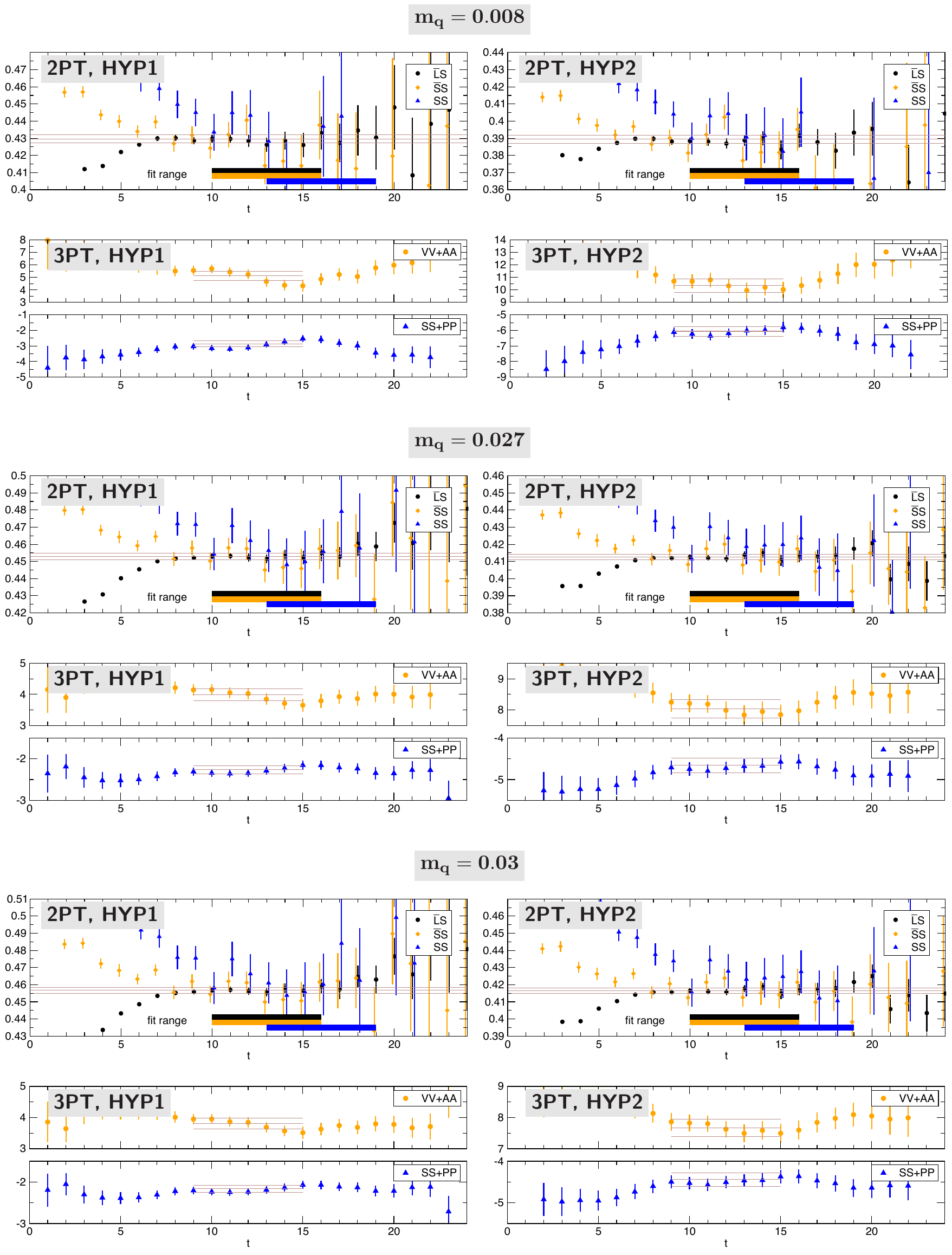}
\caption{Effective mass (two-point function) and three-point function
 plot for 32c3.
The figures show
$E_{\rm eff}=-\ln(C^{\rm XX}(t+1, 0)/C^{\rm XX}(t, 0))$
with ${\rm XX}=(\tilde{L}S, \tilde{S}S, SS)$ for 2PT,
$C_L^{SS}(t_f, t, 0)$ for 3PT VV+AA and
$C_S^{SS}(t_f, t, 0)$ for 3PT SS+PP.
Fit ranges and fit results are shown in the figures.
For three-point functions $t_f$ is fixed to be $24$.}
\label{FIG:effective_mass_plot_32c3}
\end{center}
\end{figure*}

Figs.~\ref{FIG:effective_mass_plot_24c1}--\ref{FIG:effective_mass_plot_32c3}
show effective mass plots in two-point function and three-point function plots.
The fit ranges and fit results are included in the figures.

\section{Fit range dependence}
\label{APP:Fit-range-dependence}

\begin{figure*}
\begin{center}
\includegraphics[scale=1.00, viewport = 0 0 450 590, clip]
{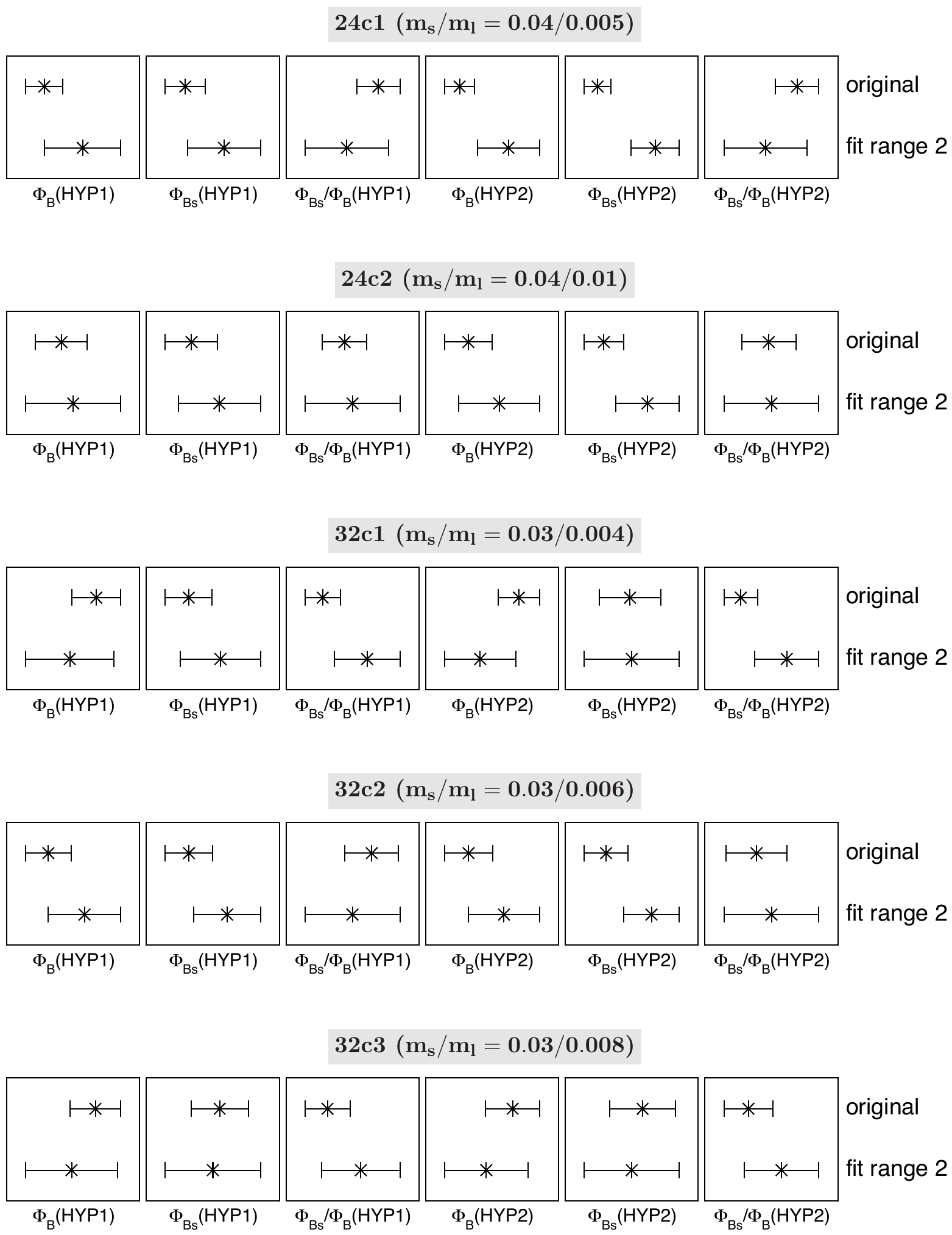}
\caption{Fit range dependence of $\Phi_B,$, $\Phi_{B_s}$ and
$\Phi_{B_s}/\Phi_B$ at each simulation point.
Horizontal labels are suppressed.
We find differences between fit range choices beyond $1$-$\sigma$
statistical error in
24c1($\Phi_B$, $\Phi_{B_s}$),
24c2($\Phi_{B_s}$),
32c1($\Phi_B$,  $\Phi_{B_s}/\Phi_B$) and
32c2($\Phi_{B_s}$).}
\label{FIG:fit_range_dependence_PB}
\end{center}
\end{figure*}

\begin{figure*}
\begin{center}
\includegraphics[scale=1.00, viewport = 0 0 450 590, clip]
{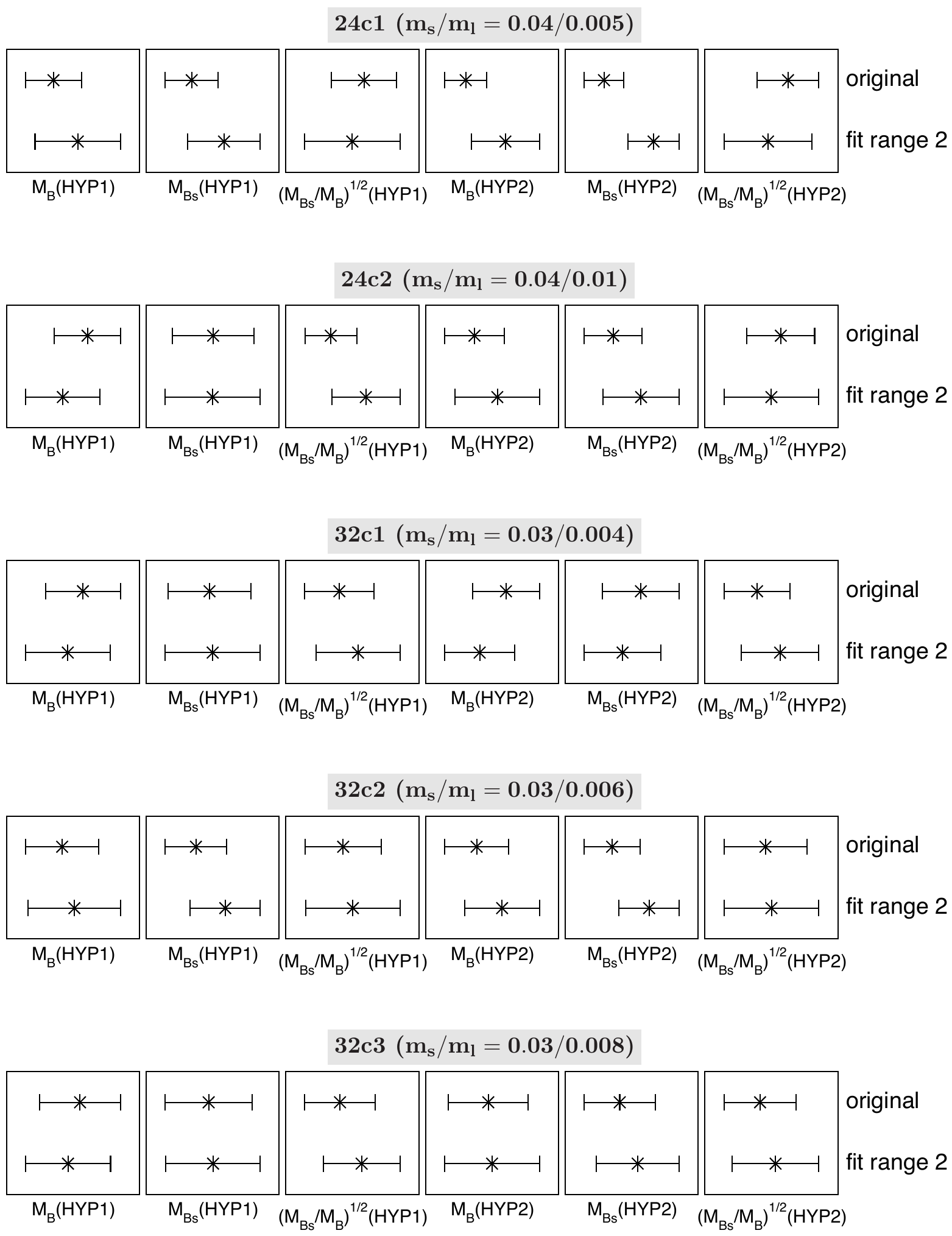}
\caption{Fit range dependence of $M_B,$, $M_{B_s}$ and
$(M_{B_s}/M_B)^{1/2}$ at each simulation point.
Horizontal labels are suppressed.
We find differences between fit range choices beyond $1$-$\sigma$
statistical error in
24c1($M_B$, $M_{B_s}$),
24c2($(M_{B_s}/M_B)^{1/2}$) and
32c2($M_{B_s}$).}
\label{FIG:fit_range_dependence_MB}
\end{center}
\end{figure*}

\begin{figure*}
\begin{center}
\includegraphics[scale=1.00, viewport = 0 0 450 590, clip]
{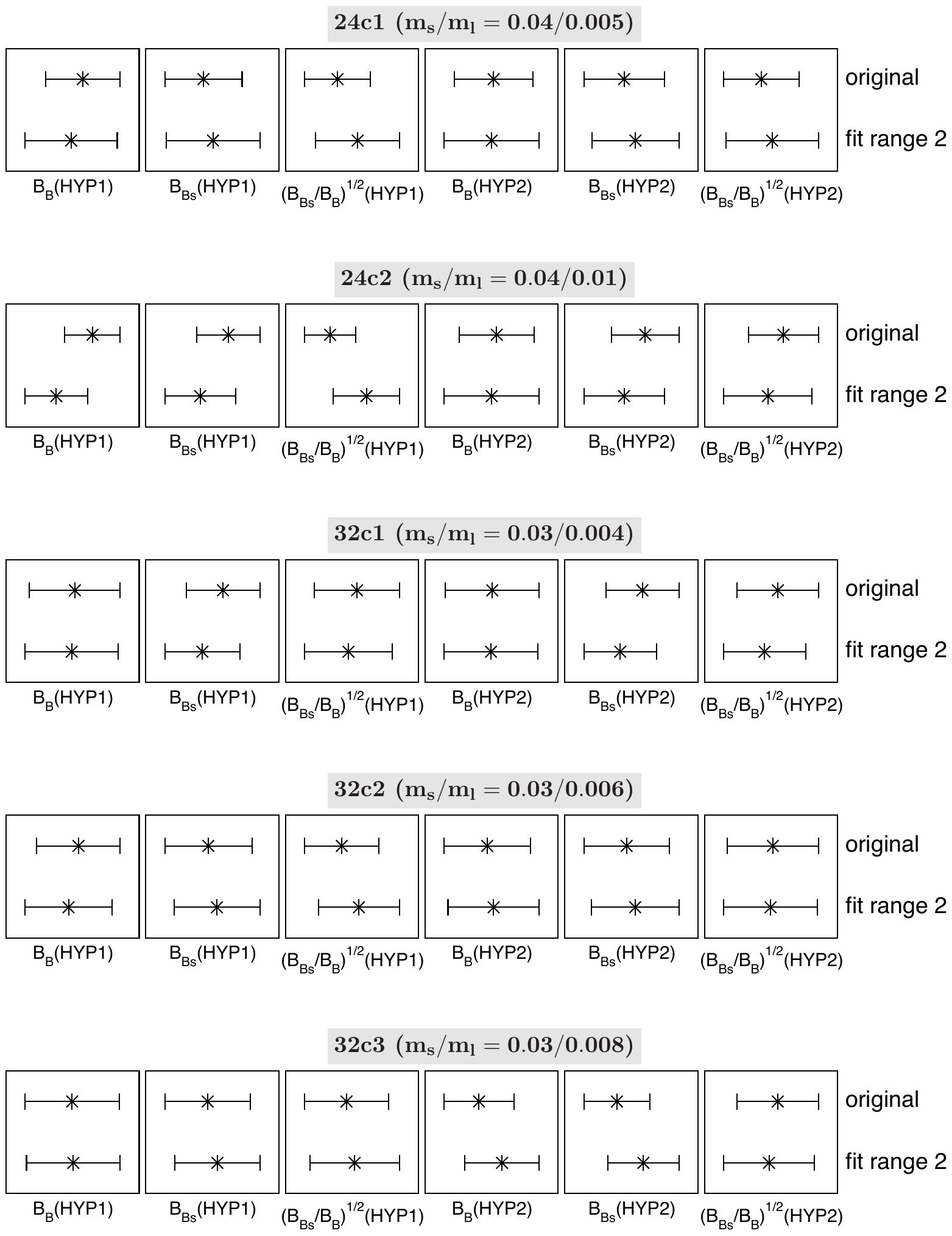}
\caption{Fit range dependence of $B_B,$, $B_{B_s}$ and
$(B_{B_s}/B_B)^{1/2}$ at each simulation point.
Horizontal labels are suppressed.
We find differences between fit range choices beyond $1$-$\sigma$
statistical error in
24c2($B_B$, $(B_{B_s}/B_B)^{1/2}$).}
\label{FIG:fit_range_dependence_BB}
\end{center}
\end{figure*}

We show fit range dependences of physical quantities at each simulation
point in Figs.~\ref{FIG:fit_range_dependence_PB},
\ref{FIG:fit_range_dependence_MB} and \ref{FIG:fit_range_dependence_BB}.
To check the dependences, we shift the minimal value of $t$ in the fit
range toward larger value by $2$ for two-point functions and shorten
the range by $2$ for three-point functions, which we name
``fit range 2'' in the figures.
To be more specific, the actual fit ranges are:
\begin{eqnarray}
 {\rm original~(24c)}&:&t=10-15~({\rm\overline{L}S}, ~{\rm\overline{S}S}), 
                        t=13-18~({\rm SS}),\nonumber\\
                      &&t=7-13~({\rm VV+AA}, {\rm SS+PP}),\nonumber\\
 {\rm original~(32c)}&:&t=10-16~({\rm\overline{L}S}, ~{\rm\overline{S}S}), 
                        t=13-19~({\rm SS}),\nonumber\\
                      &&t=9-15~({\rm VV+AA}, {\rm SS+PP}),\nonumber\\
 {\rm fit~range~2~(24c)}&:&t=12-15~({\rm\overline{L}S}, ~{\rm\overline{S}S}), 
                           t=15-18~({\rm SS}),\nonumber\\
                         &&t=8-12~({\rm VV+AA}, {\rm SS+PP}),\nonumber\\
 {\rm fit~range~2~(32c)}&:&t=12-16~({\rm\overline{L}S}, ~{\rm\overline{S}S}), 
                           t=15-19~({\rm SS}),\nonumber\\
                         &&t=10-14~({\rm VV+AA}, {\rm SS+PP}).\nonumber
\end{eqnarray}
We find disagreements between choices of fit range beyond $1$-$\sigma$
statistical error for some cases.

\end{document}